\def\mycmd{2}   
\if\mycmd1  
\documentclass[12pt, journal, compsoc, onecolumn]{IEEEtran}
\else 
\documentclass[10pt,journal,compsoc]{IEEEtran}
\IEEEoverridecommandlockouts
\fi         
\ifCLASSINFOpdf
\else
\fi

\usepackage{multicol}
\usepackage{makecell}

\usepackage{amsmath,amsfonts}
\usepackage{cases}
\usepackage{empheq}
\usepackage{algorithm}
\usepackage{algpseudocode}
\usepackage{array}
\usepackage{booktabs}
\usepackage[caption=false,font=normalsize,labelfont=sf,textfont=sf]{subfig}
\usepackage{textcomp}
\usepackage{placeins}

\usepackage{soul}
\usepackage{url}
\usepackage{verbatim}
\usepackage{graphicx}
\usepackage{cite}
\usepackage[x11names]{xcolor} 
\usepackage{mathtools}
\usepackage{enumitem}
\usepackage{multicol,multirow}
\usepackage{adjustbox}

\usepackage{tikz}
\usetikzlibrary {arrows.meta}
\newcounter{problem}
\newcounter{save@equation}
\newcounter{save@problem}

\makeatletter
\newenvironment{problem}
 {\setcounter{problem}{\value{save@problem}}%
  \setcounter{save@equation}{\value{equation}}%
  \let\c@equation\c@problem
  \renewcommand{\theequation}{$\mathcal{P}_\arabic{equation}$}%
  \subequations
  }
 {\endsubequations
  \setcounter{save@problem}{\value{equation}}%
  \setcounter{equation}{\value{save@equation}}%
 }

\DeclareMathOperator*{\argmin}{argmin}

\usepackage{titlesec}
\titleformat{\paragraph}[runin]
  {\normalfont\normalsize\bfseries}{}{0em}{}
\titlespacing{\paragraph}
  {0pt}{0.5ex plus 0.5ex minus .2ex}{1em}

\newtheorem{theorem}{Theorem}
\newtheorem{proposition}{Proposition}

\newtheorem{remark}{Remark}

\newcommand{\blue}[1]{\textcolor{blue}{#1}}
\usepackage[framemethod=tikz]{mdframed}
\usepackage{soul}
\usepackage{color}

\soulregister\cite7
\soulregister\ref7
\soulregister\eqref7
\soulregister\pageref7

\usepackage{pifont,xspace}
\newcommand{\cmark}{\ding{51}\xspace}%
\newcommand{\xmarkg}{\textcolor{lightgray}{\ding{55}}\xspace}%

\begin{document}

\newcommand{\mytitle}{Distributed Task Offloading and Resource Allocation for Latency Minimization in Mobile Edge Computing Networks}

\title{\mytitle}

\newcommand{\myauthor}{Minwoo Kim, Jonggyu Jang,~\IEEEmembership{Member,~IEEE}, Youngchol Choi, and Hyun Jong Yang,~\IEEEmembership{Member,~IEEE}}

\author{\myauthor
\thanks{
This research was supported in part by Korea Institute of Marine Science \& Technology Promotion(KIMST) funded by the Ministry of Oceans and Fisheries, Korea (RS-2022-KS221606, Development of underwater space resource creation technologies), in part by the National Research Foundation of Korea (NRF) grant funded by the Korea government (MSIT) (No. RS-2023-00250191), and in part by the MSIT(Ministry of Science and ICT), Korea, under the ITRC(Information Technology Research Center) support program(IITP-2024-2021-0-02048) supervised by the IITP(Institute for Information \& Communications Technology Planning \& Evaluation).
}
\thanks{M. Kim and J. Jang are with the Department of Electrical Engineering, Pohang University of Science and Technology (POSTECH), 
(email: \{mwkim0210,jgjang\}@postech.ac.kr).
Y. Choi is with the Korea Research Institute of Ships and Ocean Engineering (KRISO), (email: ycchoi@kriso.re.kr).
H. J. Yang is with the Department of Electrical and Computer Engineering, Seoul National University, (email: hjyang@snu.ac.kr).
\textit{Minwoo Kim and Jonggyu Jang equally contributed.
(The corresponding author is Hyun Jong Yang.)}
}
}



\IEEEtitleabstractindextext{%
\begin{abstract}
The growth in artificial intelligence (AI) technology has attracted substantial interests in latency-aware task offloading of mobile edge computing (MEC)---namely, minimizing service latency.  
Additionally, the use of MEC systems poses an additional problem arising from limited battery resources of MDs.
This paper tackles the pressing challenge of latency-aware distributed task offloading optimization, where user association (UA), resource allocation (RA), full-task offloading, and battery of mobile devices (MDs) are jointly considered. 
In existing studies, joint optimization of overall task offloading and UA is seldom considered due to the complexity of combinatorial optimization problems, and in cases where it is considered, linear objective functions such as power consumption are adopted.
Revolutionizing the realm of MEC, our objective includes all major components contributing to users' quality of experience, including latency and energy consumption.
To achieve this, we first formulate an NP-hard combinatorial problem, where the objective function comprises three elements: communication latency, computation latency, and battery usage.
We derive a closed-form RA solution of the problem; next, we provide a distributed pricing-based UA solution.
We simulate the proposed algorithm for various resource-intensive tasks.
Our numerical results show that the proposed method Pareto-dominates baseline methods.
More specifically, the results demonstrate that the proposed method can outperform baseline methods by \textbf{1.62 times shorter latency} with \textbf{41.2\% less energy consumption}.
\end{abstract}

\begin{IEEEkeywords}
Latency minimization, delay minimization, mobile edge computing, resource allocation, user association, task offloading, energy efficiency, edge AI.
\end{IEEEkeywords}
}

\maketitle

\section{Introduction}  \label{sec:introduction}

\IEEEPARstart{A}{S} artificial intelligence (AI) technology advances and support for such technology in mobile devices (MDs) continues to grow, the demand for resource-intensive tasks in edge computing is anticipated to surge.
We have already experienced the impact of emerging applications with heavy computational load such as voice-activated virtual assistants, generative model-based applications, and autonomous driving systems.
{These} applications' high demand for computing power and energy poses a significant challenge for MDs, as they are limited in processing power and battery life.
A viable solution to this problem is the implementation of \textit{task offloading} using mobile edge computing (MEC), which entails shifting local tasks to a remote edge server.
While this approach lightens the computational burden on MDs, it increases the communication demands of the network. 
For edge servers, this results in an overall increase in both computational and communication loads, leading to potential service delays.
To minimize the {latency} while maintaining sustainability of MEC system devices, it is essential to jointly consider communication, computation resources, and energy consumption.
Addressing these concerns, especially in systems heavily reliant on AI, introduces a set of novel challenges to be tackled in optimizing task offloading.

\begin{table*}
    \centering
    \caption{Comparison of the Proposed Method and Previous Works  \label{tab:related_works}}
\adjustbox{width=1\linewidth}{
\begin{tabular}{llcccccccccccccc}
\toprule
\multicolumn{2}{c}{\multirow{2}{*}{\textbf{Novelty and Related Works}}} & \multirow{2}{*}{\textbf{Ours}} & {\cite{10243579}} & \cite{10024305} & \cite{9712216} & \cite{9806318} & \cite{10102429} &  \cite{9714257} & \cite{9745305} & \cite{9496271} & \cite{9712282} & \cite{9678008} & \cite{9214878} & \cite{9525179} & \cite{9087909}  \\
& & & {('23)} & ('23) & ('23) & ('23) & ('23) &  ('23) & ('22) & ('22) & ('22) & ('22) & ('22) & ('21) & ('20) \\ 
\midrule
\multirow{6}{*}{\textbf{Architecture}} & Multi-Cell Networks & \cmark & {\cmark} & \cmark & \xmarkg & \cmark & \cmark & \xmarkg & \xmarkg & \xmarkg & \cmark & \cmark & \xmarkg & \xmarkg & \cmark \\ \cmidrule{2-16} 
\multicolumn{1}{l}{} & Multi-User Networks  & \cmark & {\cmark} & \cmark & \cmark & \cmark & \cmark & \cmark & \cmark & \cmark & \cmark & \cmark & \cmark & \cmark & \cmark \\ \cmidrule{2-16}
\multicolumn{1}{l}{} & User Association     & \cmark & {\xmarkg} & \xmarkg & \xmarkg & \cmark & \cmark & \xmarkg & \xmarkg & \xmarkg & \xmarkg & \cmark & \xmarkg & \xmarkg & \cmark \\ \cmidrule{2-16}
\multicolumn{1}{l}{} & Resource Allocation  & \cmark & {\cmark} & \cmark & \cmark & \cmark & \cmark & \cmark & \cmark & \cmark & \cmark & \cmark & \cmark & \cmark & \cmark  \\ 
\cmidrule{2-16}
\multicolumn{1}{l}{} & Full Offloading      & \cmark & {\xmarkg} & \xmarkg & \cmark & \xmarkg & \xmarkg & \xmarkg & \cmark & \cmark & \xmarkg & \xmarkg & \cmark & \xmarkg & \cmark  \\ 
\cmidrule{2-16}
\multicolumn{1}{l}{} & \makecell[{{>{}p{3.2cm}}}]{{Distributed optimization (multi-cell)}} & {\cmark} & {\xmarkg} & {\xmarkg} & {-} & {\xmarkg} & {\xmarkg} & {-} & {-} & {-} & {\cmark} & {\xmarkg} & {-} & {-} & {\xmarkg}  \\ 
\midrule
\multirow{2}{*}{\textbf{Objective}} & Energy Efficiency & \cmark & {\xmarkg} & \cmark & \cmark & \cmark & \cmark & \cmark & \xmarkg & \xmarkg & \cmark & \xmarkg & \xmarkg & \cmark & -  \\ \cmidrule{2-16}
& Delay Minimization   & \cmark & {\cmark} & \cmark & \cmark & \cmark & \cmark & \cmark & \cmark & \cmark & \cmark & \cmark & \cmark & \cmark & -  \\ \cmidrule{1-16}
\multirow{3}{*}{\textbf{Approach}} & Optimization & \cmark & {\cmark} & \xmarkg & \xmarkg & \cmark & \xmarkg & \cmark & \cmark & \cmark & \cmark & \xmarkg & \xmarkg & \cmark & - \\ \cmidrule{2-16}
& Lyapunov             & \xmarkg & {\xmarkg} & \xmarkg & \cmark & \cmark & \cmark & \xmarkg & \xmarkg & \xmarkg & \xmarkg & \xmarkg & \xmarkg & \xmarkg & \cmark   \\ \cmidrule{2-16}
& Deep Learning & \xmarkg & {\xmarkg} & \cmark & \xmarkg & \xmarkg & \cmark & \xmarkg & \xmarkg & \xmarkg & \xmarkg & \cmark & \cmark & \xmarkg & -  \\ \midrule
\multirow{2}{*}{\textbf{Performance Analysis}} & Latency Minimization & \cmark & {\cmark} & \cmark & \cmark & \xmarkg & \cmark & \cmark & \cmark & \xmarkg & \cmark & \cmark & \cmark & \xmarkg & -  \\ \cmidrule{2-16}
& Energy Consumption   & \cmark & {\xmarkg} & \cmark & \cmark & \cmark & \cmark & \xmarkg & \xmarkg & \xmarkg & \cmark & \xmarkg & \xmarkg & \cmark & -  \\ \midrule
\multirow{2}{*}{\textbf{Theoretical Analysis}} & Convergence Analysis & \cmark & {\xmarkg} & \cmark & \cmark & \xmarkg & \cmark & \cmark & \xmarkg & \xmarkg & \cmark & \cmark & \cmark & \cmark & \cmark  \\ \cmidrule{2-16}
& Upperbound Analysis  & \cmark & {\xmarkg} & \xmarkg & \xmarkg & \xmarkg & \cmark & \xmarkg & \xmarkg & \xmarkg & \xmarkg & \xmarkg & \xmarkg & \xmarkg & \cmark  \\ \bottomrule
\end{tabular}
}
\end{table*}

\subsection{Challenges of Task Offloading in MEC Systems}  \label{subsec:challenges_of_task_offloading}

\paragraph{1) The need for full offloading of AI tasks.}
In general, as mentioned in \cite{6497017}, full offloading problems can be relaxed to partial offloading problems, which simplifies discrete variables into differentiable continuous variables.
Thus, in many previous studies, partial offloading techniques are widely used to solve complex problems \cite{8789664, 10024305, 9806318}. 
However, since most AI tasks are unsuitable for partial offloading, MEC optimization for AI task offloading requires hard decisions rather than soft decisions. 
This constraint is a challenge when applying AI task offloading into complex scenarios.


\paragraph{2) Joint optimization of user association and resource allocation.}

The process of offloading local tasks in a MEC network is analogous to the conventional user association and resource allocation (UARA) problem commonly encountered in wireless networks. 
{In typical wireless networks, }a user equipment (UE) connects to a base station (BS), and resources such as bandwidth are allocated accordingly. 
{Similarly, in MEC systems, each MD associates with an edge server (ES) to receive resource allocations.}
{However, MEC systems need to manage both computational and communication resources, where the delays arise from both computation and communication stages. }
{What makes the problem worse is the diversity in computational capabilities of both local and edge devices, such as variations in floating point operations per second (FLOPS), VRAM capacity, and the number of available cores at the ESs.}
{Additionally, centralized optimization, which has been adopted in most existing works, introduces further delays due to the communication overhead that occurs during information exchange.}
Therefore, a {distributed} approach to {joint optimization of} user association (UA) and resource allocation (RA) is necessary to effectively address the implications of these dual factors.


\paragraph{3) Trade-off between {latency} and energy consumption.}

In a MEC system, local devices have the option to offload tasks to an edge server or process them locally. 
In scenarios where the edge server is heavily loaded, local computation can be advantageous in terms of latency, despite its higher energy consumption.
In such situations, it is essential to prioritize task offloading based on the remaining battery life to maintain network continuity. 
It becomes necessary to understand the relationship between the degree of prioritization and the resulting latency versus local energy consumption. 
One of the key challenges is to find a Pareto-optimal solution that balances these factors effectively.

Motivated by these complexities in MEC systems, we identify several crucial questions: 
\begin{itemize}
    \item \textit{How can the remaining battery life of mobile devices be factored into the strategy for task offloading?}
    \item \textit{What constitutes the most effective optimization strategy for distributed UARA that takes into account both communication and computation latency?}
\end{itemize}

Building on these questions, our paper focuses on the specific constraints inherent in MEC systems and formulates an optimization problem for UARA. 
We develop a novel optimization-based solution to address this challenge. 
To validate the effectiveness of our solution, we conduct simulations under conditions that closely mimic real-world scenarios. 
The results of these simulations show that our model outperforms other baseline models in terms of latency and energy efficiency, highlighting its potential applicability in practical MEC environments.

\subsection{Summary of Our Contributions} 
 \label{subsec:introduction_summary_of_our_contributions}

To the best of the authors' knowledge, our work represents the first attempt at optimizing the joint problem of UARA for communication and computation resources in the context of {\textit{distributed}} task offloading.
{
Previous works' contributions and their limitations can be summarized as follows:}
\begin{itemize}
    \item {As in Table \ref{tab:related_works}, topics regarding UARA in MEC networks have been studied intensively~\cite{10243579,10024305,9712216,9806318, 10102429,9714257,9745305,9496271,9712282,9678008,9214878,9525179,9087909}. Amongst them, only a few works have addressed UARA for full offloading in MEC networks~\cite{9712216, 9745305, 9496271, 9214878, 9087909}, and even fewer research have jointly considered energy efficiency in MEC networks~\cite{9712216, 9214878}.}
    \item {More importantly, distributed optimization, which requires extremely minimal information exchange for control information, is challenging for reducing communication overhead. Despite its importance, only one work~\cite{9712282} in Table \ref{tab:related_works} has considered distributed optimization, which does not consider full-offloading.}
    \item {One might assume that partial offloading is a more general form of full offloading, similar to how continuous variables encompass discrete variables. However, at least from an optimization standpoint, continuous-valued variables are often easier to manage because most problems can be solved using gradient-based algorithms. As a result, previous studies have tackled the full-offloading problem by relaxing it into a more tractable form with continuous-valued variables \cite{9712216}, which has loss of generality.}
\end{itemize}
{We further highlight previous studies and their contributions in Section \ref{sec:related_works}.}
Building upon these foundations, the contributions of our work are multifaceted and can be summarized as follows:
\begin{itemize}
    \item We unveil a groundbreaking task offloading strategy, optimization UARA in a novel approach. 
    This strategy not only i) prioritizes full-task offloading with a keen eye on the MD's battery level but also ii) features a fully distributed optimization of load balancing.
    \item {For the sake of full-task offloading, the formulated optimiation problem is a mixed integer problem.}
    We take on the challenge of the mixed-integer combinatorial optimization problem, which has not been solved in previous studies. 
    To solve the problem, we leverage a novel auxiliary variable-based relaxation technique that enables us to balance loads by a continuous variable, not a combinatorial one. 
    The solution we derive facilitates distributed optimization while simultaneously lowering the execution complexity.
    \item We provide a thorough analysis of both \textit{the convergence and the optimality} of the distributed optimization problem, offering insights into the efficacy of our method. 
    \item {As well as theoretical analysis, our distributed optimization method is scalable for increasing network size (number of edge devices or servers), since the computational complexity at each MD and ES is $\mathcal{O}(N_\textrm{ES})$ and $\mathcal{O}(N_\textrm{MD})$, respectively. Furthermore, the total communication overhead is \textit{linearly}-proportional to the number of edge servers. }
    \item Through experimental results, we demonstrate that the proposed scheme consumes at least 39.1\% less energy, while achieving a reduction in latency by at least 1.98 times.
    Best results in each objective show 5.29 times reduction in latency and 64.0\% reduction in energy consumption. {Also, the duality gap analysis shows that the proposed method can closely achieve its latency lower bound.}
    These findings are supported by simulations in practical environments, incorporating various AI services and load scenarios.
\end{itemize}


\begin{figure}[t]
    \centering
    \includegraphics[width=\if\mycmd1 0.6\else0.9\fi\linewidth]{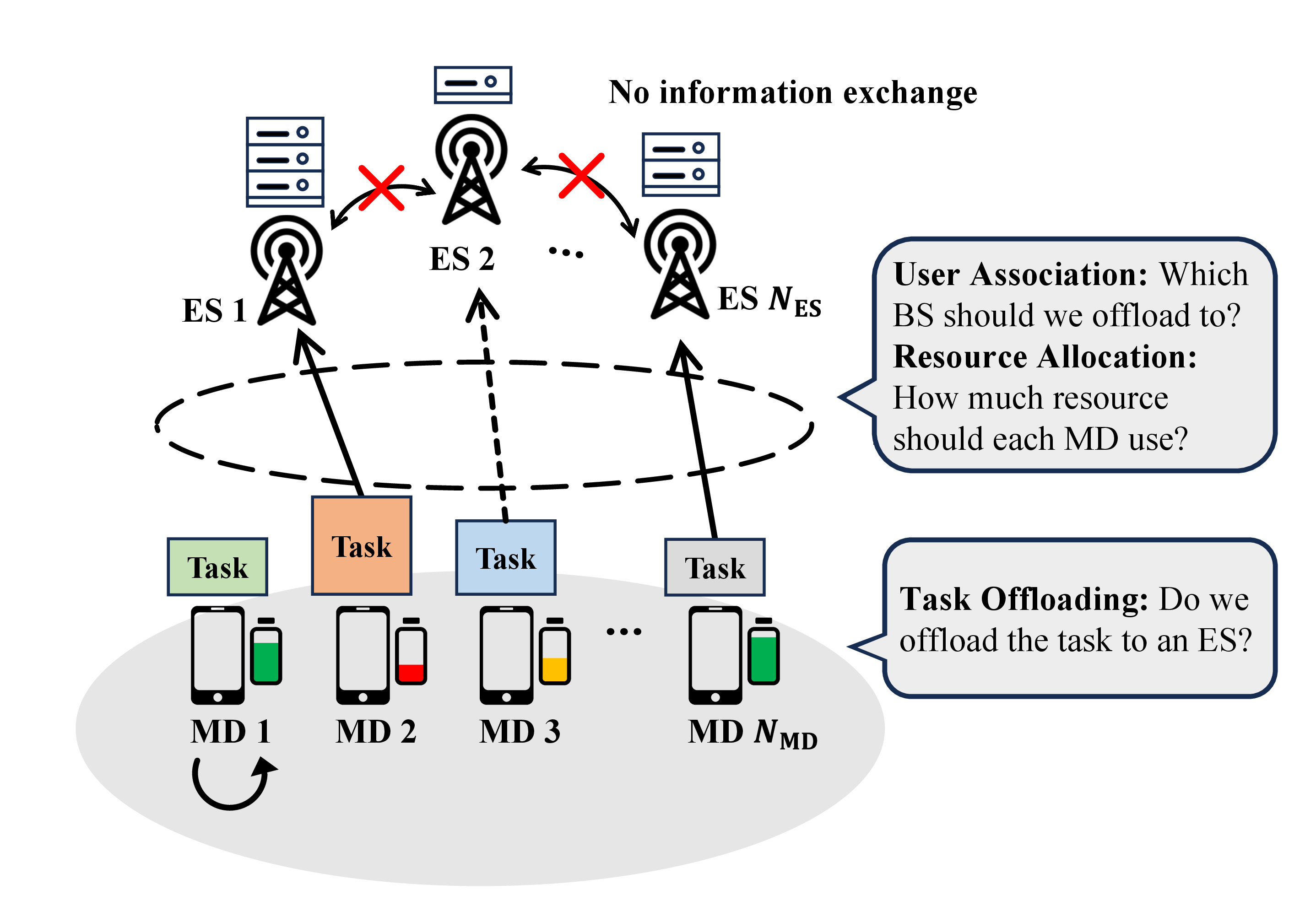}
    \caption{An illustration of the system model. $N_{\text{MD}}$ MDs with limited battery and heterogeneous tasks can associate with $N_{\text{ES}}$ ESs. Information exchange between the ESs is unnecessary.}
    \label{fig:system_model}
\end{figure}

\section{System Model and Problem Formulation} \label{sec:system_model}
In this section, we first introduce the system model overview embracing communication/computation modeling.
Then, we provide a joint UARA optimization problem for prioritized task offloading.

\subsection{System Model Overview}  \label{subsec:system_model_overview}

Figure \ref{fig:system_model} illustrates a MEC network composed of $N_{\text{ES}}$ edge servers and $N_{\text{MD}}$ MDs, where each of the edge servers has its capacity-limited computing hardware. 
The total computation capacities of MD $i$ and ES $j$ are denoted as $F_{\text{MD},i}$ and $F_{\text{ES},j}$ respectively.
We assume that each MD always has a task at hand, creating a task whenever the previous task is finished.
The generated service at MD $i$ has a computation load of $f_i$ flops and a communication load of $d_i$ bits. 
For readability, we summarize the variables used in this paper in Table \ref{tab:notations}.

In the edge server layer, each ES broadcasts its price to users in a decentralized manner. 
Based on the received prices, each MD aggregates load information from base stations to decide whether to compute locally or offload its task to a remote edge server. 
In this context, edge servers allocate computing and communication resources to minimize the delay experienced by the users they are servicing.



\paragraph{Assumptions.}
In our system model, we have postulated the following assumptions to better emulate the real-world scenario and to ease the deployment of our proposed method. 
\begin{enumerate}[label=\textit{A\arabic*})]
    \item 
    We assume that the time intervals are small enough that the channel differences between adjacent time steps are negligible.
    \item Based on the 3GPP small cell Scenario 1 \cite{3gpp.36.932}, we assume all of the ESs share the cell areas, thereby cooperating to serve users efficiently.
    \item We assume that the frequency reuse factor is $N_{\text{ES}}$, i.e., all the ESs occupy separate bandwidth.
    \begin{itemize}
        \item As our objective is to optimize offloading over the long term, we can leverage the previous time step's channel interference to facilitate dynamic offloading and make full use of bandwidth resources.
    \end{itemize}
    \item For brevity of the notation, we omit the time step index in this paper.
    The index is negligible since our focus is on long-term optimization of network load balancing.
\end{enumerate}

\subsection{Task Offloading Variables and Constraints}

In a MEC system, once a task is generated at an MD, the MD makes its offloading decision. 
Let $x_{ij}$ denote the offloading decision of MD $i$ regarding ES $j$. 
The association of MD $i$ and ES $j$ is represented by 

\begin{equation}
    x_{ij}=\begin{cases}
        1, & \text{ES $j$ servers MD $i$,}\\
        0, & \text{otherwise.}
    \end{cases}\label{x_def}
\end{equation} 

For readability, we define an augmented matrix of $x_{ij}$ as $\mathbf{X} \in \{0, 1\}^{N_{\text{MD}}\times N_{\text{ES}}}$, where $[\mathbf{X}]_{i, j}=x_{ij}$.
Since AI task cannot be partially offloaded, at most 1 ES can be associated to an MD.
Hence, the UA indication variable $x_{ij}$ is constrained as follows:
\begin{equation}
    \sum_{k\in\mathcal{E}} x_{ik}\leq 1 \text{ and } x_{ij}\in \{0, 1\},\; \forall i \in \mathcal{M},\; j \in \mathcal{E}. \label{eq:x_constraint}
\end{equation}
If $x_{ij}=0$ for all $j\in\mathcal{E}$, MD $i$ processes its task locally.

\subsection{Communication Model}

In this paper, we assume a flat power spectral density (PSD), i.e., the transmission power is uniform across all frequency resources for every MDs.
The MDs transmit the task data with full transmission power, where the transmission power of MD $i$ is denoted by $P_{i}$.
To mitigate the inter-cell interference, the bandwidth is equally divided and occupied by different edge servers.
Let us define the bandwidth of the network as $W$.
Then, the total bandwidth of ES $j$ is expressed as $W_j=W/B$, for all $j\in\mathcal{E}$. 
The allocated bandwidth is then divided and allocated to the associated MDs.
The fraction of the bandwidth allocated from ES $j$ to MD $i$ is denoted by $y_{ij}$, i.e., the bandwidth allocated to MD $i$ can be represented by $\sum_{j\in\mathcal{E}}y_{ij}W_j$.

To simplify our notation, we augment the variables $y_{ij}$ to create $\mathbf{Y}$.
Since the bandwidth allocated to each MD cannot overlap, the variable $\mathbf{Y}$ is constrained by

\begin{equation}
    \sum_{k\in\mathcal{M}} y_{kj}\leq 1 \text{ and } y_{ij}\in [0, 1],\; \forall i \in \mathcal{M},\; j \in \mathcal{E}.  \label{eq:y_constraint}
\end{equation}

\paragraph{Capacity Model.}

Let $h_{ij}$ denote the channel coefficient between MD $i$ and ES $j$. 
Again, the allocated bandwidth of each MD does not overlap. 
We can therefore define the signal-to-noise-ratio (SNR) of MD $i$ served by ES $j$ as $\text{SNR}_{ij}=\frac{|h_{ij}|^2 P_{i}}{N_0}$. 

\begin{remark}[Interference model]
Here, we operate under the assumption that the ES divides the bandwidth equally, which may not be optimal in terms of frequency efficiency. 
Nevertheless, by finely segmenting the time slots as previously designed, the interference state from the previous time slot approximates that of the current time slot. 
This allows for a distributed formulation of the Signal-to-Interference-plus-Noise Ratio (SINR) model, where the ESs share the total bandwidth.
\end{remark}

Then, the achievable rate of MD $i$ associated to an ES is defined by 
\begin{equation}\label{eq:data_rate}
    \sum_{j\in\mathcal{E}}x_{ij}y_{ij}R_{ij},
\end{equation}
where the spectral efficiency $R_{ij}$ is defined as
\begin{equation}
    W_j\log(1+\text{SNR}_{ij}).
\end{equation}

\paragraph{Latency Injected from Communication.}

The latency is an important measure for many AI tasks, especially when they require real-time computation.
In our scenario, the Latency is incurred by 2 major components: communication delay and computation delay.
Here, we examine the communication latency model.
If an MD decides to process its task locally, there exists no communication delay.
On the other hand, once an MD decides to offload a task to ES $j$, $d_i$ bits of data should be transferred to ES $j$ via a wireless channel before the service is processed. 

Using the achievable rate definition in \eqref{eq:data_rate}, we have the communication delay of MD $i$ associated to ES $j$ as follows:
\begin{equation}
     \sum_{j\in\mathcal{E}} \frac{d_ix_{ij}}{R_{ij}  y_{ij}+\epsilon} = \sum_{j\in\mathcal{E}}x_{ij}D_{\text{cm},ij}, \label{eq:comm_delay}
\end{equation}
where an arbitrarily small constant $\epsilon$ is added to the denominator as a regularization technique to prevent potential computational issues associated with division by zero.

\subsection{Computing Model}  \label{subsec:computing_model}

Here, we introduce our computing model and provide computing delay of local computation and edge computation. 
In general, local devices such as mobile devices and laptops have fewer computation cores compared to the ES (GPU server). 
For a more practical modeling of computing delay, we divide the computation of {resource-intensive} tasks into parallelizable and non-parallelizable parts.
Since our objective is to minimize end-to-end delay, if parallelizable portion is high, offloading can lead to higher delay reduction due to abundant number of processors at the edge server.
To model this delay model, we adopt Amdahl's law.
Amdahl's law describes the relationship between the latency of a task and the number of available processors, accounting for the portion of the task where parallel computing is feasible:
\begin{equation}\label{eq:amdahl}
    \textbf{Amdahl's law: } S_{\text{latency}}(s)=\frac{1}{(1-\rho)+\frac{\rho}{s}},
\end{equation}
where $S_{\text{latency}}$ signifies the acceleration in completing the entire task, $\rho$ represents the parallelizable fraction of the original task, and $s$ denotes the achievable acceleration of the parallelizable proportion.

\paragraph{Computing Latency at MDs.}

If MD $i$ has a computational task at hand, it can fully leverage its computing resource to the task. 
The total number of computing cores at MD $i$ is represented as $Z_{\text{MD}_i}$, and each of the cores has a computing capacity of $F_{\text{MD}_i}$. 
From our assumption, up to one task is processed at an MD.
Thus, the computing delay of MD $i$ with task that requires $f_i$ flops and has a parallelizable fraction of $\rho_i$ is represented by
\begin{equation}
\begin{aligned}
    D_{\text{MD},i}  & = \frac{f_i}{F_{\text{MD}_i}}(1-\rho_i) + \frac{f_i}{F_{\text{MD}_i}Z_{\text{MD}_i}}(\rho_i),  \label{eq:computing_delay}
\end{aligned}
\end{equation}
where the first term represents the delay obtained from the core-independent part of the task, and the second term represents the delay deriving from the parallelizable portion of the task.

\paragraph{Computing Latency at Edge Servers.}

The most significant difference in computation delay between local computing and task offloading is on shared computing resources. 
More simply, in each ES, the tasks offloaded from its associated MDs are simultaneously processed. 
Let $Z_{\text{ES},j}$ denote the total number of cores available at ES $j$. 
Then, if ES $j$ allocates a fraction $z_{ij}$ of computing cores to $i$, the task generated at MD $i$ is computed with $Z_{\text{ES},j}z_{ij}$ cores. 
Thus, the computing delay of the task generated at MD $i$ offloaded to ES $j$ can be represented by 
\begin{align}
    & \sum_{j\in\mathcal{E}} \left( \frac{f_i}{F_{\text{ES},j} + \epsilon } (1-\rho_i)+\frac{f_i}{F_{\text{ES},j} Z_{\text{ES},j}z_{ij}+ \epsilon}\rho_i \right)x_{ij}  \label{eq:computing_delay_2}\\
    = & \sum_{j\in\mathcal{E}}x_{ij} \left( D_{\text{ES}(s),ij} +  D_{\text{ES}(p),ij} \right)  \label{eq:comput_delay_ES}\\
    = & \sum_{j\in\mathcal{E}}x_{ij} D_{\text{ES},ij}.  \label{eq:computing_delay_4}
\end{align}
Similar to equation \eqref{eq:comm_delay}, a very small constant $\epsilon$ is added to the denominator to prevent indeterminacy occurring from division by zero.
Also, the computing resource variable $z_{ij}$ is constrained by 
\begin{equation}
    \sum_{k\in\mathcal{M}} z_{kj}\leq1 ~\text{ and }~ z_{ij}\in [0, 1],\; \forall i \in \mathcal{M},\; j \in \mathcal{E}.  \label{eq:z_constraint}
\end{equation}

\paragraph{Memory Requirement.}

Another factor to consider is the amount of memory the task consumes.
The memory the tasks use should not exceed the VRAM of the edge device. 
If we denote the proportion of the memory used to load model parameter as $M_j$, the constraints imposed on the variable can be expressed as
\begin{equation}
    \sum_{k\in\mathcal{M}} m_{i}x_{kj}\leq 1-M_j,\; \forall i \in \mathcal{M},\; j \in \mathcal{E},  \label{eq:memory_constraint}
\end{equation}
where $m_{i}$ is the linearly-increasing memory consumption term.

\begin{remark}
    When assessing the memory usage of AI models, both the number of parameters of the model and the number of activations should be examined.
    Here, we employ a model size computation method introduced in \cite{korthikanti2023reducing} to calculate the memory size of Transformer-based models.
    The memory calculating equation is as follows:
    \begin{equation}
        \text{Memory}=p\times (sbhL(10+\frac{24}{t}+5\frac{as}{ht}) + |\theta|),
    \end{equation}
    where $p$ is the precision in bytes, $s$ is the sequence length, $b$ is the batch size, $h$ is the size of the hidden dimension, $L$ is the number of transformer layers, $t$ is the tensor parallel size, $a$ is the number of attention heads, and $|\theta|$ is the number of parameters of the model.
    We use this formula to assess the memory size of the AI models based on Vision Transformer (ViT).    
\end{remark}

\subsection{Energy Consumption Model}

\paragraph{Energy Consumption at MDs.}

Most MDs have limited battery capacity while some user devices may be powered and therefore does not have finite battery capacity.
To take this scenario into account and generalize our scenario, we compose our user set $\mathcal{M}$ with both type of users, one with limited battery and the other with infinite energy.

In MDs, energy is expended on communication and computation. 
If a task is not offloaded, the device energy is spent on computation only. 
However, if offloading occurs, energy is expended on the transmission of radio waves at a specific frequency. 
We first look at the case of local computation. 
Let us define the computing energy efficiency as $\delta_{\text{cp,}i}$ (Watt/flops). 
Then, the computing energy of MD $i$ per unit time is represented by 
\begin{equation}
    E_{\text{cp},i} =\delta_{\text{cp,}i} \cdot(F_{\text{MD}_i}(1-\rho_i) + F_{\text{MD}_i}Z_{\text{MD}_i}\rho_i)(1-\sum_{j\in\mathcal{E}}x_{ij}).  \label{eq:comp_energy}
\end{equation}
Next, defining the power efficiency of transmission of MD $i$ as $\delta_\text{cm,i}$, the energy consumption of MD $i$ connected to ES $j$ for offloading can be expressed as follows:
\begin{equation}
    E_{\text{cm,}i}= \delta_\text{cm,i}\sum_{j\in\mathcal{E}}\left(x_{ij} y_{ij} W_j \right) P_{i}.  \label{eq:comm_energy}
\end{equation}

Thus, the total energy consumption at MD $i$ is denoted by 
\begin{equation}
    E_{\text{tot}, i} = E_{\text{cp},i} + E_{\text{cm},i} + E_{b,i},
\end{equation}
where $E_{b,i}$ denotes the base energy consumption required to operate MD $i$.

\paragraph{Energy consumption at Edge Servers.}

In ESs, the majority of the energy consumption is due to the computing process. 
Similar to equation \eqref{eq:comput_delay_ES}, the total energy consumption required to process the offloaded tasks from MDs is denoted as

\begin{equation}
    E_{\text{tot}, j} = \delta_{\text{cp},j} \sum_{i\in\mathcal{M}} \left(F_{\text{ES},j}(1-\rho_i) + F_{\text{ES},j}Z_{\text{ES},j}z_{ij}\rho_i \right),
\end{equation}
where we omit the base energy consumption for brevity.

\subsection{Problem Formulation}  \label{subsec:problem_formulation}

We first introduce the objective function of our problem formulation. 
We aim to minimize end-to-end delay while reducing energy consumption of low-battery-level MDs. 
Thus, the objective function is formulated by a balanced form of average end-to-end delay and battery usage.
\paragraph{Average Latency.}
Here, for brevity of the notation, we use sum-delay of the tasks generated from MDs as follows:

\begin{equation}
\begin{aligned}
    & \sum_{i\in \mathcal{M}}\sum_{j\in \mathcal{E}} x_{ij}\left( D_{\text{cm},ij}+ D_{\text{ES},ij} \right) \\ 
    & + \sum_{i\in\mathcal{M}}\left( 1 - \sum_{j\in\mathcal{E}} x_{ij} \right)D_{\text{MD},i} \label{eq:sum_delay}.
\end{aligned}
\end{equation}

\begin{figure}
\centering
\includegraphics[width=0.99\linewidth]{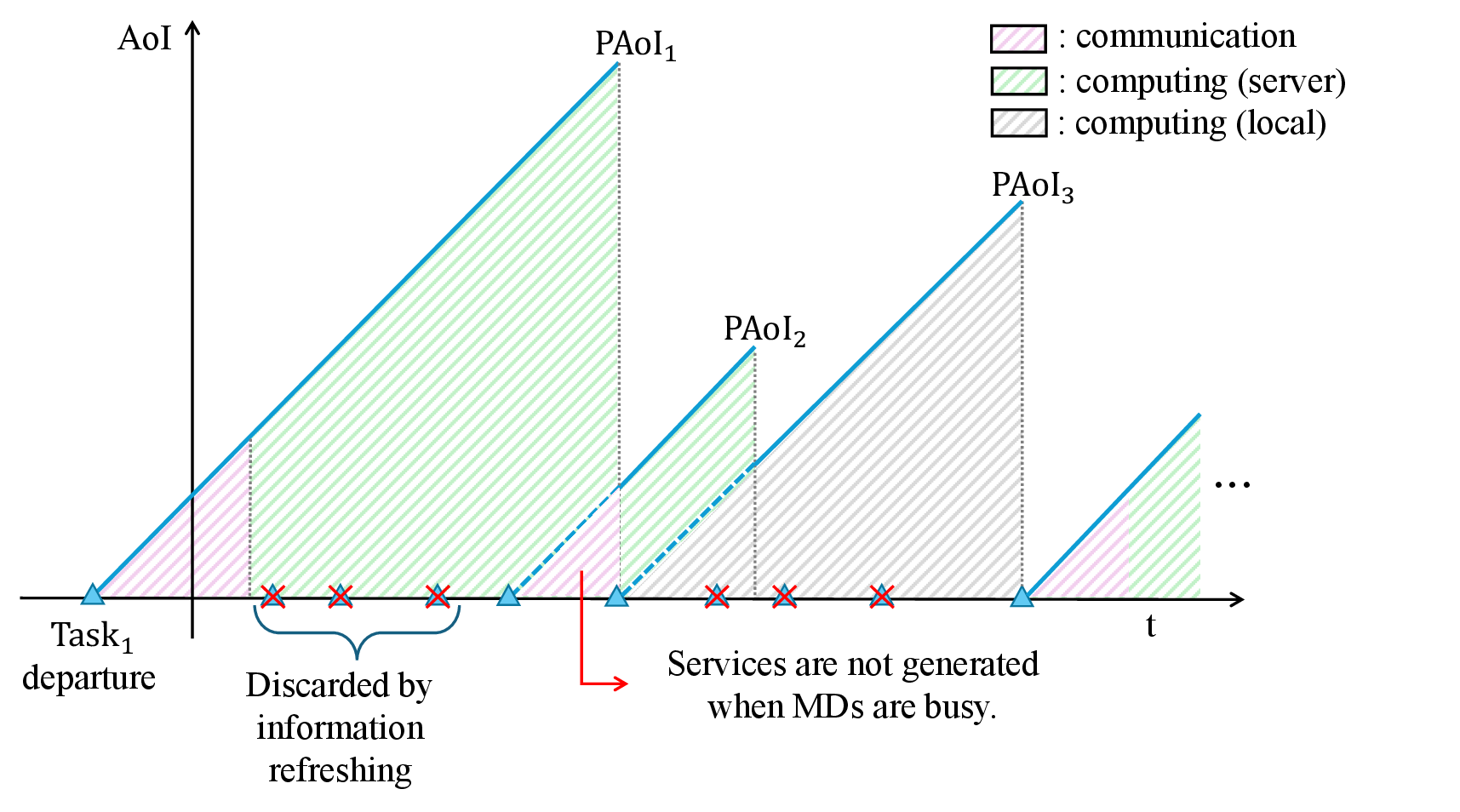}
\caption{Illustration of the time-evolution of the AoI in our system. The communication and computation periods are depicted with purple and green slashes, respectively. The local computing is depicted with gray slashes.}
\label{fig:AoI_illustration}
\end{figure}

\begin{remark}[Objective function from the viewpoint of AoI]
\label{remark:AoI}
    In Fig. \ref{fig:AoI_illustration},  we have depicted an illustration of the time-evolution of the AoI in our system. 
    We focus on the peak AoI (PAoI) represented in \cite{9380899}. 
    The PAoI is formulated as 
    \begin{equation}
        \Delta=\lim_{N\rightarrow \infty} \frac{1}{N} T_n,
    \end{equation}
    where $N$ denotes the number of processed tasks and $T_n$ denotes the peak of the age of $n$-th process.
    As depicted in Fig. \ref{fig:AoI_illustration}, the peak age is proportional to the average delay of each service. 
    Hence, with the linear age ($\propto t$), our objective function complies with the concept of AoI with linear age and PAoI.
\end{remark}

\paragraph{Penalized Term for Battery Usage for Low-Battery level MDs.}
We now formulate the energy consumption of local computation.
The energy penalty of local processing is denoted as 
\begin{equation}\label{eq:objective_1}
    G_i = \delta_{\text{cp}, i} f_i- \delta_{\text{cm}, i}\frac{d_i}{R_{ij}}.
\end{equation}
Then, the energy consumption part of the objective function is formulated by 
\begin{equation}\label{eq:objective_2}
    \max_{\mathbf{X}} \sum_{i\in\mathcal{M}}\left( 1 - \sum_{j\in\mathcal{E}} x_{ij} \right) \frac{G_i}{B_i},
\end{equation}
where $B_i$ is the remaining battery of MD $i$.
We take the fractional representation of the two variables to enhance the consideration of device energy, particularly in situations where the energy levels are low.

By combining the objective functions \eqref{eq:objective_1} and \eqref{eq:objective_2} with the constraints presented in the previous sections, the UARA optimization problem is formulated as 

\begin{problem} \label{eq:P1}
    \begin{alignat}{2}
        & \text{\ref{eq:P1}: } \min_{\mathbf{X}, \mathbf{Y}, \mathbf{Z}} & \quad &    \sum_{i\in \mathcal{M}}\sum_{j\in \mathcal{E}} x_{ij}\left( D_{\text{cm},ij}+ D_{\text{ES},ij} \right) \nonumber \\ 
        & & & + \sum_{i\in \mathcal{M}} \left(1-\sum_{j\in \mathcal{E}} x_{ij}\right) \left( D_{\text{MD},i} - \alpha \frac{G_i}{B_i} 
        \right) \label{P1:obj} \\
        & ~~~~~~~~~ \text{s.t. } & & \eqref{eq:x_constraint}, \eqref{eq:y_constraint}, \eqref{eq:comm_delay}, \eqref{eq:computing_delay} - \eqref{eq:memory_constraint}, \eqref{eq:comp_energy}, \eqref{eq:comm_energy},
    \end{alignat}
\end{problem}
where $\alpha$ is the balance coefficient between MD energy preservation and average latency.
{
Specifically, the energy-related term $\frac{G_i}{B_i}$ is integrated into the time-domain objective function with the help $\alpha$.
Here, $\alpha$ can be regarded as a unit conversion factor, converting energy consumption into delay.
Additionally, $\alpha$ serves as a weight term of energy expenditure.
By adjusting the value of $\alpha$, the network can regulate all the MDs' local energy consumption.}
In our problem, we do not incorporate energy conservation of ESs as an objective for minimization, given that each ES is typically connected to an energy source.
Consequently, we prioritize on energy conservation of MDs with the risk of battery depletion due to limited battery capacity.

In Problem \ref{eq:P1}, the main challenge is that the problem is in a non-linear form in terms of all unknown variables $\mathbf{X}, \mathbf{Y}$ and $\mathbf{Z}$.
Also, the binary variable $\mathbf{X}$ is a combinatorial variable, coupled with $\mathbf{Y}$, which is continuous.
This exacerbates the problem, as the problem becomes non-linear and a mixed-integer problem.

\section{Optimization-based Offloading Scheme} \label{sec:problem_sol}

\begin{figure}[!ht]
    \centering
    \includegraphics[width=\if\mycmd1 0.77\else 1\fi\linewidth]{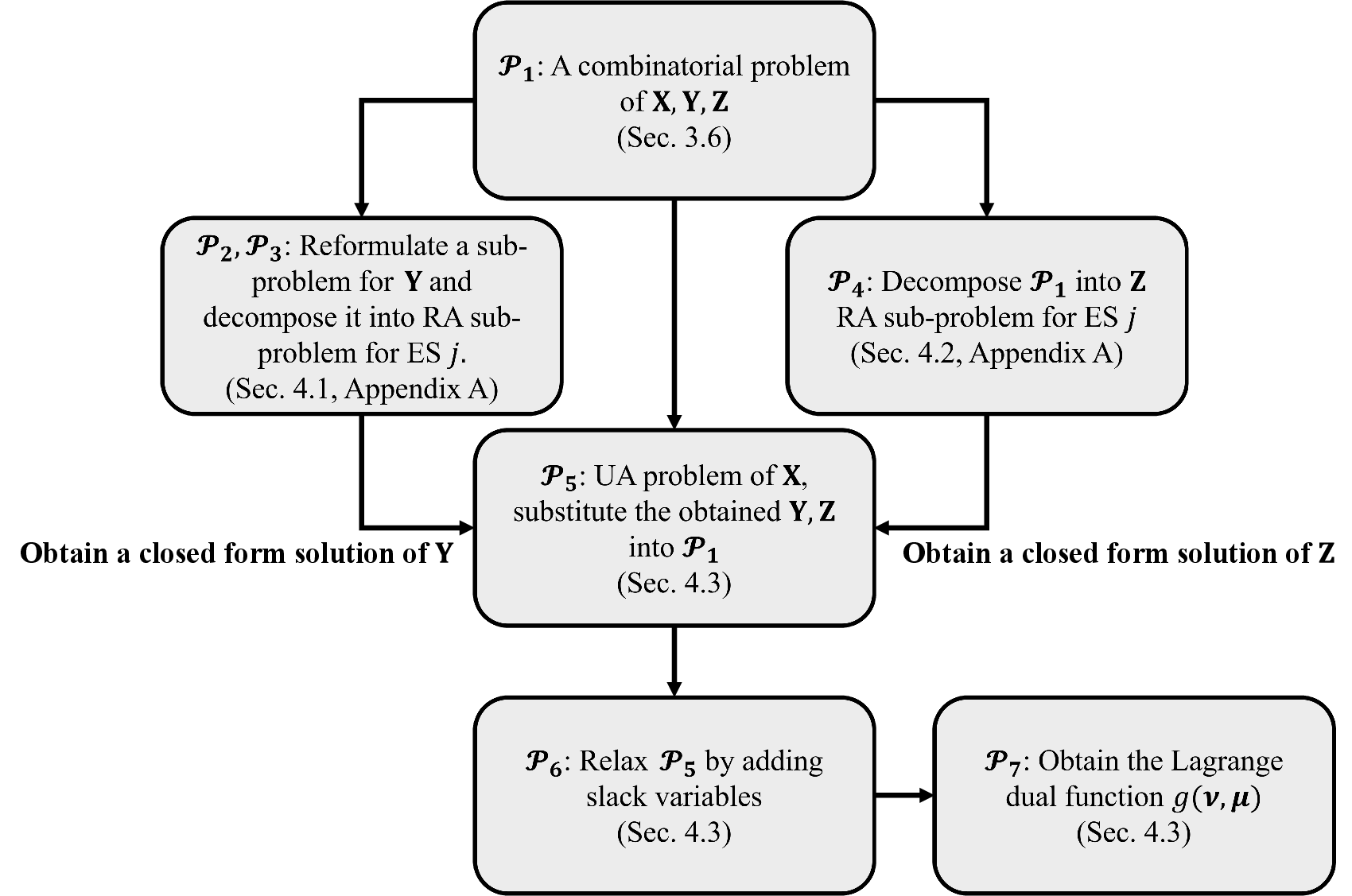}
    \caption{Schematic of the algorithms for the latency and energy consumption minimization problem in the MEC system.}
    \label{fig:schematic}
\end{figure}

In this section, we propose a distributed task offloading framework via UARA optimization under MEC networks.
To this end, we first solve the RA part of the original problem \ref{eq:P1} for given UA variable $\mathbf{X}$, where closed-form solutions are obtained for both communication and computing RA variable.
Next, we obtain the UA solution $\mathbf{X}$ for the optimal $\mathbf{Y}$ and  $\mathbf{Z}$.
The resulting solution minimizes the objective function, which is a balanced function of average latency and energy preservation.
Furthermore, we analyze the optimality and convergence of the proposed algorithm in the next section.


\subsection{Communication Resource Allocation With Fixed User Association}\label{subsec:RA_commun}

We first focus on solving for the variable $\mathbf{Y}$ in the context of Problem \ref{eq:P1}.
For brevity of the explanation, we formulate a sub-problem, in which the other variables ($\mathbf{X}$ and $\mathbf{Z}$) are fixed\footnote{Here, although the variables $\mathbf{X}$ and $\mathbf{Z}$ are fixed, we can obtain a closed-form solution of $\mathbf{Y}$. 
That is to say, we can plug-in the optimal variable $\mathbf{Y}$ into Problem \ref{eq:P1} and solve the problem only for the variable $\mathbf{X}$.
Furthermore, the variable $\mathbf{Z}$ is independent with the variable $\mathbf{Y}$ for fixed $\mathbf{X}$. }.
Hence, Problem \ref{eq:P1} can be reformulated to a subproblem for $\mathbf{Y}$ as
\begin{problem} \label{eq:P2}
    \begin{alignat}{2}
        & \text{\ref{eq:P2}: } \min_{\mathbf{Y}} & \quad &  \sum_{i\in\mathcal{M}}\sum_{j\in\mathcal{E}} \frac{d_ix_{ij}}{R_{ij}  y_{ij}+\epsilon} \\
        & ~~~~~~~~~ \text{s.t. } & & \sum_{k\in\mathcal{M}} y_{kj} \le 1, ~~ \forall j\in \mathcal{E} \\ 
        & & & y_{ij} \ge 0, ~~ \forall (i,j)\in\mathcal{M}\times\mathcal{E}.
    \end{alignat}
\end{problem}

Since all the constraints and objective function in Problem \ref{eq:P2} are decomposable for each ES $j$, we can relax the problem into sub-problems for each ES $j$ as follows:
\begin{problem} \label{eq:P3}
    \begin{alignat}{2}
        & \text{\ref{eq:P3}: } \min_{\mathbf{Y}} & \quad &  \sum_{i\in\mathcal{M}_j} \frac{d_i}{R_{ij}  y_{ij}} \\
        & ~~~~~~~~~ \text{s.t. } & & \sum_{k\in\mathcal{M}_j} y_{kj} \le 1 \\ 
        & & & y_{ij} \ge 0, ~~ \forall i\in\mathcal{M}_j,
    \end{alignat}
\end{problem}
where $\mathcal{M}_j$ is a subset of $\mathcal{M}$ such that $\mathcal{M}_j=\{i\in \mathcal{M}|x_{ij}=1\}$.

We obtain the solution to the Problem \ref{eq:P3} using the Karush-Kuhn-Tucker (KKT) conditions.
For brevity of the notation, we present the details of the derivation in Appendix \ref{app:RA_cm}, which can be found in the supplemental material. 
From the solution of Problem \ref{eq:P3}, the optimal bandwidth allocation is given as 
\begin{align} \label{eq:optimal_y}
    y_{ij}={}&\left(\sqrt{d_i/R_{ij}} \right) \Big/ \left(\sum_{k\in\mathcal{M}_j}\sqrt{d_k/R_{kj}}\right) \\
    ={}&\left(\sqrt{d_i/R_{ij}} \right) \Big/ \left(\sum_{k\in\mathcal{M}}\sqrt{d_k/R_{kj} } x_{kj}\right).
\end{align}

\subsection{Computing Resource Allocation With Fixed User Association}\label{subsec:RA_comput}

Similar to the bandwidth resource allocation from the previous section, the computing resource allocation problem can be decomposed for each ES $j$. 
The decomposed computing resource allocation sub-problem for ES $j$ is represented by 
\begin{problem} \label{eq:P4}
    \begin{alignat}{2}
        & \text{\ref{eq:P4}: } \min_{\mathbf{Z}} & \quad &  \sum_{i\in\mathcal{M}_j} \left( \frac{f_i(1-\rho_i)}{F_{\text{ES},j} } +\frac{f_i\rho_i}{F_{\text{ES},j} Z_{\text{ES},j}z_{ij}} \right) \\
        & ~~~~~~~~~ \text{s.t. } & & \sum_{k\in\mathcal{M}_j} z_{kj} \le 1 \\ 
        & & & z_{ij} \ge 0, ~~ \forall i\in\mathcal{M}_j.
    \end{alignat}
\end{problem}

We use KKT conditions to find the optimal solution to Problem \ref{eq:P4}.
We omit the solution in this section, and we elaborate the solution in Appendix \ref{app:RA_cm} (available at the supplmental material). 
The optimal solution for the computing resource allocation is denoted as follows:
\begin{align}\label{eq:optimal_z}
    z_{ij} 
    ={}& \sqrt{\frac{f_i \rho_i}{F_{\text{ES},j} Z_{\text{ES},j}}}\Big/ \left( \sum_{k\in\mathcal{M}}\sqrt{\frac{f_k \rho_k}{F_{\text{ES},j} Z_{\text{ES},j}}}x_{kj} \right).
\end{align}

\subsection{User Association}
In Sections \ref{subsec:RA_commun} and \ref{subsec:RA_comput}, we obtain the optimal solutions of $\mathbf{Y}$ and $\mathbf{Z}$ for given UA variable $\mathbf{X}$. 
Here, we aim to solve the optimization problem \ref{eq:P1}, with the solutions for RA variables $\mathbf{Y}$ and $\mathbf{Z}$ are provided in \eqref{eq:optimal_y} and \eqref{eq:optimal_z}, respectively.

First, for the brevity of the notation, we rewrite the problem \ref{eq:P1} by substituting the optimal RA solutions into the problem.
The problem can be reformulated as
\begin{problem} \label{eq:P5}
    \begin{alignat}{2}
        & \text{\ref{eq:P5}: } \min_{\mathbf{X}} & \quad &  \sum_{j\in\mathcal{E}} \left(\sum_{i\in\mathcal{M}}\sqrt{\hat{D}_{\text{ES}(s),ij}}x_{ij}\right)^2 \nonumber \\
        & & & ~~~~~ + \sum_{j\in\mathcal{E}}\left(\sum_{i\in\mathcal{M}}\sqrt{\hat{D}_{\text{cm},ij}}x_{ij}\right)^2\label{eq:p5_objective}   \\ 
        & & & ~~~~~ + \sum_{j\in\mathcal{E}}\sum_{i\in\mathcal{M}}c_{ij}x_{ij}  \nonumber \\
        & ~~~~~~~~~ \text{s.t. } & & \sum_{j\in\mathcal{E}}x_{ij} \le 1, ~~\forall i\in\mathcal{M} \\ 
        & & & x_{ij}\in\{0,1\}, ~~ \forall (i,j)\in\mathcal{M}\times\mathcal{E},
    \end{alignat}
\end{problem}
where $c_{ij} = - D_{\text{ES}(p),ij} -D_{\text{MD}, i} + \alpha \frac{G_i}{B_i}$, $\hat{D}_{\text{ES}(s),ij}=\frac{f_i\rho_i}{F_{\text{ES},j}Z_{\text{ES},j}}$, and $\hat{D}_{\text{cm},ij} =  d_i/R_{ij}$

A well-designed heuristic method could also obtain a near-optimal solution; however, it would require heavy information exchange between the MDs in order to achieve good performance.
As for our scheme, we show that an optimal solution is obtainable without such communication overhead.

Before solving the problem; the challenge is that the first and second terms in Equation \eqref{eq:p5_objective} are non-linear to the binary variable $x_{ij}$ and cannot be decomposed. 
To obtain a solution, we may require computation-intensive exhaustive search, the computational complexity of which is $\mathcal{O}(N_{\text{ES}}^{N_{\text{MD}}+1}N_{\text{MD}})$.
To resolve this issue, we propose a method to add two slack variables: i) $a_j=\left(\sum_{i\in\mathcal{M}}\sqrt{\hat{D}_{\text{cm},ij}}x_{ij}\right)^2$ and ii) $b_j=\left(\sum_{i\in\mathcal{M}}\sqrt{\hat{D}_{\text{ES}(s),ij}}x_{ij}\right)^2$, where $\mathbf{a} = [a_1, \cdots, a_{N_{\text{ES}}}]$ and $\mathbf{b} = [b_1, \cdots, b_{N_{\text{ES}}}]$.
With these slack variables, we can reformulate the problem \ref{eq:P5} without loss of generality, as follows:

\begin{problem} \label{eq:P6}
    \begin{alignat}{2}
        & \text{\ref{eq:P6}: } \min_{\mathbf{a}, \mathbf{b}, \mathbf{X}} & \quad &  
        \sum_{j\in\mathcal{E}} (a_j + b_j)+ \sum_{i\in\mathcal{M}}\sum_{j\in\mathcal{E}}c_{ij}x_{ij} \label{eq:P6_obj} \\
        & ~~~~~~~~~ \text{s.t. } & & \sum_{j\in\mathcal{E}}x_{ij} \le 1, ~~\forall i\in\mathcal{M} \\ 
        & & & x_{ij}\in\{0,1\}, ~~ \forall (i,j)\in\mathcal{M}\times\mathcal{E} \\ 
        & & & \sqrt{a_j}=\sum_{i\in\mathcal{M}}\sqrt{\hat{D}_{\text{cm},ij}}x_{ij} \label{eq:P6_const_d}\\ 
        & & & \sqrt{b_j}=\sum_{i\in\mathcal{M}}\sqrt{\hat{D}_{\text{ES}(s),ij}}x_{ij}. \label{eq:P6_const_e}
    \end{alignat}
\end{problem}

Now, the objective function of the problem given by \eqref{eq:P6_obj} is linear. 
Furthermore, $x_{ij}$ related terms in the constraints are all linear, which enables us to find the optimum $\mathbf{X}$ by minimizing the Lagrangian of problem \ref{eq:P6}.
Interestingly, the association variable $x_{ij}$ of MD $i$ can be computed in MD $i$ locally without exchanging information with other MDs. 
Henceforth, in Lagrangian duality approach, if the Lagrangian multiplier of the dual problem can be computed distributedly, the algorithm can be implemented in a fully distributed manner.
In the remainder of this section, we solve the problem in the following steps:
\begin{enumerate}
    \item Minimization of Lagrangian of Problem \ref{eq:P6} with respect to variables $\mathbf{X}$, $\mathbf{a}$, and $\mathbf{b}$. ($\mathcal{L}(\mathbf{X}, \mathbf{a}, \mathbf{b})$.)
    \item Obtain dual objective function $g(\boldsymbol{\mu},\boldsymbol{\nu})$ of Problem \ref{eq:P6} by substituting the optimal variables (from step 1) into the problem. 
    \item Solve the Lagrangian dual function by maximizing the dual objective function. 
\end{enumerate}

\subsubsection{Step 1: Minimize Lagrangian}

In the first step of the Lagrangian duality approach, we obtain the Lagrangian dual objective function of the problem \ref{eq:P6}.
To this end, we formulate problem \ref{eq:P6} into the Lagrangian form as follows:
\begin{equation}
\begin{aligned}
    & \mathcal{L}(\mathbf{X},\mathbf{a}, \mathbf{b}) = \sum_{j\in\mathcal{E}} (a_j + b_j)
    + \sum_{i\in\mathcal{M}} \sum_{j\in\mathcal{E}} c_{ij}x_{ij}  \\
    &\hspace{50pt minus 1fil} + \sum_{j\in\mathcal{E}} \mu_{j}\left( -\sqrt{a_j} + \sum_{i}\sqrt{\hat{D}_{\text{cm},ij}} x_{ij} \right) \label{p3:lagrangian} \\
    &\hspace{50pt minus 1fil} + \sum_{j\in\mathcal{E}}\nu_{j}\left( -\sqrt{b_j} + \sum_i \sqrt{\hat{D}_{\text{ES}(s),ij}}x_{ij} \right),
\end{aligned}
\end{equation}
where $\mu_j$ and $\nu_j$ are the Lagrangian multipliers corresponding to the constraints \eqref{eq:P6_const_d} and  \eqref{eq:P6_const_e}.
For simplicity, we define augmented vector notation of Lagrangian multipliers by $\boldsymbol{\mu} = [\mu_1, \cdots, \mu_{N_{\text{ES}}}]$ and $\boldsymbol{\nu} = [\nu_1, \cdots, \nu_{N_{\text{ES}}}]$.
We can now obtain the Lagrangian dual objective function by 
\begin{equation}\label{eq:dual_definition}
    g(\boldsymbol{\mu},\boldsymbol{\nu})  = \min_{\mathbf{X},\mathbf{a},\mathbf{b}}\mathcal{L}(\mathbf{X},\mathbf{a},\mathbf{b}).
\end{equation}
Because the primal variables $\mathbf{a}$, $\mathbf{b}$, and $\mathbf{X}$ in Lagrangian are not coupled, we minimize those variables separately. 

\paragraph{Step 1-1: Minimizers $\mathbf{a}^*$ and $\mathbf{b}^*$ for Lagrangian.} 

Because the Lagrangian \eqref{p3:lagrangian} is strongly convex, the minimizers $\mathbf{a}$ and $\mathbf{b}$ can be obtained by first-order optimality condition.
Denoting the minimizers by $\mathbf{a}^*=[a_1^*, \cdots, a_{N_{\text{ES}}}^*]$ and $\mathbf{b}^*=[b_1^*, \cdots, b_{N_{\text{ES}}}^*]$, we have the following condition:
\begin{align}
    \frac{\partial \mathcal{L}}{\partial a_j}=1-\frac{\mu_j}{2\sqrt{a_j}} = 0 &\Longrightarrow a_j^* = \frac{\mu_j^2}{4} \\
    \frac{\partial \mathcal{L}}{\partial b_j} = 1-\frac{\nu}{2\sqrt{b_j}} = 0 &\Longrightarrow b_j^* = \frac{\nu_j^2}{4}.
\end{align}

\paragraph{Step 1-2: Minimizer $\mathbf{X}^*$ for Lagrangian.} 
Let us define the optimal $\mathbf{X}$ minimizing Lagrangian $\mathcal{L}$ as $\mathbf{X}^*$.
Although the variable $\mathbf{X}$ is combinatorial, the minimizer $\mathbf{X}^*$ for the Lagrangian can be easily obtained since $x_{ij}\in\{0,1\}$ and $\sum_{j\in\mathcal{E}}x_{ij}\le 1$. 
More specifically, $\mathbf{X}^*$ can be obtained by selecting index $j$ (ES $j$) with the minimum coefficient, whereas none of the ESs are chosen if the minimum coefficient is larger than zero, i.e.,
\begin{equation}
\label{eq:optimal_x}
    x_{ij}^{*}=\begin{cases}
        1, &j=\argmin(\mu_j\sqrt{\hat{D}_{\text{cm},ij}}+\nu_j\sqrt{\hat{D}_{\text{ES}(s),ij}} \\ 
        & \hspace{30pt}+ c_{ij}), \\
        0, &\text{otherwise.}
    \end{cases}
\end{equation}

\subsubsection{Step 2: Formulate Lagrangian dual problem}

Substituting the minimizers $\mathbf{X}^*$, $\mathbf{a}^*$, and $\mathbf{b}^*$ into the Lagrangian, we rewrite the Lagrangian dual objective function of \eqref{eq:dual_definition} as
\begin{equation}
    \begin{aligned}\label{eq:dual_obj_def}
        & g(\boldsymbol{\mu},\boldsymbol{\nu}) = -\sum_{j\in\mathcal{E}}\left(\frac{\mu_j^2}{4}+\frac{\nu_j^2}{4}\right) \\ 
        & + \sum_{i\in\mathcal{M}}\min_{j\in\mathcal{E}}(\mu_j\sqrt{\hat{D}_{\text{cm},ij}}+\nu_j\sqrt{\hat{D}_{\text{ES}(s),ij}} +c_{ij}).
    \end{aligned}
\end{equation}

Since the minimum of affine functions is concave, the new dual objective function \eqref{eq:dual_obj_def} is a concave function.
Finally, we find the optimum $\boldsymbol{\mu}$ and $\boldsymbol{\nu}$ by maximizing the Lagrangian dual objective function as follows: 
\begin{problem} \label{eq:P7}
    \begin{alignat}{2}
        & \text{\ref{eq:P7}: } \max_{\boldsymbol{\mu},\boldsymbol{\nu}} & \quad &  
        g(\boldsymbol{\mu},\boldsymbol{\nu}).
    \end{alignat}
\end{problem}

\begin{algorithm}[t]
\caption{Pricing-Based Distributed Task Offloading Algorithm for Latency Minimization}
\label{alg:proposed}
\begin{algorithmic}[1]
\\ \textbf{Initialisation: }
\\ ~~~~ $-$ Randomly initialize $\nu_j$ and $\mu_j$ for all $j\in\mathcal{E}$.
\\ ~~~~ $-$ Initialize step size $\eta_1$ and $\eta_2$.
\For {$t=0$ to $T$}
    \State ES $j$ for all $j\in\mathcal{E}$ broadcast prices $\mu_j$ and $\nu_j$ to users.  \Comment{Broadcast Price }
    \For {$i\in\mathcal{M}$}
        \If {MD $i$ generates a service}
            \State Update $x_{ij}$, $j\in\mathcal{E}$ by \eqref{eq:optimal_x}. \Comment{Task offloading}
        \EndIf
    \EndFor
    \For {$j\in\mathcal{E}$}
        \State \textit{\# Locally update the lines 13 - 15 at ES $j$ }
        \State Update $y_{ij}$ and $z_{ij}$ $\forall i\in\mathcal{M}_j$ by \eqref{eq:optimal_y} and \eqref{eq:optimal_z}. 
        \State Update $\mu_j$ and $\nu_j$ by \eqref{eq:sub-grad-descent}. \Comment{Update price}
    \EndFor
\EndFor 
\end{algorithmic}
\end{algorithm}

\subsubsection{Step 3: Solve Lagrangian dual problem}

In order to find the optimal solutions $\boldsymbol{\mu}^*$ and $\boldsymbol{\nu}^*$ of Problem \ref{eq:P7}, we use the sub-gradient descent method, which is widely used in the optimization of non-differentiable functions. 
In fact, the dual function is sub-differentiable because the dual function $g(\boldsymbol{\mu},\boldsymbol{\nu})$ is the minimum of affine functions of $\boldsymbol{\mu}$ and $\boldsymbol{\nu}$.
Thus, we design the sub-gradient descent algorithm by 
\begin{equation}
\label{eq:sub-grad-descent}
    \begin{aligned}
        \mu_j \leftarrow \mu_j + \eta_1\left( \sum_{i\in\mathcal{M}}\sqrt{\hat{D}_{\text{cm},ij}}x_{ij}^*-\frac{\mu_j}{2}\right) \\ 
        \nu_j \leftarrow \nu_j + \eta_2 \left( \sum_{i\in\mathcal{M}}\sqrt{\hat{D}_{\text{ES}(s),ij}}x_{ij}^*-\frac{\nu_j}{2}\right),
    \end{aligned}
\end{equation}
where $\eta_1$ and $\eta_2$ denote the step sizes for $\boldsymbol{\mu}$ and $\boldsymbol{\nu}$, respectively. 
As previously stated, the update procedure in \eqref{eq:sub-grad-descent} can be done with local information only; thus, the proposed method can optimize the load of the network in a distributed manner.

{
Now, we analyze the information exchange of the network.
The information exchange necessary for the proposed method can be outlined as follows:}
\begin{itemize}
    \item {\textbf{Control information collection from the MDs to the ESs:}}
    {From the update equation {\protect\eqref{eq:sub-grad-descent}}, the information required for ES $j$ to determine its price is $\sum_{i\in\mathcal{M}}\sqrt{\hat{D}_{\textrm{cm},ij}}x_{ij}^*$ and $\sum_{i\in\mathcal{M}}\sqrt{\hat{D}_{\textrm{ES}(s),ij}}x_{ij}^*$.}
    {Those values are already accessible at each ES, as ES  $j$ can obtain the delay term of the serving users ($x_{ij}^*=1$) without additional information exchange.}
    \item {\textbf{ED-side actions:} The task of each MD is to simply compare the weighted delay for ESs and determine their $x_{ij}$ values, as outlined in \eqref{eq:optimal_x}. Subsequently, the remaining communication load is negligible since the MD only needs to send two integers \textit{when the MEC network is initialized} -- each representing the number of bits and the number of flops of the task.}
    \item {\textbf{Information exchange between the ESs:}
    No information exchange between the ESs is needed, as the price of an ES is calculated based on the load from the allocated MDs only.}
    \item {\textbf{Price broadcasting from the ESs to the MDs:}
    As the price each ES needs to broadcast is a single floating point number, only a short communication is necessary.}
\end{itemize}

In Algorithm \ref{alg:proposed}, we present a detailed step-by-step implementation of the proposed scheme.

\section{Theoretical Analysis}  \label{sec:theoretical_analysis}

In Section \ref{sec:problem_sol}, we proposed a joint task offloading and resource allocation scheme for long-term delay minimization. 
The proposed method can be implemented in multiple edge servers without cooperation.
In this section, we analyze the optimality and convergence of the proposed algorithm. 

\subsection{Optimality}

\begin{figure}[t]
    \centering
    \adjustbox{width=\if\mycmd1 0.7\else0.85\fi\linewidth}{
    \begin{tikzpicture}[scale=1, domain=0:10, samples=100]
    \draw[thick, ->] (-1.1,0) -- (10.1,0) node[right] {\large $\mu_j$};
    \draw[thick, ->] (0,-0.6) -- (0,5.1) node[above] {};
    \draw[thick, ->] (0,6) -- (0,11.1) node[above] {\large $g(\boldsymbol{\mu},\boldsymbol{\nu})$};
    \draw[thick, ->] (-1.1,6.5) -- (10.1,6.5) node[right] {\large $\mu_j$};

    \def\s{0.8};  
    \def\l{2};  

    \def\a{8};
    \def\b{3.03};
    \def\c{9};

    \def\d{4.5};
    \def\e{3.2};
    \def\f{7};

    \def\g{3.5};
    \def\h{3.2};
    \def\j{7};

    \def\k{0};
    \def\m{3.03};
    \def\n{9};

    \draw[dashed, very thick, color=cyan!40!white, domain=0-\l/6:\l+\l/2] plot(\x, {\b+sqrt(\c*\c-\x*\x+2*\a*\x-\a*\a)});
    \draw[dashed, very thick, color=cyan!40!white, domain=\l-\l/2:2*\l+\l/2] plot(\x, {\e+sqrt(\f*\f-\x*\x+2*\d*\x-\d*\d)});
    \draw[dashed, very thick, color=cyan!40!white, domain=2*\l-\l/2:3*\l+\l/2] plot(\x, {\h+sqrt(\j*\j-\x*\x+2*\g*\x-\g*\g)});
    \draw[dashed, very thick, color=cyan!40!white, domain=3*\l-\l/2:3*\l] plot(\x, {\m+sqrt(\n*\n-\x*\x+2*\k*\x-\k*\k)});

    \draw[very thick, color=cyan!70!gray, domain=0:\l] plot(\x, {\b+sqrt(\c*\c-\x*\x+2*\a*\x-\a*\a)});
    \draw[very thick, color=cyan!70!gray, domain=\l:2*\l] plot(\x, {\e+sqrt(\f*\f-\x*\x+2*\d*\x-\d*\d)});
    \draw[very thick, color=cyan!70!gray, domain=2*\l:3*\l] plot(\x, {\h+sqrt(\j*\j-\x*\x+2*\g*\x-\g*\g)});
    \draw[very thick, color=cyan!70!gray, domain=3*\l:4*\l+\l/4] plot(\x, {\m+sqrt(\n*\n-\x*\x+2*\k*\x-\k*\k)});

    \filldraw[cyan!70!gray] (2,9.738) circle (2pt);
    \filldraw[cyan!70!gray] (4,10.182) circle (2pt);
    \filldraw[cyan!70!gray] (6,9.738) circle (2pt);

    \foreach \i in {1,2,3}
    {
        \draw[dashed, color=black!70!white] (\l*\i,4-\i) -- (\l*\i,11.0) node[right] {};
    }

    \draw[-{Latex[length=3mm]}, very thick, color=cyan!50!black] (8.9,10.1) -- (7.3,8.6);
    \node[text width=3cm, color=cyan!50!black] at (9.7,10.5)
    {\Large $g(\boldsymbol{\mu}, \boldsymbol{\nu})$};

    \draw[very thick, color=blue!40!gray, domain=0:9.5] plot (\x, \x/2) node[right=0.15cm] {\large};
    
    \draw[-{Latex[length=3mm]}, very thick, color=blue!40!gray] (9.8,3.7) -- (8.5,4);
    \node[text width=3cm, color=blue!40!gray] at (11.5,3.5)
    {\Large $\sum\limits_{j\in\mathcal{E}}\frac{\mu_j}{2}$};

    \foreach \x in {1,2,3}
        \draw[dashed, very thick, green!40!gray] (\l*\x,5-\x -0.5) -- (\l*\x,4-\x -0.5);
    
    \foreach \x in {1,2,3,4}
        \draw[very thick, color=green!40!gray] (\l*\x-2,5-\x -0.5) -- (\l * \x,5-\x -0.5);
    \draw[very thick, green!40!gray]  (8,1 -0.5) -- (9,1 -0.5) node[right=0.15cm] {\large};
    
    \draw[-{Latex[length=3mm]}, very thick, color=green!40!gray] (10.5,1.7) -- (9,0.6);
    \node[text width=3cm, color=green!40!gray] at (10,2)
    {\Large $\sum\limits_{i\in\mathcal{M}}\sqrt{D_{\text{ES}(p), ij}}x^*_{ij}$};

    \filldraw[very thick, green!40!gray] (0,4 -0.5) circle (3pt);
    \foreach \x in {1, 2, 3}
    {
        \filldraw[very thick, green!40!gray, fill=white] (2*\x,4-\x -0.5) circle (3pt);
        \filldraw[very thick, green!40!gray] (2*\x,5-\x -0.5) circle (3pt);
    }    

    \draw[dashed, very thick] (4, 2) ellipse (0.5 and 1);
    
    \draw[-{Latex[length=3mm]}, very thick] (4,-0.4) -- (4,0.9);
    \node[text width=8cm] at (5,-0.8) 
    {\large The solution may lie on a non-differential \\ point of function $g(\boldsymbol{\mu},\boldsymbol{\nu})$};

    \end{tikzpicture}
    }
    \caption{Intuition of the situation that the solution lies on a non-differentiable point.}
    \label{fig:intuition-constraint-mismatch}
\end{figure}

From Sections \ref{subsec:RA_commun} and \ref{subsec:RA_comput}, the RA solutions \eqref{eq:optimal_y} and \eqref{eq:optimal_z} are globally optimal for given $\mathbf{X}$. 
Therefore, there is no optimality loss in the RA solutions. 
However, in the optimization of $\mathbf{X}$, there may exist a gap between the dual solution and primal solution.
Here, we analyze the duality gap of the UA problem and the convergence of the proposed scheme. 

Before analyzing the optimality, we clarify that the constraints on slack variables \eqref{eq:P6_const_d} and \eqref{eq:P6_const_e} may not be satisfied if $\mathbf{X}$ is recovered by \eqref{eq:optimal_x}. 
Figure \ref{fig:intuition-constraint-mismatch} illustrates the intuition that the solution cannot always satisfy the slack constraints. 
As depicted in the figure, the gradients of $g(\boldsymbol{\mu}, \boldsymbol{\nu})$ respect to $\boldsymbol{\mu}$ and $\boldsymbol{\nu}$ are represented by 
\begin{equation}
\begin{aligned}
        \frac{\partial g}{\partial \mu_j} = \sum_{i\in\mathcal{M}}\sqrt{\hat{D}_{\text{ES}(s),ij}}x_{ij}^*-\frac{\mu_j}{2} \\ 
        \frac{\partial g}{\partial \nu_j} = \sum_{i\in\mathcal{M}}\sqrt{\hat{D}_{\text{ES}(p),kj}}x_{ij}^*-\frac{\nu_j}{2}.
\end{aligned}    
\label{eq:gradient}
\end{equation}

Simply put, because the dual function $g$ is concave, if the dual function is differentiable at all points, the solution must satisfy 
\begin{equation}
\begin{aligned}
    \frac{\mu_j}{2} = \sum_{i\in\mathcal{M}}\sqrt{\hat{D}_{\text{ES}(s),ij}}x_{ij}^* ~\text{and}~
    \frac{\nu_j}{2} = \sum_{i\in\mathcal{M}}\sqrt{\hat{D}_{\text{ES}(p),kj}}x_{ij}^*. 
\end{aligned}    
\label{eq:gradient_condition}
\end{equation}

The condition \eqref{eq:gradient_condition} is satisfied if the two terms, i) monotonically decreasing affine term and ii) monotonically increasing step-like term\footnote{As the value of $\nu_{1,j}$ increases, the indicator $x_{ij}$ changes one to zero, i.e., the services are offloaded to other ESs.}, are equal.

However, as depicted in Fig. \ref{fig:intuition-constraint-mismatch}, since $\mathbf{X}$ is in a discrete space, the solution may lie on a non-differentiable point. 
The green-colored fist term is discontinuous at the points where the output of the $\argmin$ operator in \eqref{eq:optimal_x} changes. 
That is, the objective function of Problem \textbf{\ref{eq:P7}} can be non-differentiable for $\boldsymbol{\mu}$ or $\boldsymbol{\nu}$. 
The black dashed circle in Fig. \ref{fig:intuition-constraint-mismatch} indicates an example where the solution is located at a non-differentiable point. 
This means that the constraints \eqref{eq:P6_const_d} and \eqref{eq:P6_const_e} cannot be satisfied. 
Whether the constraints \eqref{eq:P6_const_d} and \eqref{eq:P6_const_e} are satisfied or not, we can recover the variable $\mathbf{X}$ using \eqref{eq:optimal_x}; however, the average latency term in the objective function of Problem \ref{eq:P6} is different from the actual latency, as written below:
\begin{itemize}
    \item The objective function of the primal problem \ref{eq:P6}: $\sum_{j\in\mathcal{E}} (a_j + b_j)+ \sum_{i\in\mathcal{M}}\sum_{j\in\mathcal{E}}c_{ij}x_{ij}$.
    \item Actual delay with penalized energy consumption corresponding to $\mathbf{X}$: 
    $$
    \begin{aligned}
        &\sum_{j\in\mathcal{E}} \left(\sum_{i\in\mathcal{M}}\sqrt{\hat{D}_{\text{ES}(s),ij}}x_{ij}\right)^2  + \sum_{j\in\mathcal{E}}\left(\sum_{i\in\mathcal{M}}\sqrt{\hat{D}_{\text{cm},ij}}x_{ij}\right)^2\\
        &+ \sum_{i\in\mathcal{M}}\sum_{j\in\mathcal{E}}c_{ij}x_{ij}.
    \end{aligned}
    $$
\end{itemize}

Hereby, the query is that the solution may be no longer optimal for Problem \textbf{\ref{eq:P6}}. 
In Theorem \ref{thm:duality}, it is shown that the gap between the obtained solution and the globally optimal solution is bounded. 

\begin{theorem}[Weak duality condition]\label{thm:duality}
Let us define the UA variable recovered by the dual solution to Problem \textbf{\ref{eq:P7}} as $\hat{\mathbf{X}}$, $\hat{\mathbf{a}}$, and $\hat{\mathbf{b}}$. 
Then, the performance gap between the solution $\hat{\mathbf{X}}$ and the globally optimal solution is bounded by $\mathcal{O}\left(\left(\Delta_j^{(1)}\right)^2 +\left(\Delta_j^{(2)}\right)^2\right)$,
where $\Delta_j^{(1)} = \sqrt{a_j}-\left(\sum_{i\in\mathcal{M}}\sqrt{\hat{D}_{\text{cm},ij}}x_{kj}\right)$ and $\Delta_j^{(2)} = \sqrt{b_j}-\left(\sum_{i\in\mathcal{M}}\sqrt{\hat{D}_{\text{ES}(s),ij}}x_{ij}\right)$.
\end{theorem}

\begin{IEEEproof}
For the proof of the proposition, please consult Appendix \ref{sec:appendix_A} in the supplementary materials.
\end{IEEEproof}

\paragraph{Extension to Machine Learning ($L_2$ Loss Minimization Approach).}

We have $a_j = \mu_j^2 / 4$ and $b_j = \nu_j^2/4$, where $a_j,b_j\ge0$. 
In machine learning approaches for task offloading \cite{9738455}, the learning algorithm aims to minimize a loss function.
If we set the loss function for minimizing the error terms as $\ell = \left(\Delta_j^{(1)}\right)^2 +\left(\Delta_j^{(2)}\right)^2$, we can rewrite the gradient descent step in \eqref{eq:sub-grad-descent} as follows:
\begin{equation}
\begin{aligned}
    \mu_j \leftarrow \mu_j + \eta_1 \frac{\partial\ell}{\partial\mu_j} \\ 
    \nu_j \leftarrow \nu_j + \eta_2 \frac{\partial\ell}{\partial\nu_j}.
\end{aligned}
\end{equation}

\subsection{Convergence Analysis}  \label{subsec:convergence_analysis}

It is well known that the sub-gradient descent method generally converges to the optimal point if the problem is a convex function.
Here, we show that the proposed scheme converges within  $\mathcal{O}(1/\epsilon^2)$ gradient descent steps.

\begin{proposition}[Convergence of the proposed scheme]\label{prop:A}
Let us define the sequence of Lagrange multipliers $(\boldsymbol{\mu},\boldsymbol{\nu})$ updated by \eqref{eq:sub-grad-descent} as $(\boldsymbol{\mu}^{(0)},\boldsymbol{\nu}^{(0)})$, $(\boldsymbol{\mu}^{(1)},\boldsymbol{\nu}^{(1)})$, ..., $(\boldsymbol{\mu},\boldsymbol{\nu})$ for arbitrary $k$. We define the best solution until the $T$-th step as $g_{\mathrm{max},k}=\max_{t=0,...,T}g(\boldsymbol{\mu},\boldsymbol{\nu})$. 
Then, for any $\epsilon>0$, there exist a constant step size $(\eta_1,\eta_2)$ and optimization step $T\in\mathcal{O}(1/\epsilon^2)$ satisfying the following inequality: 
\begin{equation}
    g^* - g_{\mathrm{max},T} \le \epsilon,
\end{equation}
where $g^*=\max_{\boldsymbol{\mu},\boldsymbol{\nu}}g(\boldsymbol{\mu},\boldsymbol{\nu})$.
\end{proposition}
\begin{IEEEproof}
The proof is shown in Appendix \ref{sec:appendix_B}, which is located in the supplementary material.
\end{IEEEproof}

\subsection{{Computational Complexity}}
\label{subsec:computational_complexity}

{
As previously mentioned, the proposed algorithm requires $\mathcal{O}(1/\epsilon^2)$ steps to converge, with the total computational complexity of each step being $\mathcal{O}(N_{\text{ES}} N_{\text{MD}})$. 
Therefore, for optimization with a fixed $\epsilon$, the proposed algorithm necessitates total $\mathcal{O}(\frac{1}{\epsilon^2}N_{\text{ES}} N_{\text{MD}})$ floating-point operations. By virtue of the distributed optimization, each of MDs and ESs has computational complexities of $O(N_{\text{ES}})$ and $O(N_{\text{MD}})$ in each step, respectively.
In contrast, the heuristic exhaustive search method has a complexity of $\mathcal{O}(N_{\text{ES}}^{N_{\text{MD}+1}} N_{\text{MD}})$.
In practical scenarios, the number of MDs typically exceeds 20, significantly increasing the computational complexity.
Thus, the proposed algorithm effectively reduces computational complexity, making it more efficient and scalable. 
This efficiency makes it well-suited for practical applications in large-scale MEC systems.
}

\section{Performance Evaluation}\label{sec:performance_evaluation}

In this section, we evaluate the proposed UARA offloading scheme in realistic AI-native MEC systems.
Our scheme is tested and compared in diverse network scenarios.
Our network scenarios include different network loads, different number of MDs and ESs and different simulation parameters.
We first introduce our simulation environment in subsection \ref{subsec:simulation_env}, and show the performance of our scheme in subsection \ref{subsec:eval_results}.
{For reproducibility of the results, our implementation code is available at https://github.com/mwkim00/mec\_uara.}

\subsection{Simulation Environment}  \label{subsec:simulation_env}

\begin{table}[]
\centering
    {\tiny
    \centering
    \caption{Simulation Settings}
    \adjustbox{width=\if\mycmd1 0.5\else 1.0\fi\linewidth}{
    \begin{tabular}{cc}
        \toprule
        Parameters & Values \\
        \midrule
        Number of MDs, $N_{\text{MD}}$ & 20 to 160 \\
        Number of ESs, $N_{\text{ES}}$ & {4 to 20} \\
        Bandwidth (MHz) & 10 \\
        \makecell{Maximum transmission power (dBm)} &  30 \\
        Noise power (dBm/Hz) & -174 \\
        Pathloss (dB) & $41+28\log_{10}(d)$ \\
        Coefficient of energy consumption for computing, $\delta_{\text{cp}}$ & $10^9$ \\
        Coefficient of energy consumption for communication, $\delta_{\text{cm}, i}$ & 2.6 \\
        Weight for energy penalized term, $\alpha$ & 0 to 100\\
        \bottomrule
    \end{tabular}
    }
    \label{tab:parameter_settings}
    }
\end{table}

We use the 3GPP small cell simulation document to model our channel \cite{3gpp.36.932,3gpp_38_214}.
The parameters are listed in detail in Table \ref{tab:parameter_settings}.
The cell consists of two type of users: clustered users and uniformly distributed users.
Clustered users typically increase the overall latency of the network, as the users in close proximity are more likely to be associated to the same ES.
Still, clustered users are considered in order to better mimic the actual network scenarios.
These different types of users allow us to simulate a more realistic cell network environment.
Additionally, in order to reproduce a more task-intensive network, we prompt our users to generate tasks right after the completion of their previous tasks, which increases the chance of the network saturating.

The number of MDs ranges from 20 to 160.
We use various types of MDs, including energy-limited mobile devices and energy-unlimited local computers.
For energy-limited mobile devices, we use Samsung's Galaxy S23, Huawei's Mate 60, and Apple's iPhone 14, some of the latest mobile devices available in the market.
As for the energy-unlimited computers, we use Apple's iMac with M1 processor.
{We use 4 to 20 number of ESs for our simulation.
Each ES is equipped with computing resources that outperforms MD's computing power.}
We use a combination of NVIDIA's RTX 2080, RTX 3090 and A6000 as the ES's computing device.

As for the tasks that the users generate, we use both the language-based AI tasks and computer vision tasks.
Typically, large language models (LLMs) that deal with language tasks require large computing power and small input size.
In contrast, computer vision models require comparably small computing resources, while the communication loads are relatively big due to larger input data.
We combine these two tasks in different ratio.
For language tasks, we employ Llama 2 model with approximately 7 billion parameters.
As for vision tasks, we use a combination of convolutional neural network (CNN)-based models and ViT-based models. 
Specifically, we adopt 2 residual net (ResNet) models, 2 MobileNet models, ViT-based side adapter network (SAN) and pyramid scene parsing network (PSPNet).
Using these tasks, we simulate three types of networks: network with high communication load, network with heavy computational load and a network with balanced load.
In a network with high communication load, the tasks predominantly consist of vision tasks.
The task parameters are listed in Table \ref{tab:task_parameter}.
The table values are based on the simulation results and calculations conducted by the authors.
As for a network with high computational load, the tasks that the users generate is mainly comprised of language tasks.

\renewcommand{\arraystretch}{1.2}

\begin{table}[t]
    \centering
    \caption{Communication and Computation Loads of the AI Tasks}
    \adjustbox{width=\if\mycmd1 0.6\else 1\fi\linewidth}{
    \begin{tabular}{c c c}
        \toprule
        Tasks & Communication [bits] & Computational [flops] \\
        \midrule
        Llama (7 billion) & $4.1\times 10^3$ & $5.0\times 10^{13}$ \\
        ResNet18 & $6.0\times 10^6$ & $4.2\times 10^9$ \\
        ResNet50 & $6.0\times 10^6$ & $1.8\times 10^9$ \\
        MobileNetV2 & $3.2\times 10^7$ & $3.0\times 10^8$ \\
        MobileNetV3 & $3.2\times 10^7$ & $8.0\times 10^{12}$ \\
        SAN & $9.6\times 10^7$ & $7.2\times 10^{13}$ \\
        PSPNet & $3.2\times 10^7$ & $5.2\times 10^{13}$ \\
        \bottomrule
    \end{tabular}
    }
    \label{tab:task_parameter}
\end{table}

We compare our results with some reference agents, which uses different policy schemes named as
\begin{itemize}
    \item \textbf{Random scheme,} where the MDs choose the device (either one of $N_{\text{ES}}$ ESs or locally) to compute randomly.
    The probability distribution of the selection follows a uniform distribution.
    \item \textbf{Max SINR scheme \cite{3gpp_36_872},} where the MDs determines which ES to associate based on the channel quality between the ESs and the device.
    The channel quality is assessed based on the uplink SINR.
    Hence, following this scheme, the MDs will achieve low communication delay.
    \item \textbf{Max Compute scheme~\cite{9705104,9547734}}, which enables MDs to determine the computing device depending on the computational load on the ESs.
    Computing delay is a major component of the overall latency of the system, alongside communication delay.
    To devise a scheme that reduces the computational delay, we revise the approaches used in previous works~\cite{9705104,9547734}.
    \item \textbf{Combined scheme}, which is a mixture of the max SINR scheme and the max compute scheme.
    The MDs employing combined offloading scheme determine the computing device based on both the communication channel information and the computing resource information.
\end{itemize}
For all schemes, we allow the the MDs to compute their task locally with probability $\epsilon$.
This allows for fair comparison between the heuristic schemes and the proposed scheme, as the proposed scheme is allowed to compute locally.
Enabling local computation also allows us to analyze the effects of local computation ratio.
In our simulation, we use $\epsilon=0.2$, as it leads to approximately uniform offloading between the ESs and the local device.

\subsection{Evaluation Results} \label{subsec:eval_results}

In this section, we conduct simulations to assess the effectiveness of our proposed model. 

\begin{figure*}[!ht]
\centering
\subfloat[]{\includegraphics[width=2.4in]{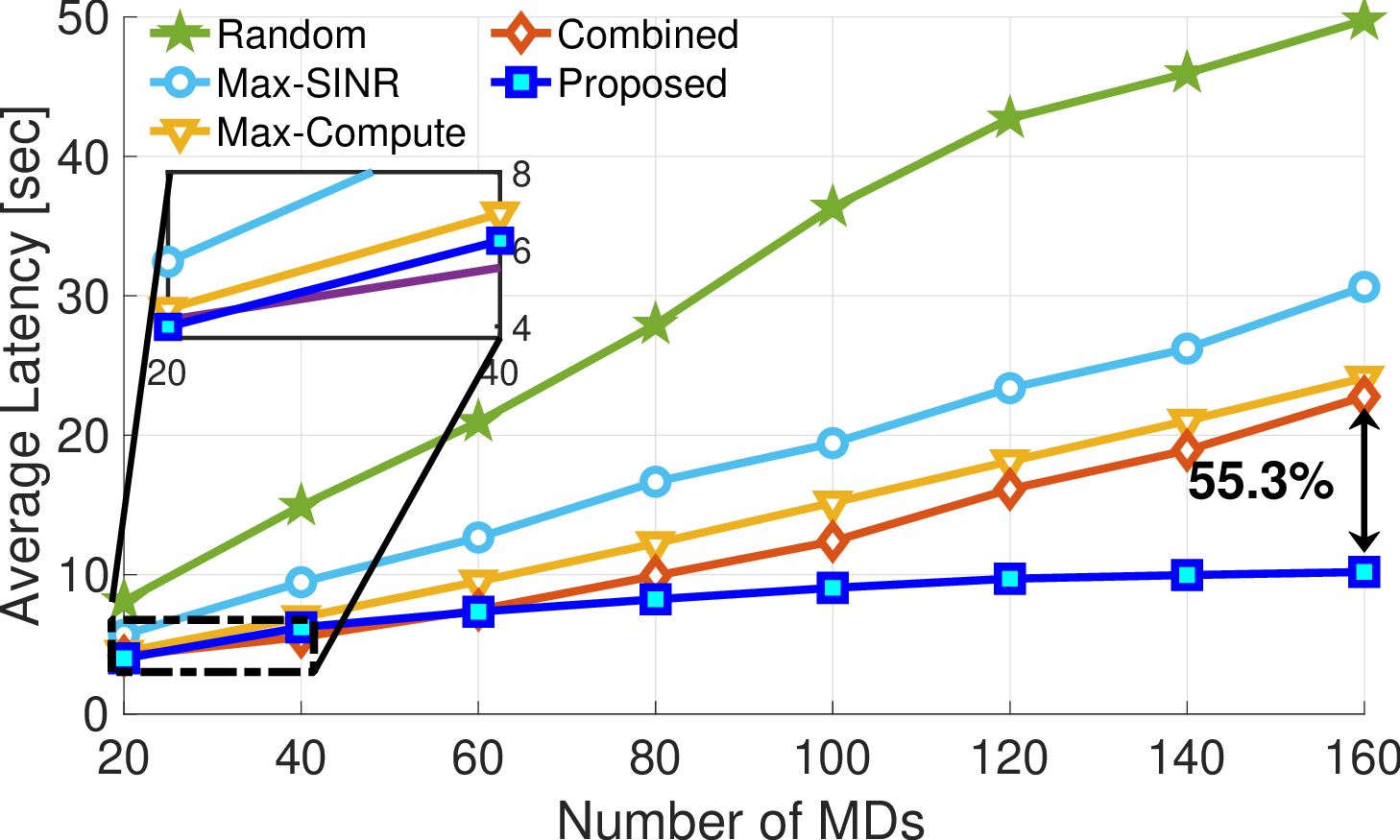}
\label{subfig:latency_base}}
\subfloat[]{\includegraphics[width=2.4in]{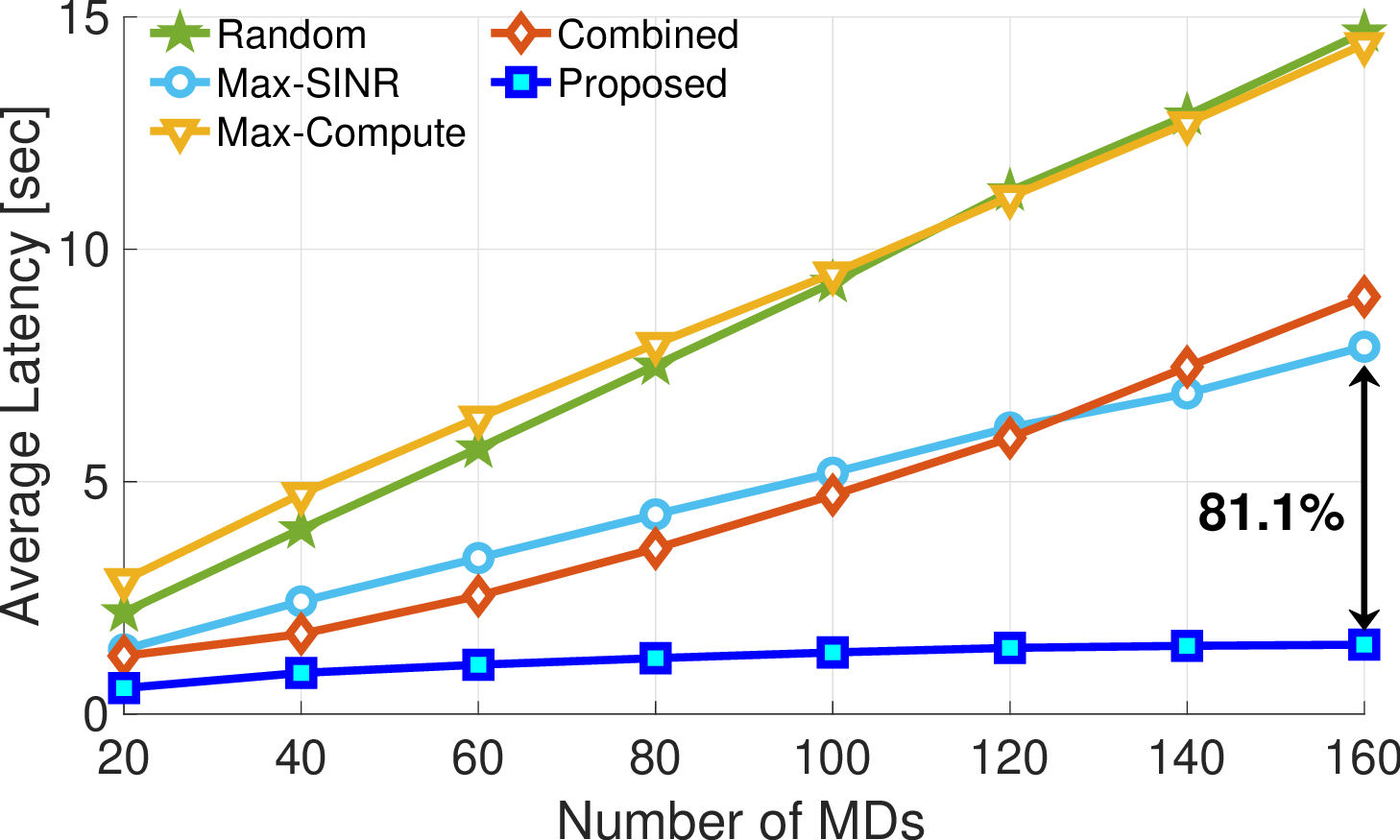}
\label{subfig:latency_highcommun}}
\subfloat[]{\includegraphics[width=2.4in]{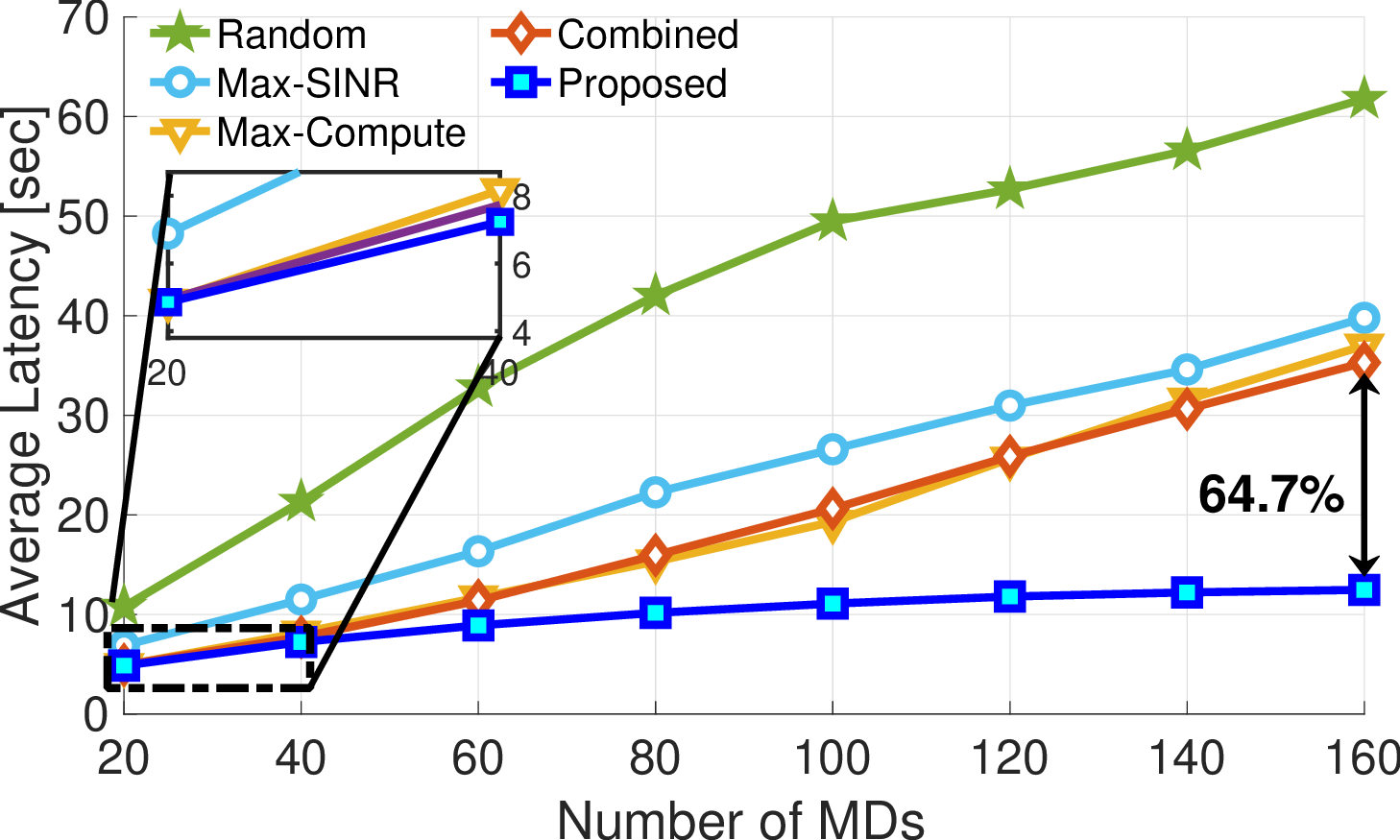}
\label{subfig:latency_highcompute}}
\caption{The system latency under different network scenarios of a MEC network where $N_{\text{ES}}=4$ and $\alpha=1$: (a) Network with average communication and computational load; (b) High communication load network; (c) High computational load network.}
\label{fig:latency}
\end{figure*}

\subsubsection{Latency Analysis} \label{subsec:latency_analysis}

{A network employing distributed optimization, including the proposed method, is expected to show robustness with regard to the environment as scalability is crucial in distributed optimization.
We test each scheme in different network environments to investigate each scheme's scalability.}
Figure \ref{fig:latency} illustrates the latency of the proposed and the reference schemes with respect to the number of MDs, which ranges from 20 to 160.
Our scheme is tested in three network scenarios: network with balanced load, communication-intensive network, and computational-intensive network.
The simulation results show that for all network scenarios, the proposed achieves the lowest latency.
More specifically, in a network with 160 MDs, the proposed scheme achieves 55.3\%, 81.1\%, and 64.7\% lower latency compared to the best performing reference scheme in each of the network scenario.
From the figures, we can see that the proposed scheme achieves a latency that is almost constant compared to the baseline schemes, which grow linearly.
From Fig. \ref{subfig:latency_highcommun}, which depicts the latency of the schemes in a network with high communication load, we observe that the max SINR scheme outperforms other baselines due to its performance in dealing with communication delay.
However, it does not excel the performance of the proposed scheme. 
Fig. \ref{subfig:latency_highcompute} shows the latency of the schemes in a computation-intensive network. 
Similar to Fig. \ref{subfig:latency_highcommun}, max compute scheme achieves the lowest latency among the baseline schemes.
Nonetheless, it still does not exceed the performance of the proposed scheme.
The reason for the proposed scheme's exceptional performance is based on the simultaneous consideration of communication/computational latency and flexible offloading policy.
How the proposed scheme outperforms other baselines is better illustrated in Figs. \ref{fig:highcommun_latency_compare} and \ref{fig:highcompute_latency_compare}.

\if\mycmd1
    \begin{figure}
        \centering
        \subfloat[]{\includegraphics[width=0.45\linewidth]{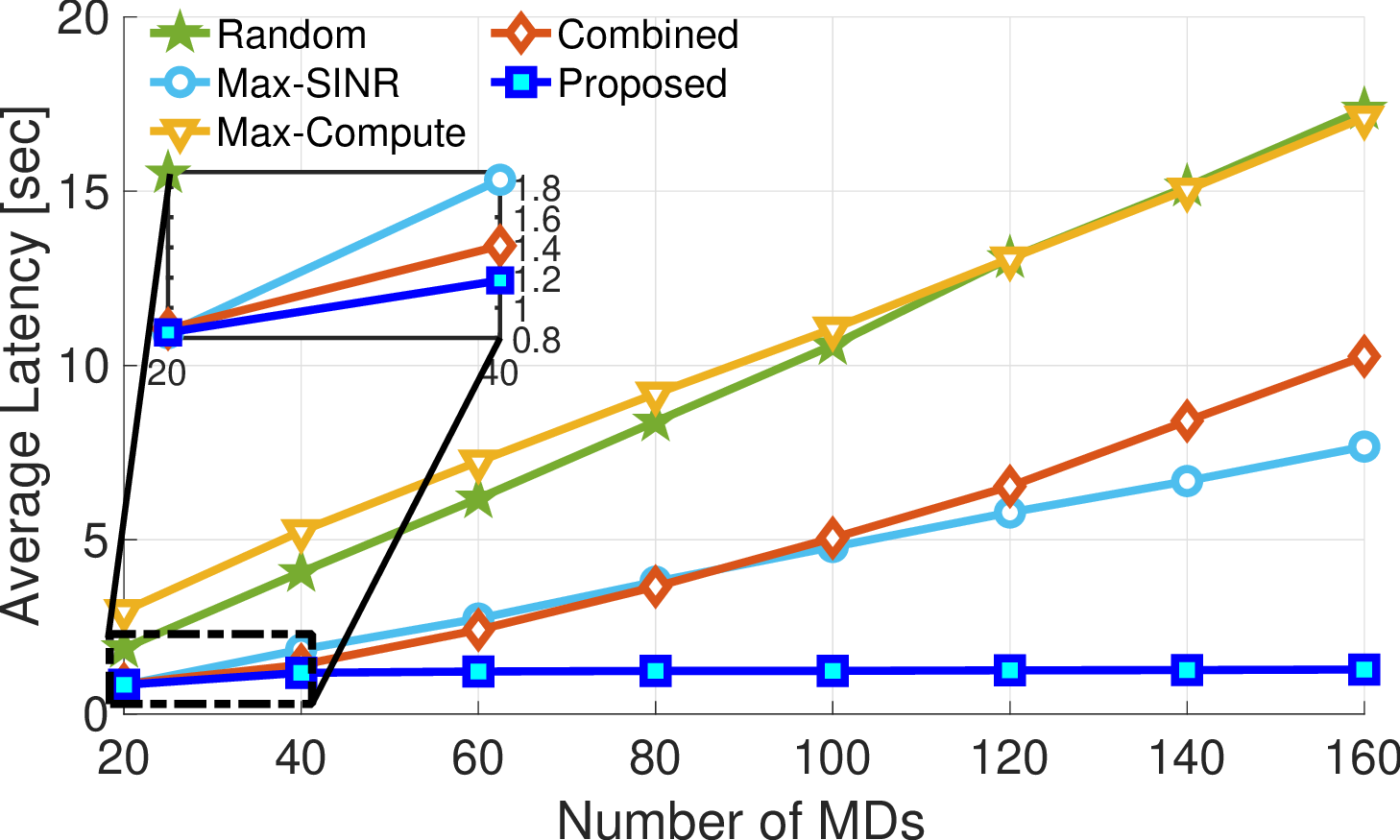}\label{subfig:highcommun_latency_bs_4}}
        \subfloat[]{\includegraphics[width=0.45\linewidth]{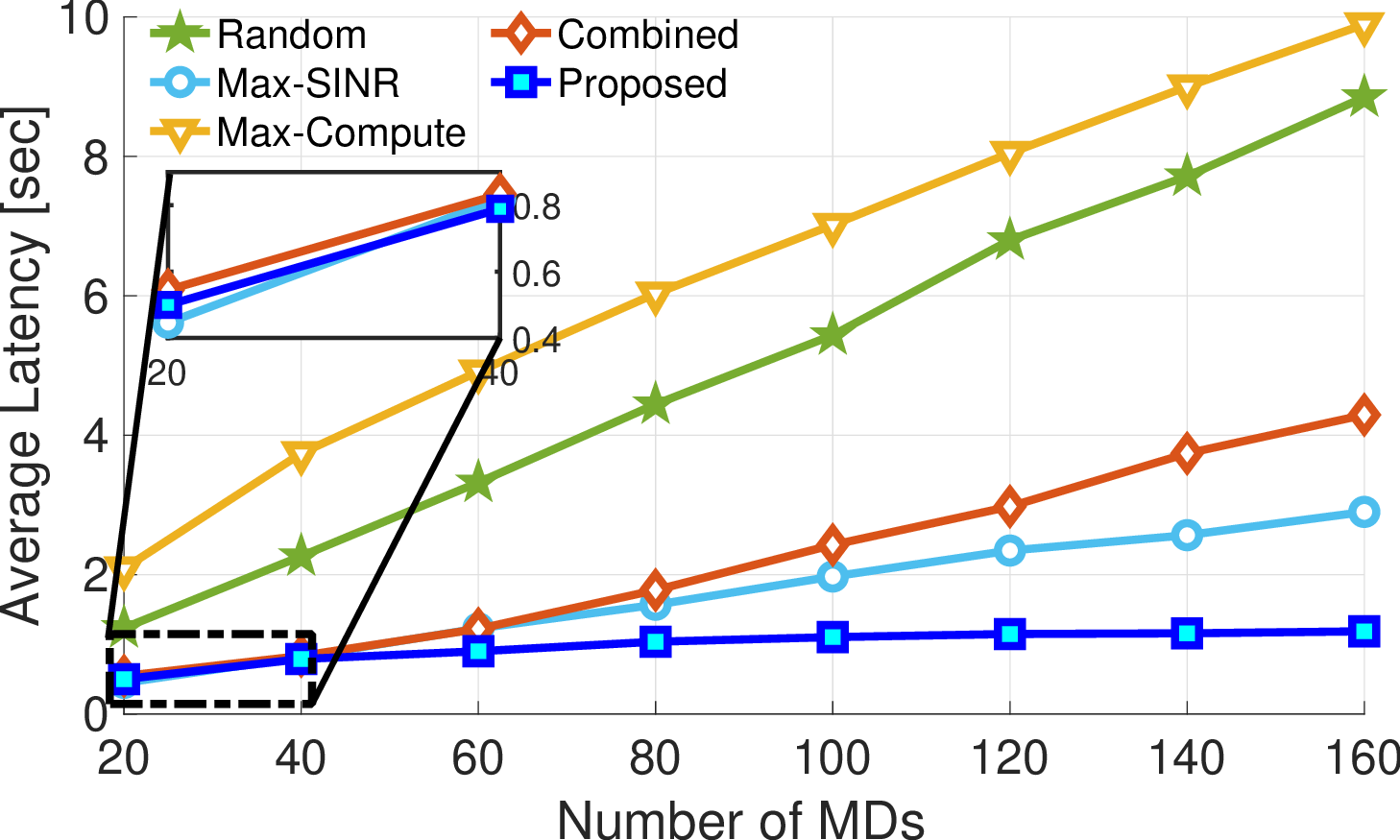}\label{subfig:highcommun_latency_bs_8}}
        \caption{Communication latency of networks with \textbf{high communication load}, where $\alpha=1$: (a) Communication latency of a network with $N_{\text{ES}}=4$; (b) Communication latency of a network with $N_{\text{ES}}=8$.}
        \label{fig:highcommun_latency_compare}
    \end{figure}
\else
    \begin{figure}
        \centering
        \subfloat[]{\includegraphics[width=\if\mycmd1 0.6\else 0.75\fi\linewidth]{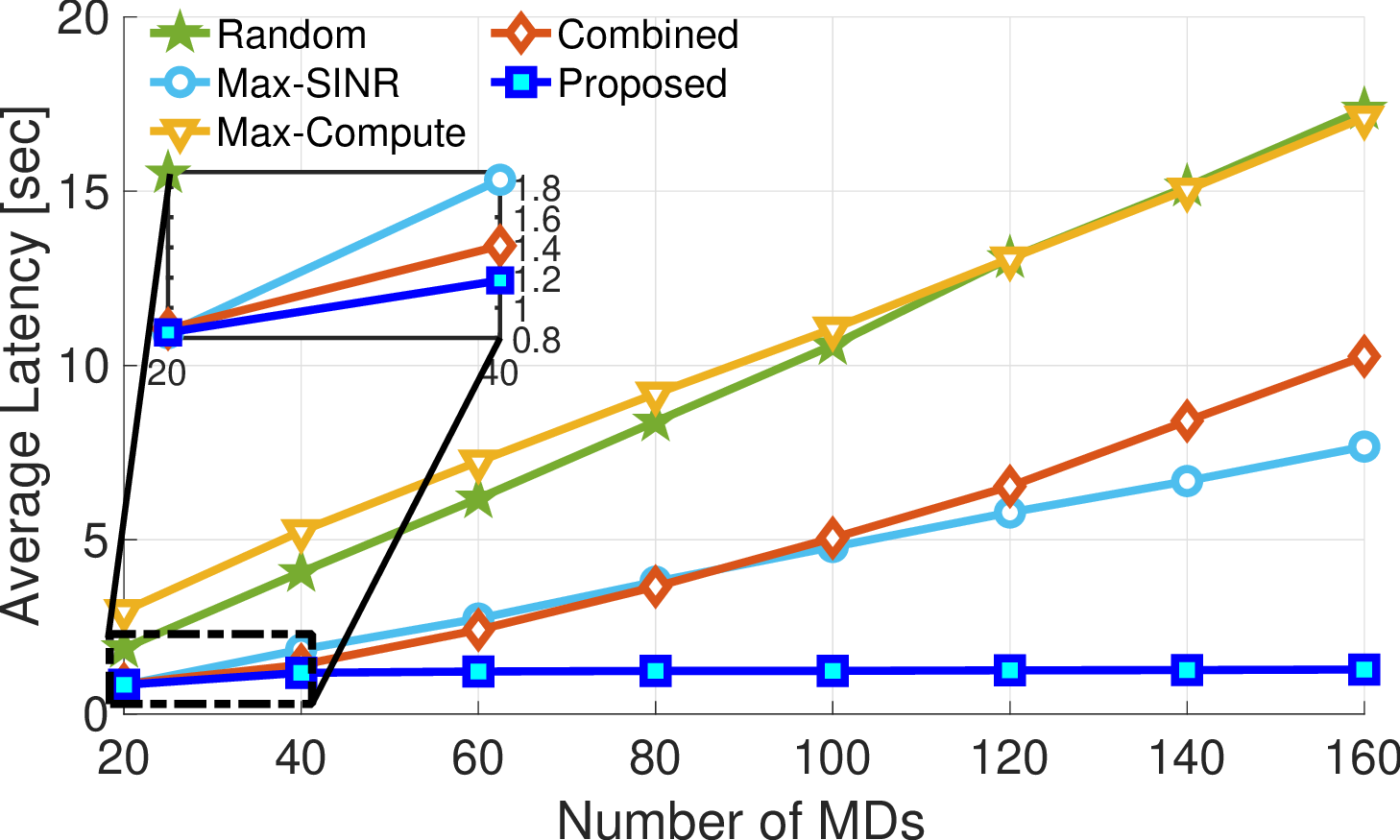}\label{subfig:highcommun_latency_bs_4}}
        \hfil
        \subfloat[]{\includegraphics[width=\if\mycmd1 0.6\else 0.75\fi\linewidth]{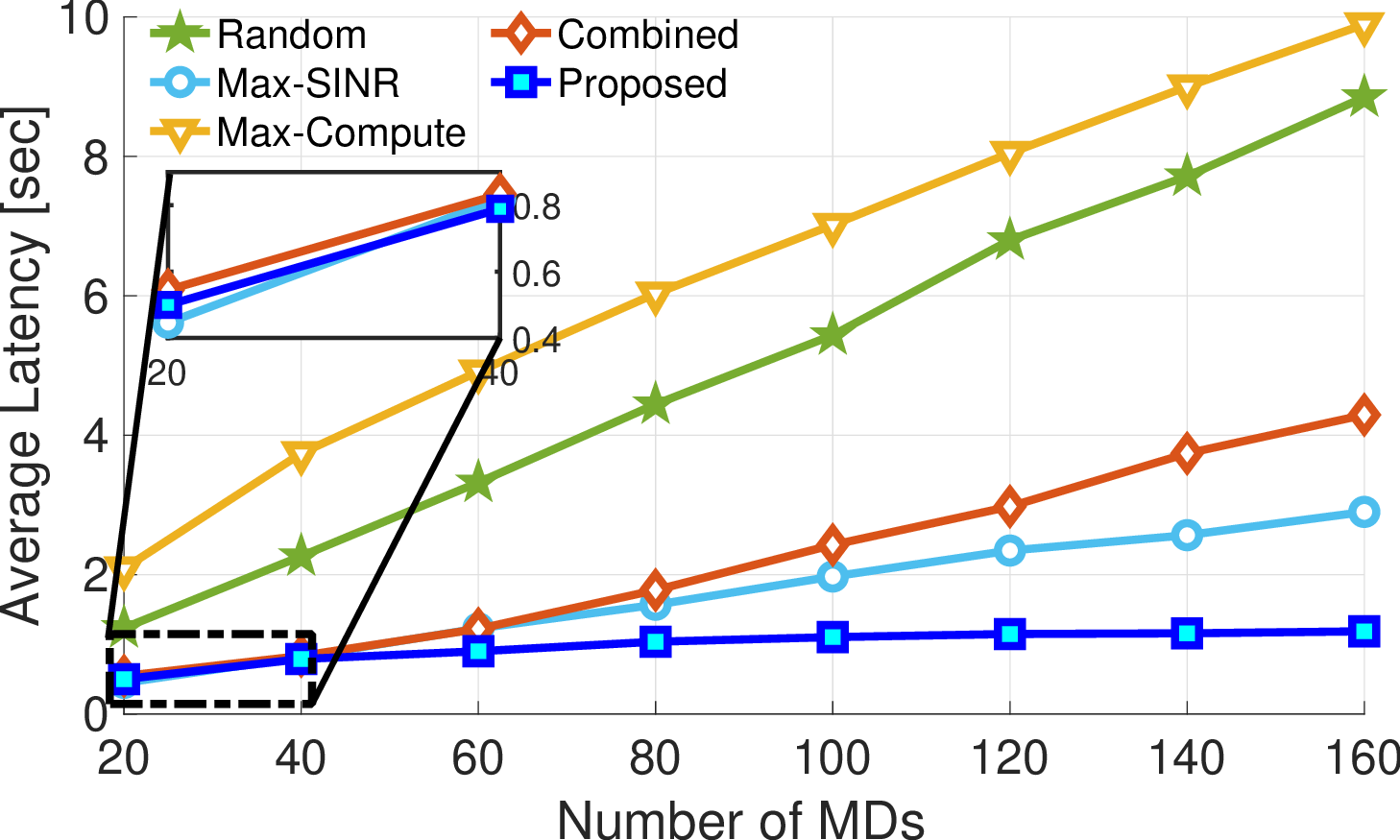}\label{subfig:highcommun_latency_bs_8}}
        \caption{Communication latency of networks with \textbf{high communication load}, where $\alpha=1$: (a) Communication latency of a network with $N_{\text{ES}}=4$; (b) Communication latency of a network with $N_{\text{ES}}=8$.}
        \label{fig:highcommun_latency_compare}
    \end{figure}
\fi

\if\mycmd1
    \begin{figure}
        \centering
        \subfloat[]{\includegraphics[width=0.45\linewidth]{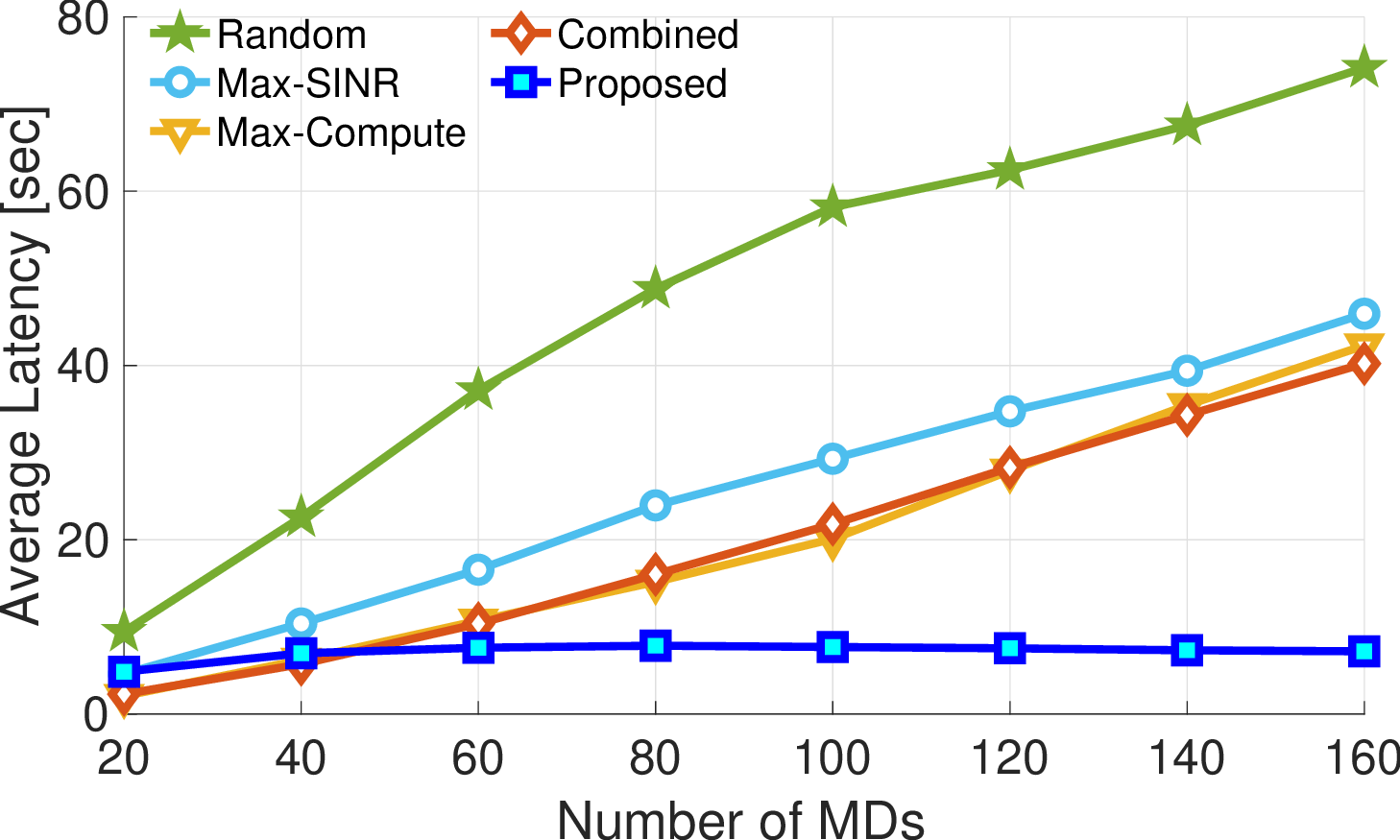}\label{subfig:highcompute_latency_bs_4}}
        \subfloat[]{\includegraphics[width=0.45\linewidth]{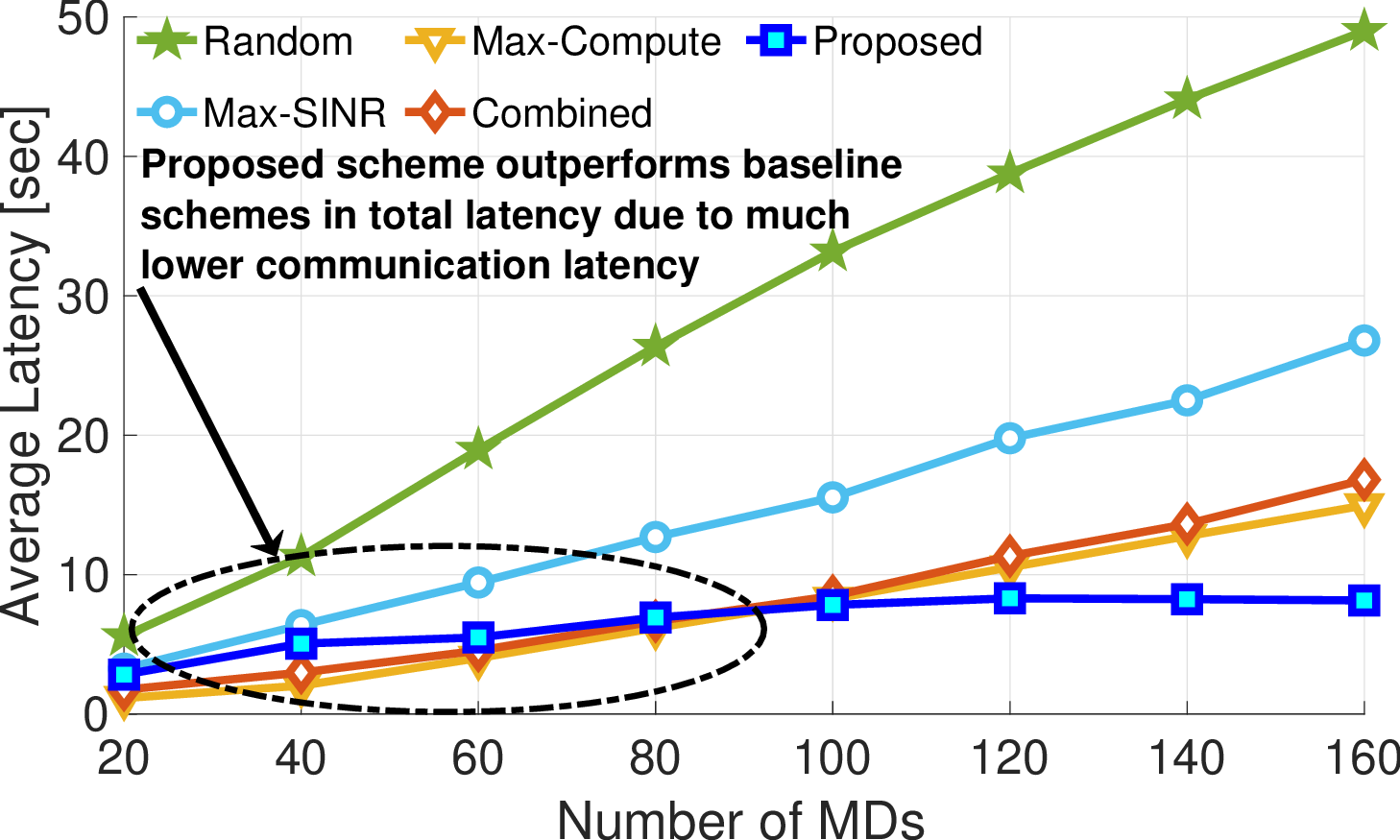}\label{subfig:highcompute_latency_bs_8}}
        \caption{Computational latency of networks with \textbf{high computation load}, where $\alpha=1$: (a) Computational latency of a network with $N_{\text{ES}}=4$; (b) Computational latency of a network with $N_{\text{ES}}=8$.}
        \label{fig:highcompute_latency_compare}
    \end{figure}
\else
    \begin{figure}
        \centering
        \subfloat[]{\includegraphics[width=\if\mycmd1 0.6\else 0.75\fi\linewidth]{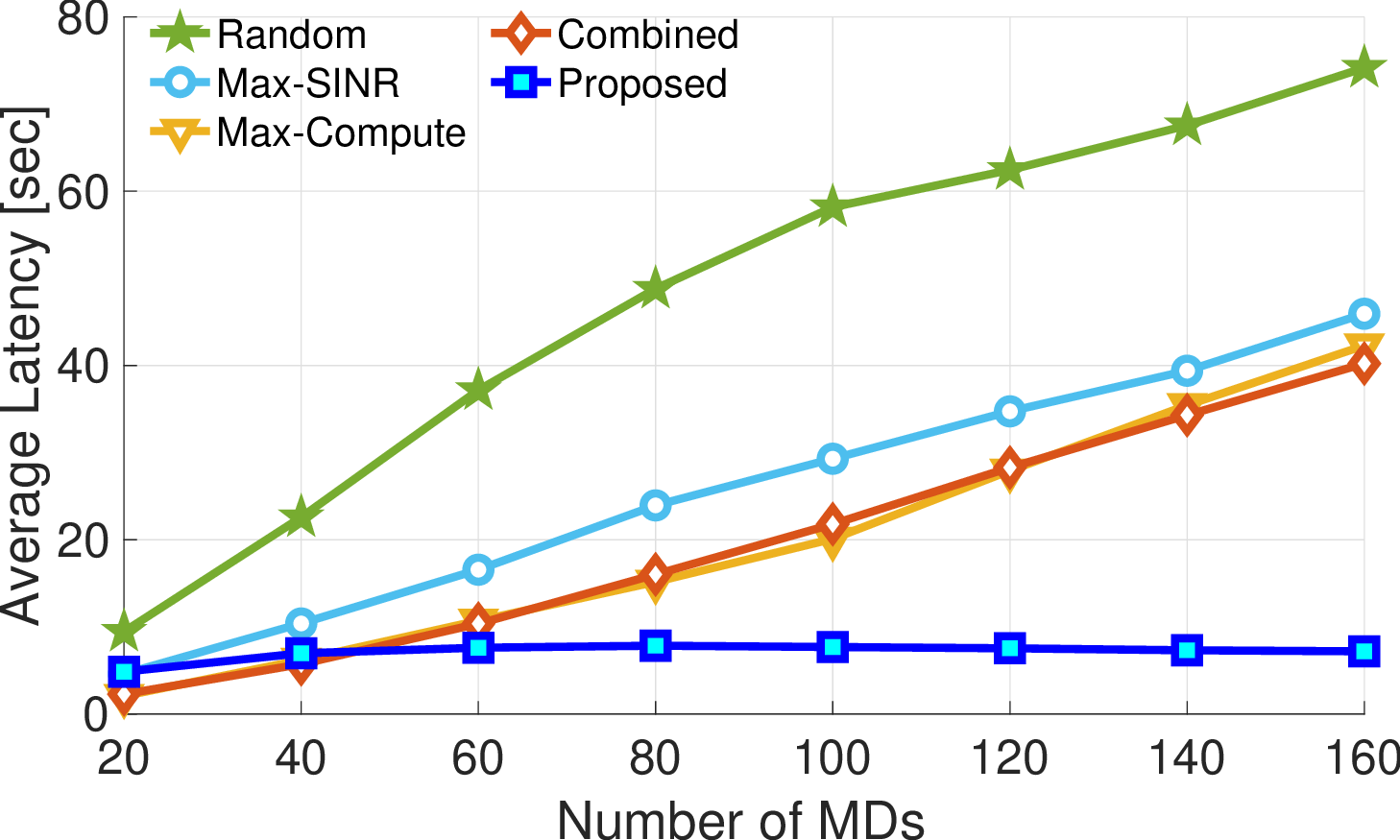}\label{subfig:highcompute_latency_bs_4}}
        \hfil
        \subfloat[]{\includegraphics[width=\if\mycmd1 0.6\else 0.75\fi\linewidth]{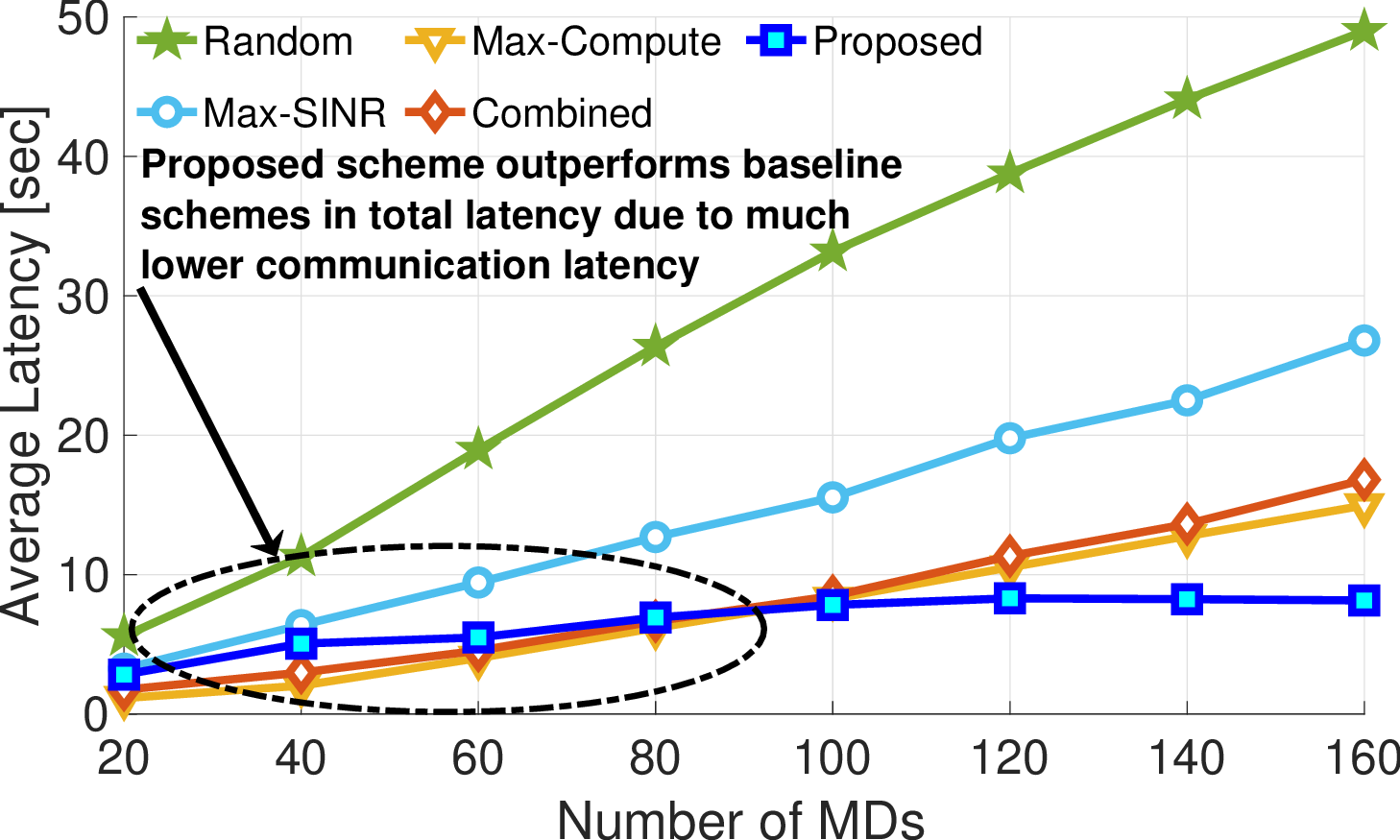}\label{subfig:highcompute_latency_bs_8}}
        \caption{Computational latency of networks with \textbf{high computation load}, where $\alpha=1$: (a) Computational latency of a network with $N_{\text{ES}}=4$; (b) Computational latency of a network with $N_{\text{ES}}=8$.}
        \label{fig:highcompute_latency_compare}
    \end{figure}
\fi

Figures \ref{fig:highcommun_latency_compare} and \ref{fig:highcompute_latency_compare} show the major latency component of different network scenarios as the number of users increases.
Specifically, Fig. \ref{fig:highcommun_latency_compare} illustrates the latency of users in a network with heavy communication load.
As for the baseline schemes, the performance degrades severely as the available number of ESs decreases from 8 to 4.
However, the proposed scheme maintains similar latency for all scenarios, showing that the algorithm is comparatively invariant to the amount of available resources.
In a network with high communication load, the communication latency term becomes the dominating factor in the overall latency of equation (\ref{eq:sum_delay}).
Hence, the optimization process of the proposed method is focused on optimizing the communication latency.
As a result, the proposed scheme achieves the lowest communication latency.
Our method achieves latency lower than the max SINR method, which can be regarded as a theoretical short-term lower bound of the communication latency.

On the other hand, in a network experiencing heavy computational load (Fig. \ref{fig:highcompute_latency_compare}), our scheme dynamically focuses on reducing the computational latency.
Similar to the communication-intensive network, this is due to the fact that computational delay is the dominating term of the latency formula.

\if\mycmd1
    \begin{figure}
        \centering
        \subfloat[]{\includegraphics[width=0.45\linewidth]{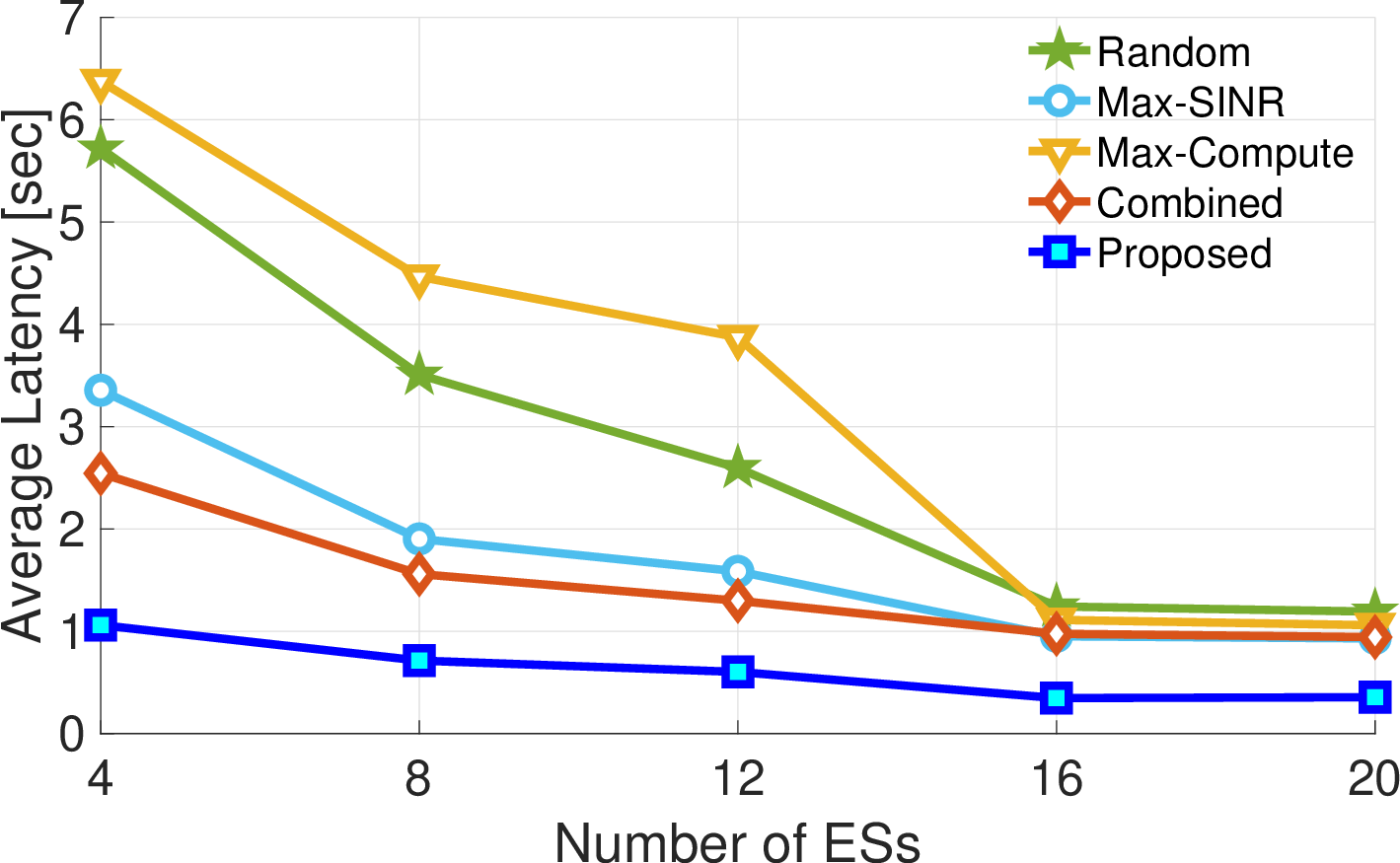}\label{subfig:highcommun_latency_ue_60}}
        \subfloat[]{\includegraphics[width=0.45\linewidth]{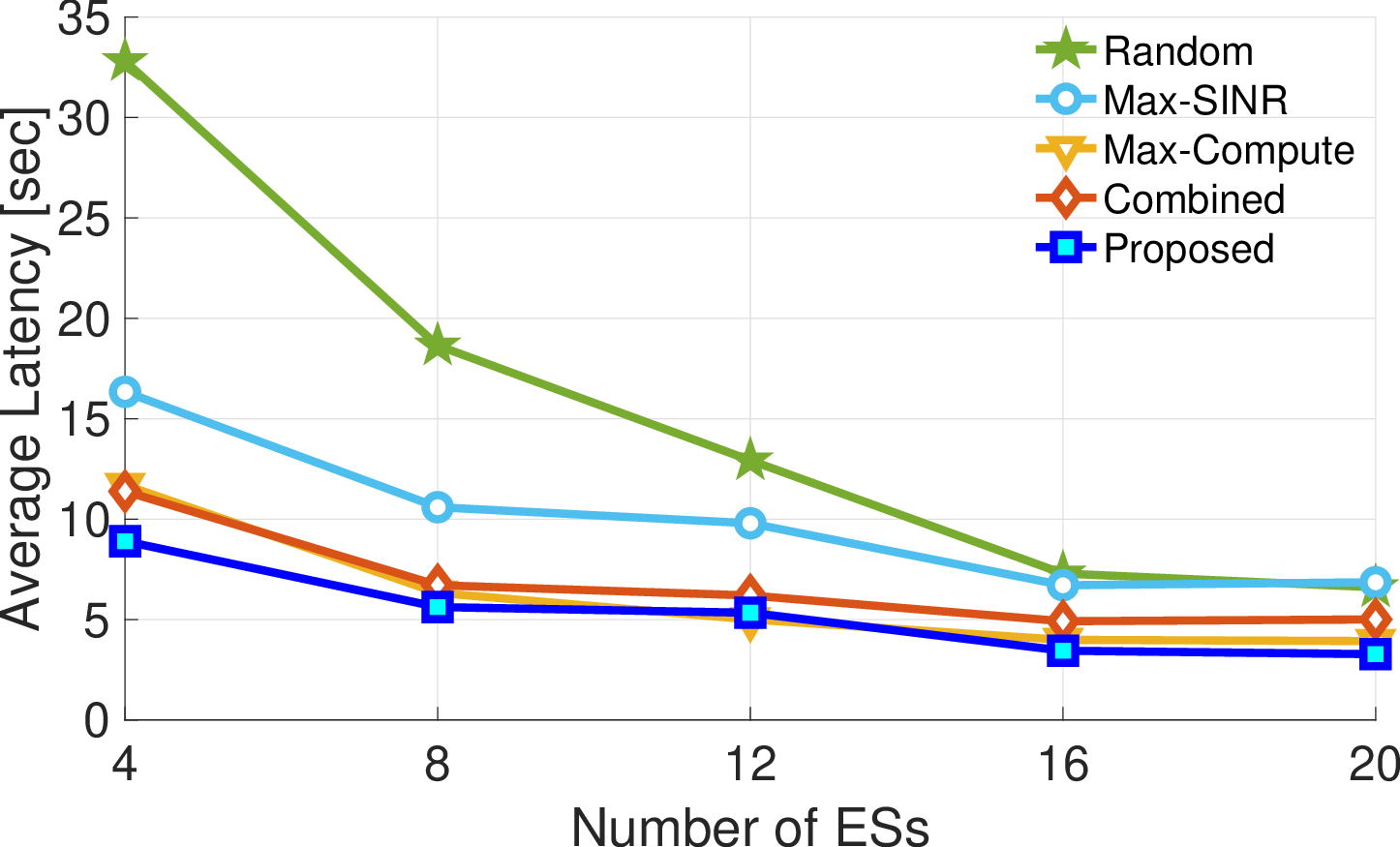}\label{subfig:highcompute_latency_ue_60}}
        \caption{\blue{The system latency of networks where $N_{\text{MD}}=60$ and $\alpha=1$: (a) Network with high communication load; (b) Network with high computational load.}}
        \label{fig:latency_compare_fixed_md}
    \end{figure}
\else
    \begin{figure}
        \centering
        \subfloat[]{\includegraphics[width=\if\mycmd1 0.6\else 0.75\fi\linewidth]{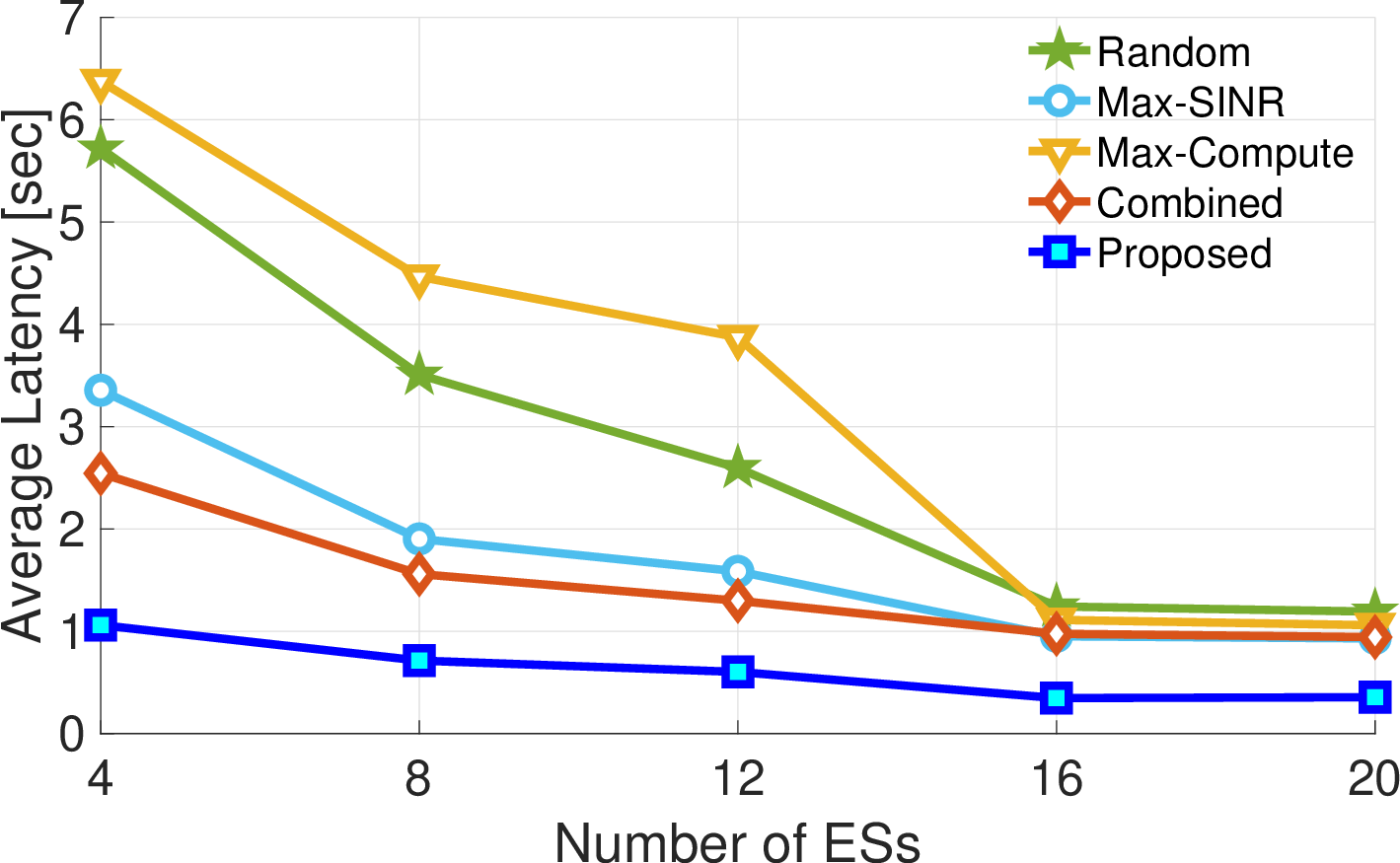}\label{subfig:highcommun_latency_ue_60}}
        \hfil
        \subfloat[]{\includegraphics[width=\if\mycmd1 0.6\else 0.75\fi\linewidth]{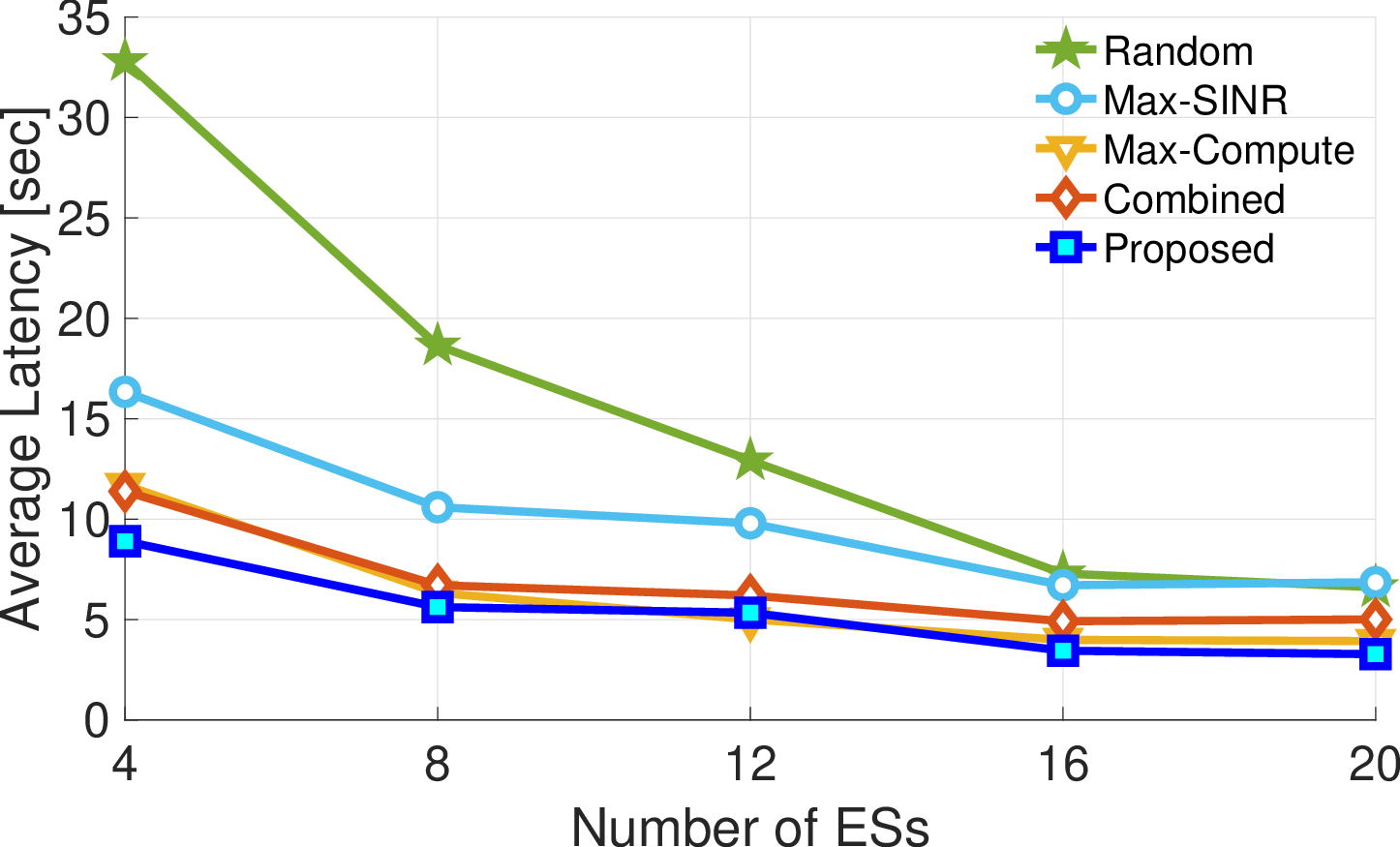}\label{subfig:highcompute_latency_ue_60}}
        \caption{{The system latency of networks where $N_{\text{MD}}=60$ and $\alpha=1$: (a) Network with high communication load; (b) Network with high computational load.}}
        \label{fig:latency_compare_fixed_md}
    \end{figure}
\fi


{
Previously, we have observed how different method performs as the number of MDs increases.
Next, we investigate the scalability of the method concerning the number of ESs.
Figure \ref{fig:latency_compare_fixed_md} exhibits the latency of each method relative to the number of ESs.
Starting from $N_{\text{ES}}=4$, we gradually increase the number of ESs up to 20 to analyze how the proposed method scales.
The results indicate that as the number of ESs increases, the latency of all the methods gradually decreases. 
More importantly, although necessities of optimization is decreased as more ESs can reduce load their individual load on computing/computation, the proposed method consistently outperforms the baseline methods. 
This demonstrates that the proposed method is robust to the change of the number of ESs, showing good scalability property while achieving the lowest latency among the schemes.
}

The results indicate that the proposed scheme is able to achieve a low latency, regardless of the network environment.
The proposed scheme outperforms other baseline schemes, even the schemes that specialized in specific network environments.

\subsubsection{Energy Analysis} \label{subsec:energy_analysis}

\begin{table}[]
\centering
\caption{Energy consumption per task of each scheme in different network environment with $N_{\text{ES}}=12$. Local energy consumption is in mWh, and edge energy consumption is in Wh.}
\adjustbox{width=\if\mycmd1 0.7\else1 \fi\linewidth}{
\begin{tabular}{cc|cc|cc|cc}
\toprule
\multirow{2}{*}{Network type} &
  \multirow{2}{*}{Scheme} &
  \multicolumn{2}{c|}{$N_{\text{MD}}$=40} &
  \multicolumn{2}{c|}{$N_{\text{MD}}$=100} &
  \multicolumn{2}{c}{$N_{\text{MD}}$=160} \\ \cmidrule{3-8}
 &             & \multicolumn{1}{c|}{Local} & Edge & \multicolumn{1}{c|}{Local} & Edge & \multicolumn{1}{c|}{Local} & Edge \\ 
 \midrule
\multirow{5}{*}{\begin{tabular}[c]{@{}c@{}}Communication\\ -intensive\end{tabular}} &
  Random &
  \multicolumn{1}{c|}{0.280} &
  24.6 &
  \multicolumn{1}{c|}{0.231} &
  28.1 &
  \multicolumn{1}{c|}{0.232} &
  31.0 \\ 
 & Max SINR    & \multicolumn{1}{c|}{0.172} & 19.5 & \multicolumn{1}{c|}{0.128} & 22.4 & \multicolumn{1}{c|}{0.128} & 25.2 \\ 
 & Max Compute & \multicolumn{1}{c|}{0.184} & 27.0 & \multicolumn{1}{c|}{0.260} & 29.0 & \multicolumn{1}{c|}{0.303} & 30.8 \\ 
 & Combined    & \multicolumn{1}{c|}{0.177} & 23.6 & \multicolumn{1}{c|}{0.183} & 26.4 & \multicolumn{1}{c|}{0.211} & 28.7 \\ 
 & Proposed    & \multicolumn{1}{c|}{\textbf{\underline{0.062}}} & 8.42 & \multicolumn{1}{c|}{\textbf{\underline{0.061}}} & 8.97 & \multicolumn{1}{c|}{\textbf{\underline{0.077}}} & 9.44 \\ 
 \midrule
\multirow{5}{*}{\begin{tabular}[c]{@{}c@{}}Computation\\ -intensive\end{tabular}} &
  Random &
  \multicolumn{1}{c|}{0.311} &
  132 &
  \multicolumn{1}{c|}{0.361} &
  138 &
  \multicolumn{1}{c|}{0.441} &
  141 \\ 
 & Max SINR    & \multicolumn{1}{c|}{0.301} & 126  & \multicolumn{1}{c|}{0.324} & 149  & \multicolumn{1}{c|}{0.365} & 182  \\ 
 & Max Compute & \multicolumn{1}{c|}{0.338} & 131  & \multicolumn{1}{c|}{0.275} & 167  & \multicolumn{1}{c|}{0.272} & 199  \\ 
 & Combined    & \multicolumn{1}{c|}{0.315} & 109  & \multicolumn{1}{c|}{0.303} & 138  & \multicolumn{1}{c|}{0.332} & 166  \\ 
 & Proposed    & \multicolumn{1}{c|}{\textbf{\underline{0.129}}} & 89.9 & \multicolumn{1}{c|}{\textbf{\underline{0.182}}} & 94.8 & \multicolumn{1}{c|}{\textbf{\underline{0.241}}} & 100  \\ 
 \bottomrule
 \multicolumn{8}{l}{* \underline{\textbf{bold \& underline}}: best performance}
\end{tabular}
}
\label{tab:energy_usage}
\end{table}

Now, we analyze the energy consumption results of the proposed scheme and the baseline schemes.
The energy consumption per task under different network scenarios is shown in Table \ref{tab:energy_usage}.
The values indicate that our method achieves the lowest local energy consumption per task for both the communication-intensive network and computation-intensive network.
In each network environment, our method consumed up to 64.0\% and 57.1\% less energy respectively compared to the best performing baseline model.
The remarkable energy conservation performance while achieving low latency lies in the strategic offloading policy.
As for the communication-intensive networks, since tasks impose high communication load, increasing the local computation ratio proves to be a viable solution.
Therefore, the proposed scheme achieves an outstanding performance by modulating its offloading ratio accordingly.
Conversely, in computation-intensive networks, where tasks have high computation load with relatively low communication load, increasing the offloading ratio could lead to better performance in both energy conservation and latency.
Hence, the proposed method achieves the best performance by adaptively regulating its offloading ratio.

\subsubsection{Pareto Graph Analysis}

\if\mycmd1
    \begin{figure}[!t]
        \centering
        \subfloat[]{\includegraphics[width=0.45\linewidth]{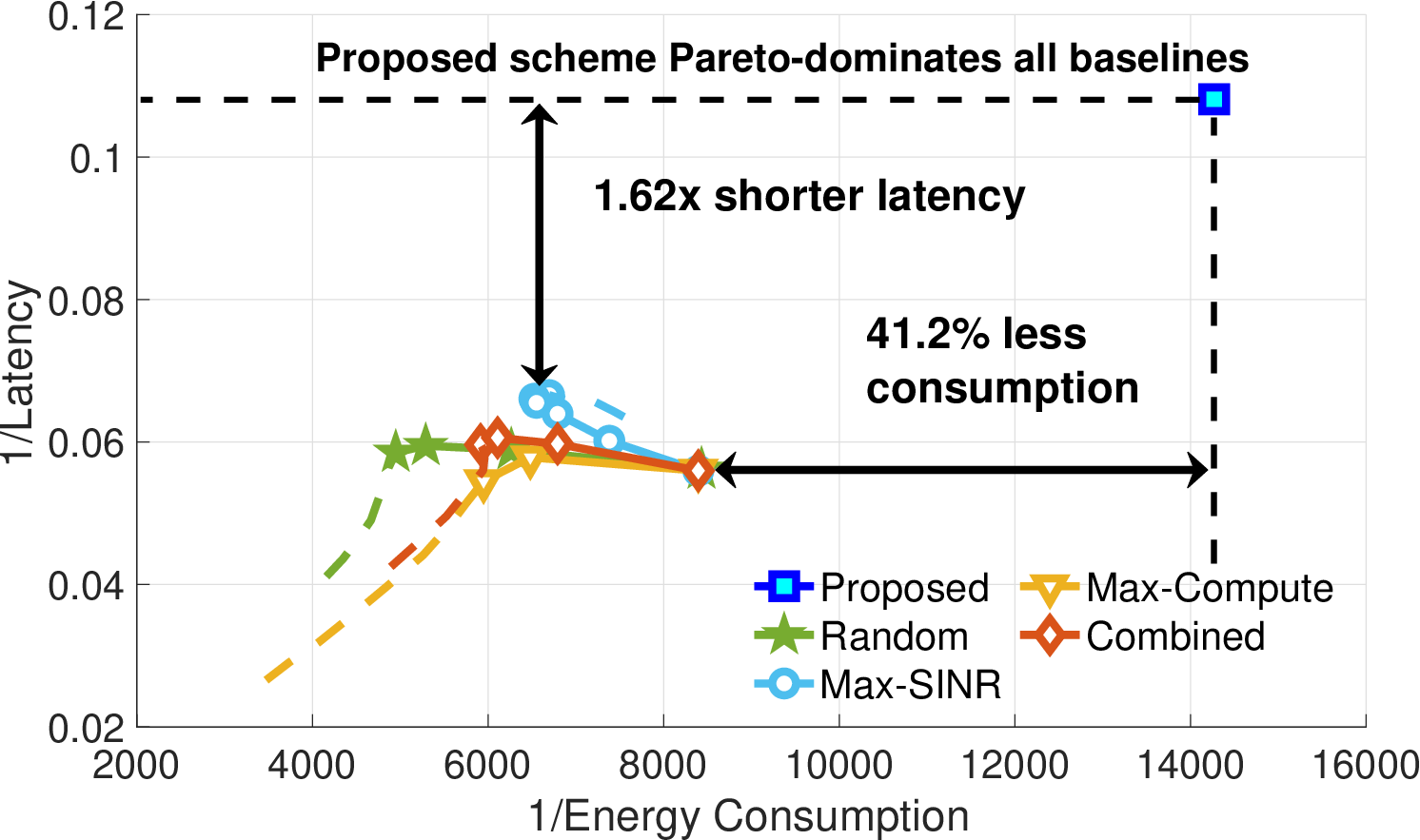}\label{subfig:highcommun-pareto}}
        \subfloat[]{\includegraphics[width=0.45\linewidth]{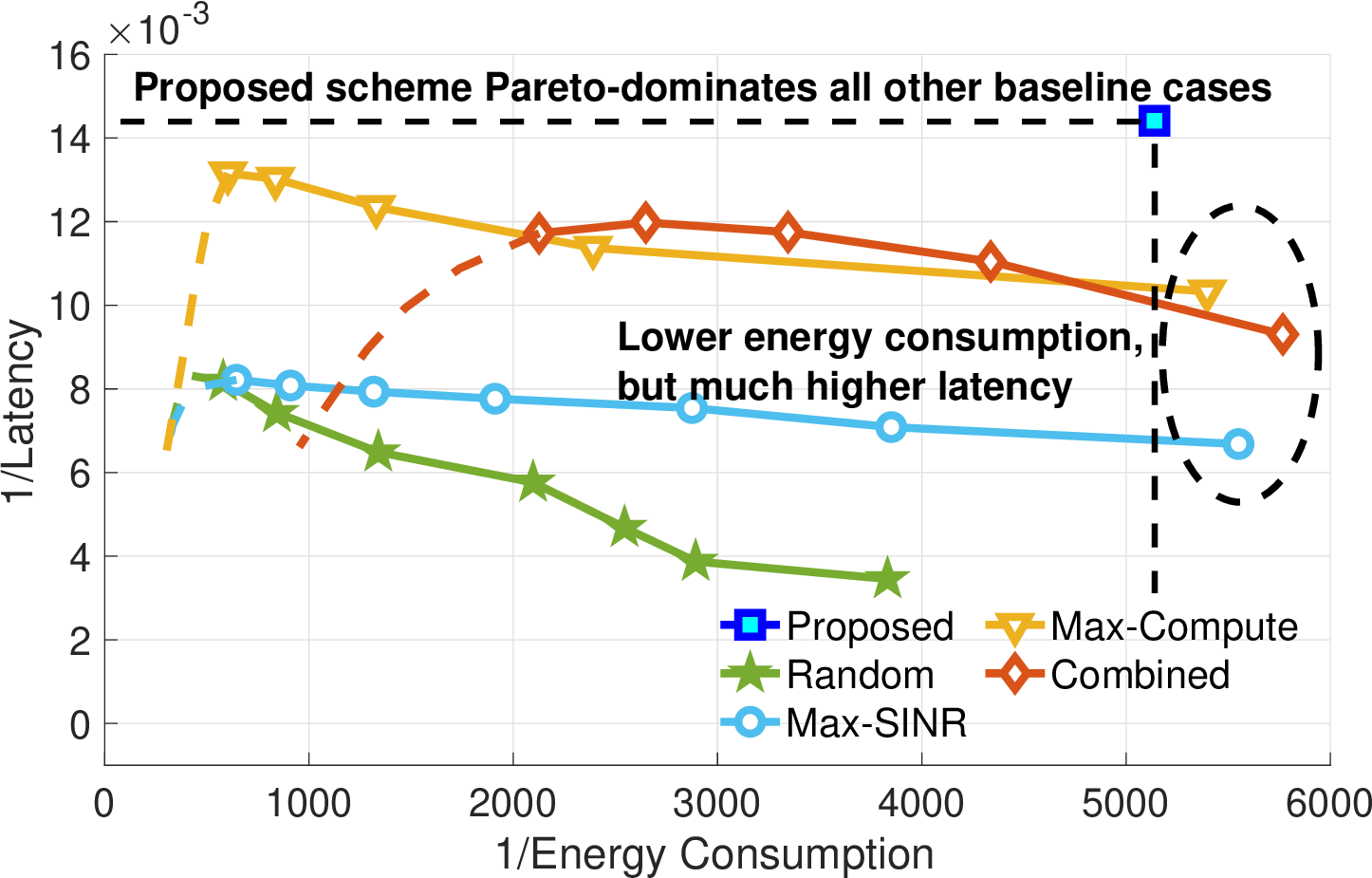}\label{subfig:highcompute-pareto}}
        \caption{Pareto efficiency graphs of different networks with 80 MDs and 8 ESs, as $\epsilon$ ranges from $0.0$ to $1.0$: (a) Network with high communication load; (b) Network with high computational load.}
        \label{fig:pareto}
    \end{figure}
\else
    \begin{figure}[!t]
        \centering
        \subfloat[]{\includegraphics[width=\if\mycmd1 0.6\else 0.75\fi\linewidth]{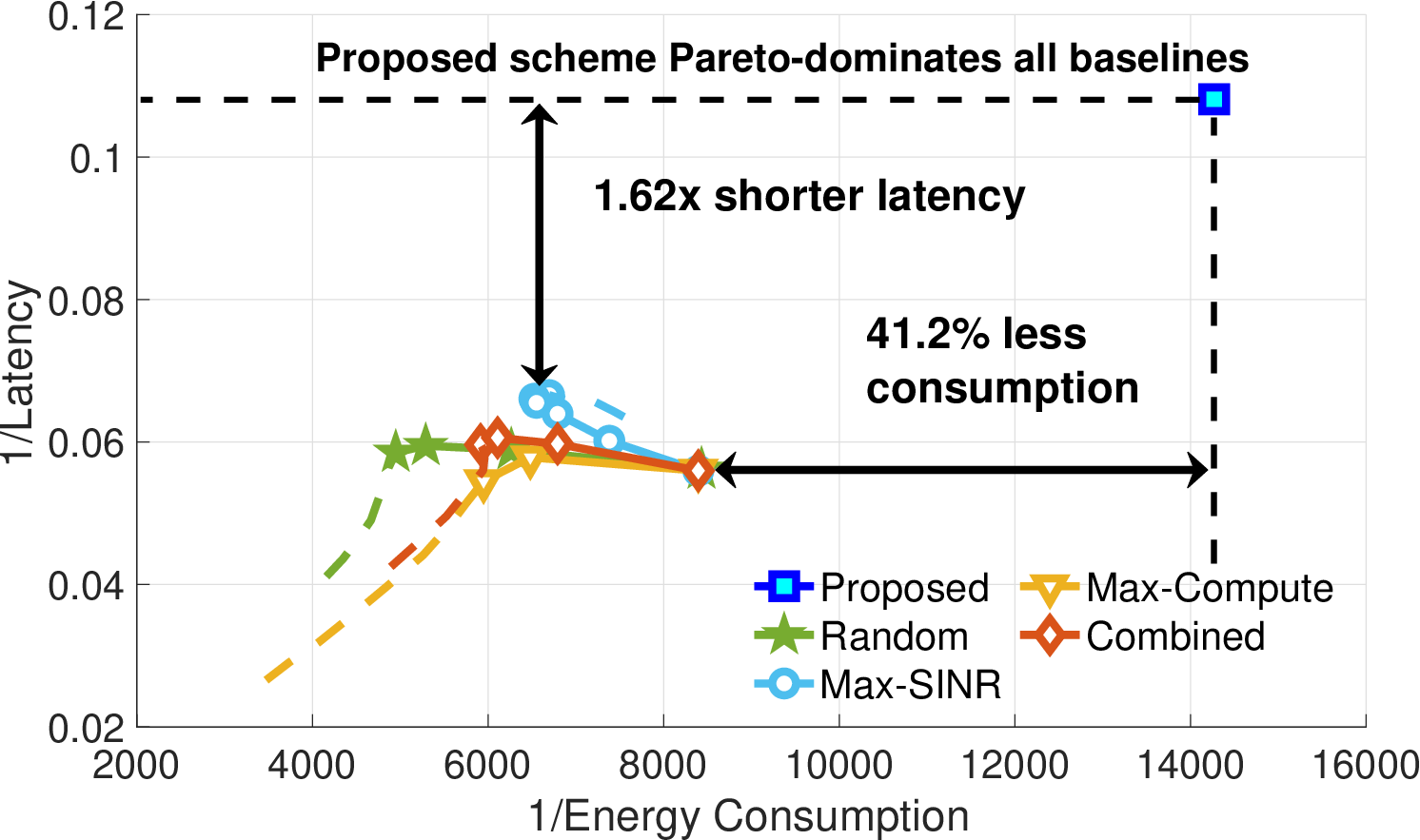}\label{subfig:highcommun-pareto}}
        \hfil
        \subfloat[]{\includegraphics[width=\if\mycmd1 0.6\else 0.75\fi\linewidth]{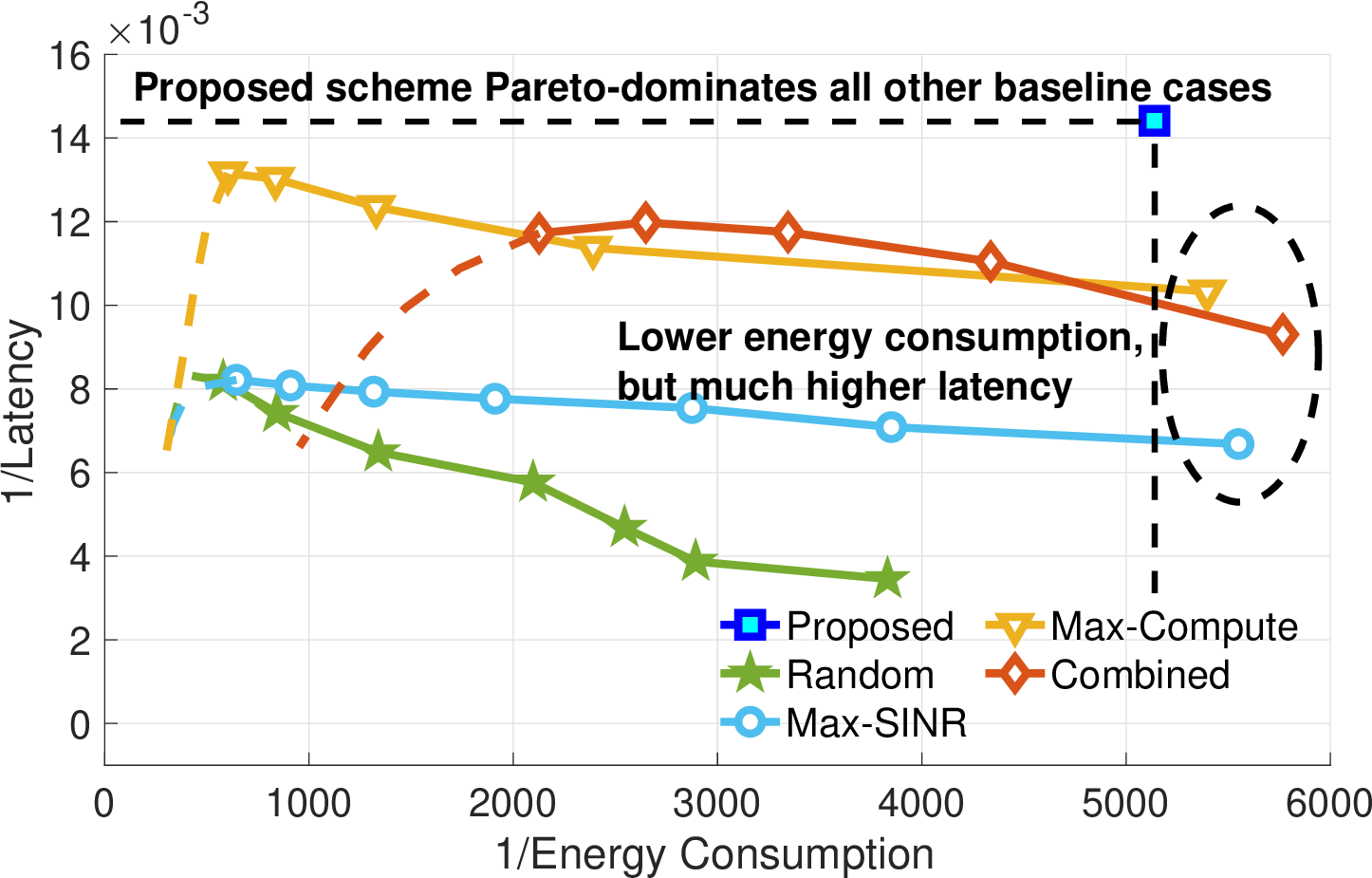}\label{subfig:highcompute-pareto}}
        \caption{Pareto efficiency graphs of different networks with 80 MDs and 8 ESs, as $\epsilon$ ranges from $0.0$ to $1.0$: (a) Network with high communication load; (b) Network with high computational load.}
        \label{fig:pareto}
    \end{figure}
\fi
From Table \ref{tab:energy_usage}, we analyzed how offloading ratio could effect the latency and the energy conservation performance.
This leads to a question: \textit{Could the baseline schemes outperform the proposed scheme if they increase their local computation ratio?}
To answer this question, we plot a Pareto efficiency graph of the reciprocals of latency and energy consumption of the models. 
Pareto efficiency graph illustrates a trade-off relationship between two or more values.
We compare the schemes' performance by examining their locations on the Pareto graph.
Since better latency and energy consumption leads to larger reciprocals of their values, if the proposed scheme is located at the upper right of the baseline schemes, this indicates that the proposed scheme excels over the baselines in both latency and energy consumption.
Additionally, if the proposed scheme surpasses the baseline in all axes, it indicates that the proposed scheme Pareto-dominates the baseline, signifying its superiority in all aspects.

To evaluate all possible cases of the baseline models, we simulate the network numerous times, while gradually increasing the offloading ratio $\epsilon$ from 0.0 to 1.0, with an increment of 0.1.
As for the proposed model, we used $\alpha=1$.
As can be observed in Fig. \ref{subfig:highcommun-pareto}, in a system with heavy communication load, the proposed scheme's performance surpasses other baseline schemes in both latency and energy efficiency.
In other words, our scheme Pareto-dominates all baseline schemes.
More specifically, the proposed scheme achieved up to 1.62 times shorter latency compared to the best performing baseline models, while simultaneously consuming 41.2\% less energy.
The result indicates that the baseline schemes cannot outperform the proposed scheme, regardless of the offloading ratio.
In Fig. \ref{subfig:highcompute-pareto}, some baseline schemes outperformed our scheme in terms of energy efficiency.
However, our scheme still outperformed the baseline in latency, achieving up to 2.16 times shorter latency while using at most 10.9\% more energy.
We also remark that the proposed scheme can also achieve such good energy consumption by increasing $\alpha$.

We note that in Fig. \ref{fig:pareto}, we only emphasized the points of the agent performance that is Pareto-efficient, i.e., we only dramatized the best performances of each agent.
The dashed line segments are affected by other factors such as task overload delay, but is negligible since the change of offloading ratio leads to better performance of both aspects of latency and energy efficiency.


\subsubsection{Effects of unit-changing parameter $\alpha$}\label{subsec:mu}

In our problem formulation, we introduced a new variable $\alpha$ that balances MD's energy consumption and latency.
Using $\alpha$ as a unit conversion parameter, we were able to jointly optimize energy usage and system latency. 
We now investigate the influence of $\alpha$ on the system's latency and energy consumption.

$\alpha$ is a weight that is multiplied to the energy penalty term of a device.
The effect that the leftover energy has on the problem will increase as $\alpha$ increases.
More intuitively, the proposed scheme will choose to compute on an edge device more often when $\alpha$ is large, in order to preserve energy.

\if\mycmd1
    \begin{figure}[t]
        \centering
        \subfloat[]
        {\includegraphics[width=0.45\linewidth]{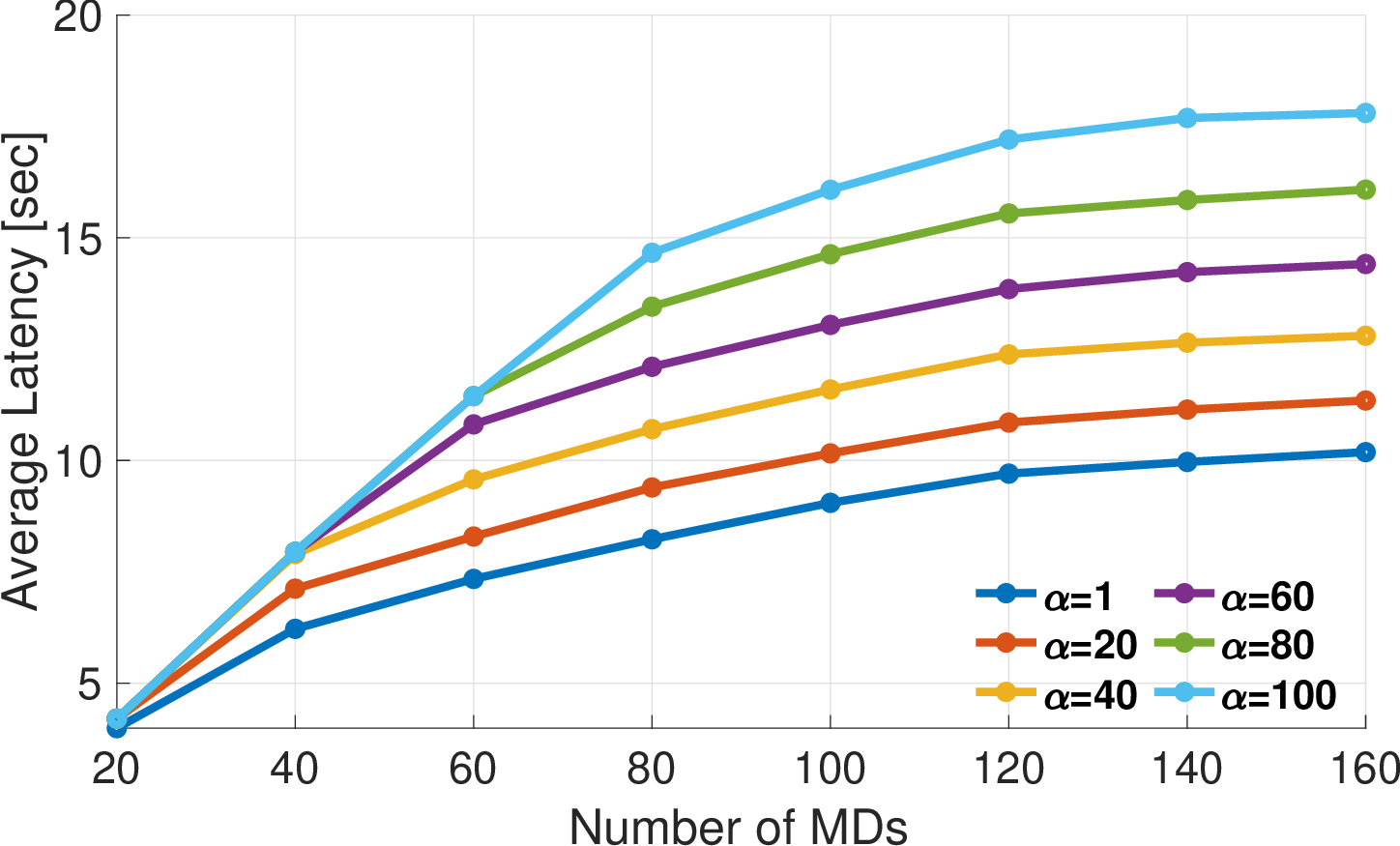}
        \label{subfig:mu_latency}}
        \subfloat[]
        {\includegraphics[width=0.45\linewidth]{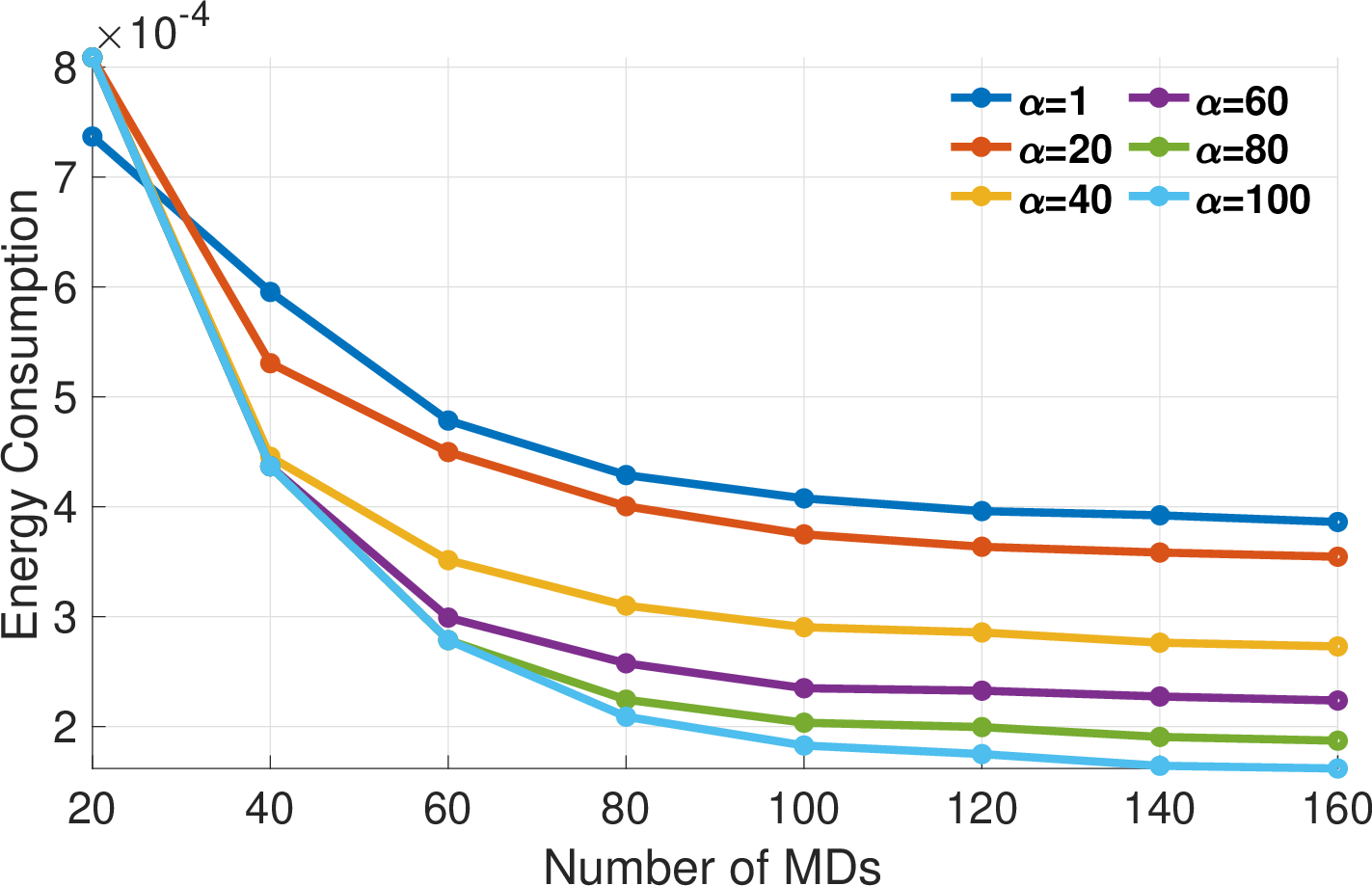}
        \label{subfig:mu_energy}}
        \caption{Average latency and energy consumptions of the network for various $\alpha$ ($N_{\text{ES}}=4$): \protect\subref{subfig:mu_latency} Average latency for different $\alpha$ values. Latency rises with an increase of $\alpha$ due to the escalation of local computation; (b) Local energy consumption for different $\alpha$ values. Energy consumption decreases as $\alpha$ rises, due to the increase of local computation.}
        \label{fig:mu_stats}
    \end{figure}
\else
    \begin{figure}[t]
        \centering
        \subfloat[]
        {\includegraphics[width=\if\mycmd1 0.6\else 0.75\fi\linewidth]{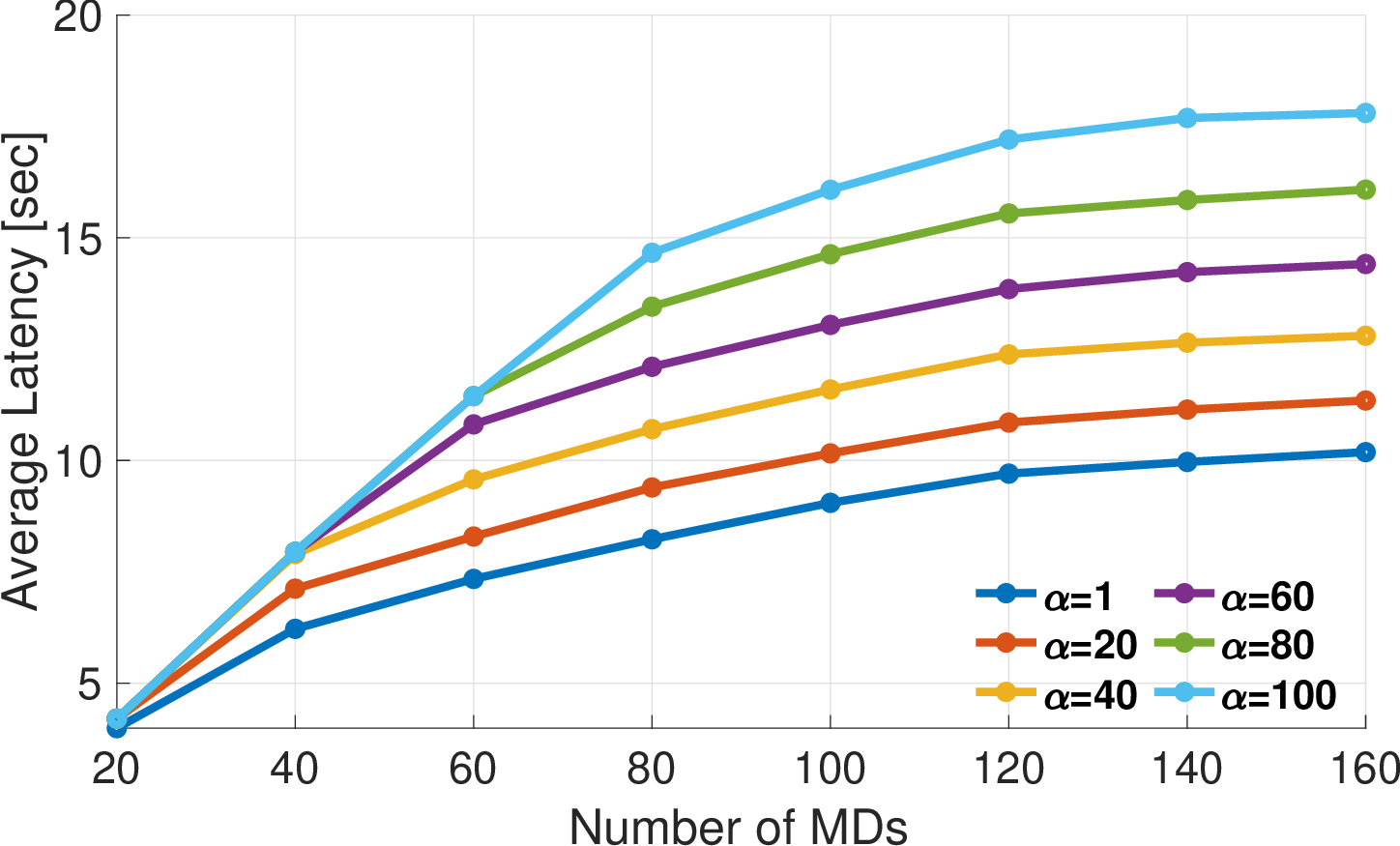}
        \label{subfig:mu_latency}}
        \hfil
        \subfloat[]
        {\includegraphics[width=\if\mycmd1 0.6\else 0.75\fi\linewidth]{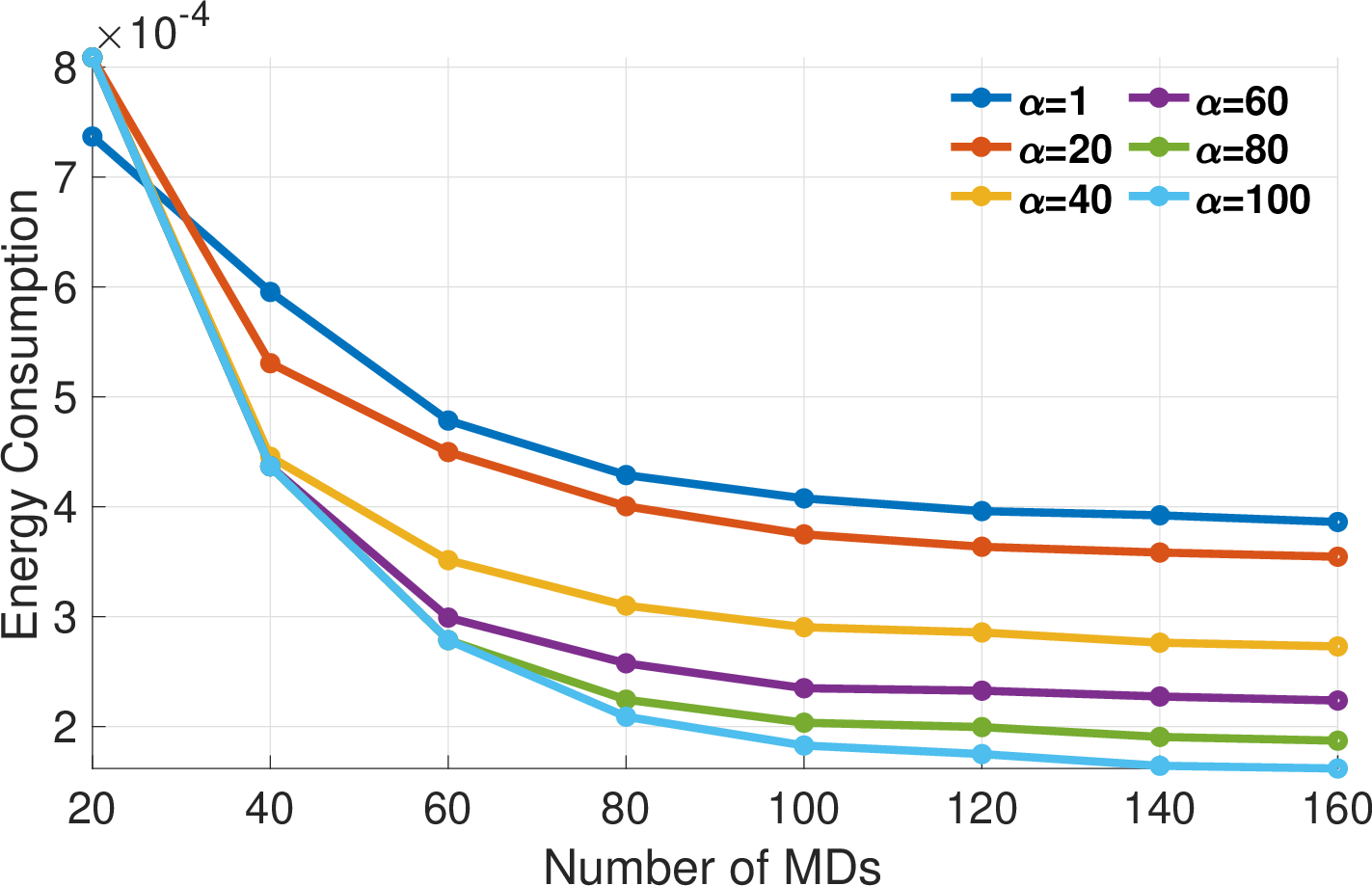}
        \label{subfig:mu_energy}}
        \caption{Average latency and energy consumptions of the network for various $\alpha$ ($N_{\text{ES}}=4$): \protect\subref{subfig:mu_latency} Average latency for different $\alpha$ values. Latency rises with an increase of $\alpha$ due to the escalation of local computation; (b) Local energy consumption for different $\alpha$ values. Energy consumption decreases as $\alpha$ rises, due to the increase of local computation.}
        \label{fig:mu_stats}
    \end{figure}
\fi

The effect of $\alpha$ on the system latency is depicted in Fig. \ref{subfig:mu_latency}.
As predicted, the lowest latency is achieved when $\alpha$ is at its minimum. 
In networks with small number of users, the effects of $\alpha$ is not as severe as in networks with large number of users, as can be seen from the small gap between the latency.
This is because the users experience sufficient amount of communication and computational resources when $N_{\text{MD}}$ is small.
Hence, the latency discrepancy between the systems is marginal when $N_{\text{MD}}$ is small, and the gap grows as the number of MDs increases.
Figure \ref{subfig:mu_energy} visualizes the relationship between the number of MDs and the local energy consumption.
Since higher value of $\alpha$ decreases the number of local computations, energy consumption naturally decreases as $\alpha$ increases.

In both of the figures of Fig. \ref{fig:mu_stats}, the inclination of the graphs tend to converge after a certain number of users.
The reason can be found from the convergence of the offloading ratio.
As the number of MDs increases, the saturation rate of the ESs rises as there are more tasks to compute.
But if the number of MDs is sufficiently large, the advantages of edge computing diminishes, as edge computing experiences bigger latency due to the overload of ES computation.
Therefore, the offloading ratio converges for large MDs, leading to a saturation of both the average latency and energy consumption of each MD.




\subsubsection{{Effects of step size $\eta$ on convergence}}
 \label{subsec:lr_convergence}

{
In Section \ref{subsec:convergence_analysis}, we have analyzed how the sub-gradient descent update step size $\eta$ could affect the convergence of the solution.
More specifically, our analysis suggests that by choosing the optimal step size, the optimality gap will monotonically decrease as the iteration $T$ increases.
Here, we experimentally visualize how different step sizes affect the update process of Lagrangian multipliers $\boldsymbol{\mu}$ and $\boldsymbol{\nu}$.
Finally, we validate that different step sizes could lead to different convergence properties of the proposed scheme.
}

\if\mycmd1
    \begin{figure}[!t]
        \centering
        \subfloat[]{\includegraphics[width=0.45\linewidth]{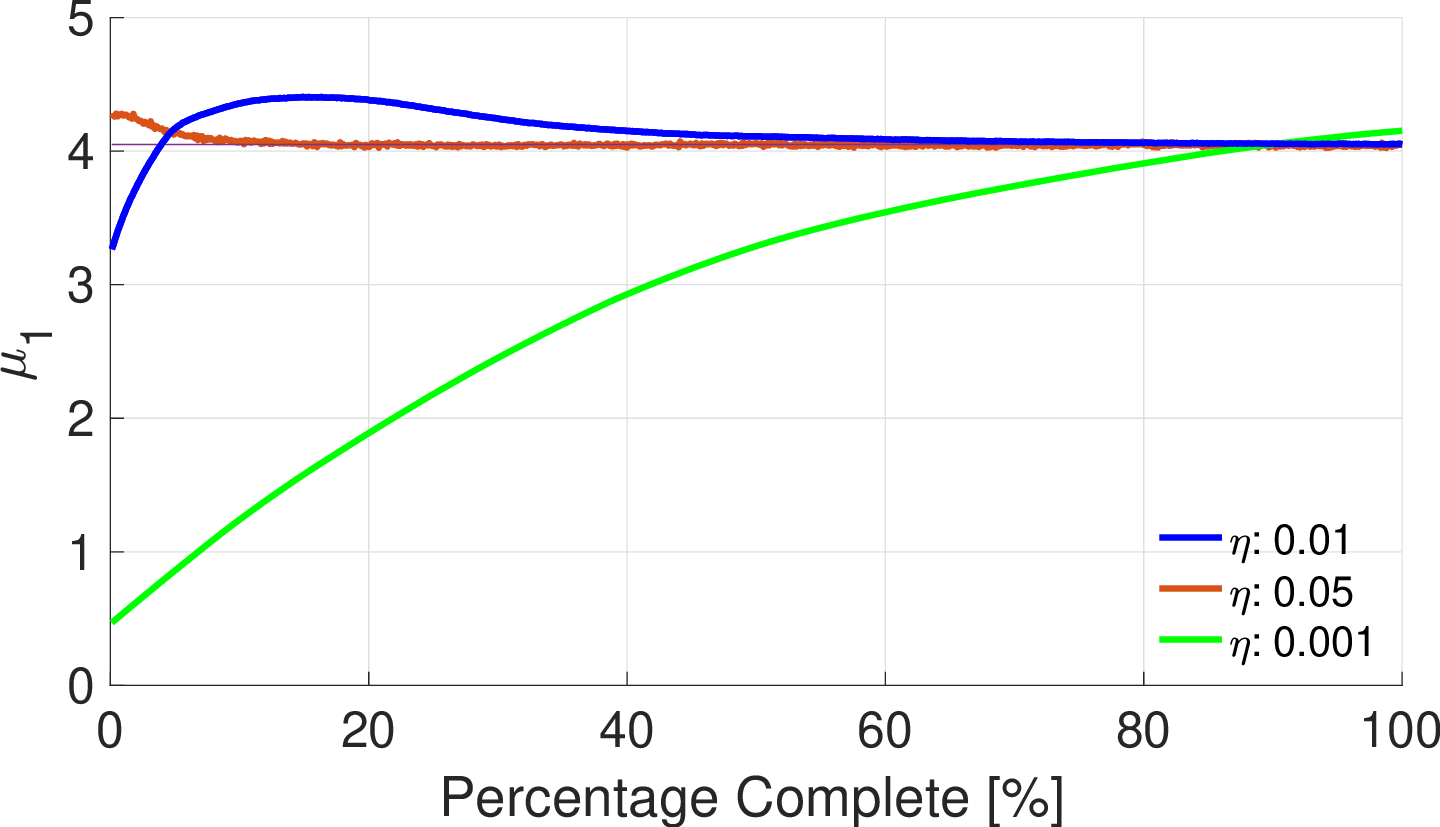}\label{subfig:mu_comparison}}
        \subfloat[]{\includegraphics[width=0.45\linewidth]{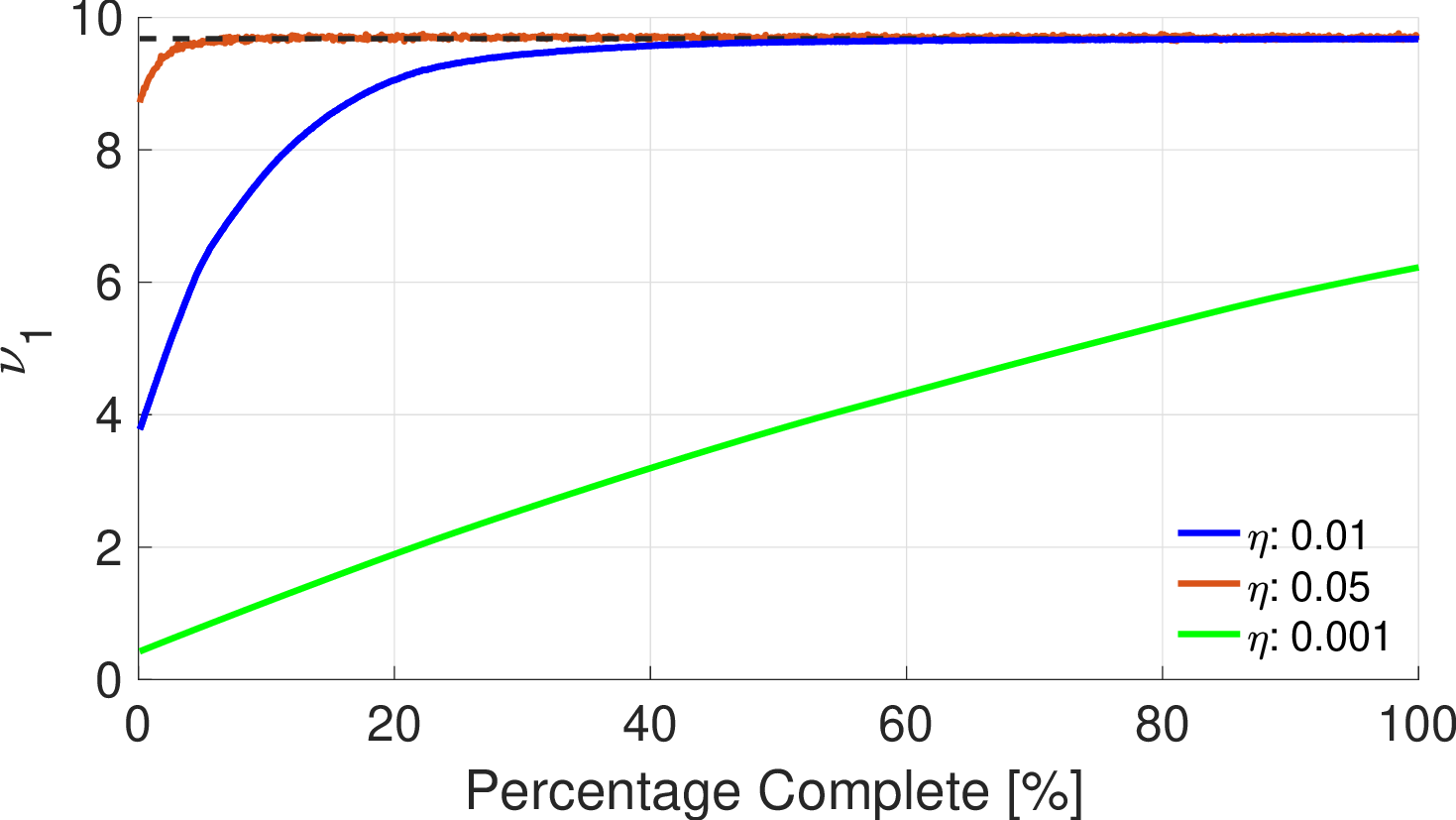}\label{subfig:nu_comparison}}
        \caption{{Convergence graph of the Lagrangian multipliers $\boldsymbol{\mu}$ and $\boldsymbol{\nu}$ at different step sizes, where the network consists of 80 MDs and 4 ESs: (a) Lagrange multiplier $\mu_1$ for ES 1; (b) Lagrange multiplier $\nu_1$ for ES 1.}}
        \label{fig:mu_nu}
    \end{figure}
\else
    \begin{figure}[!t]
        \centering
        \subfloat[]{\includegraphics[width=\if\mycmd1 0.6\else 0.75\fi\linewidth]{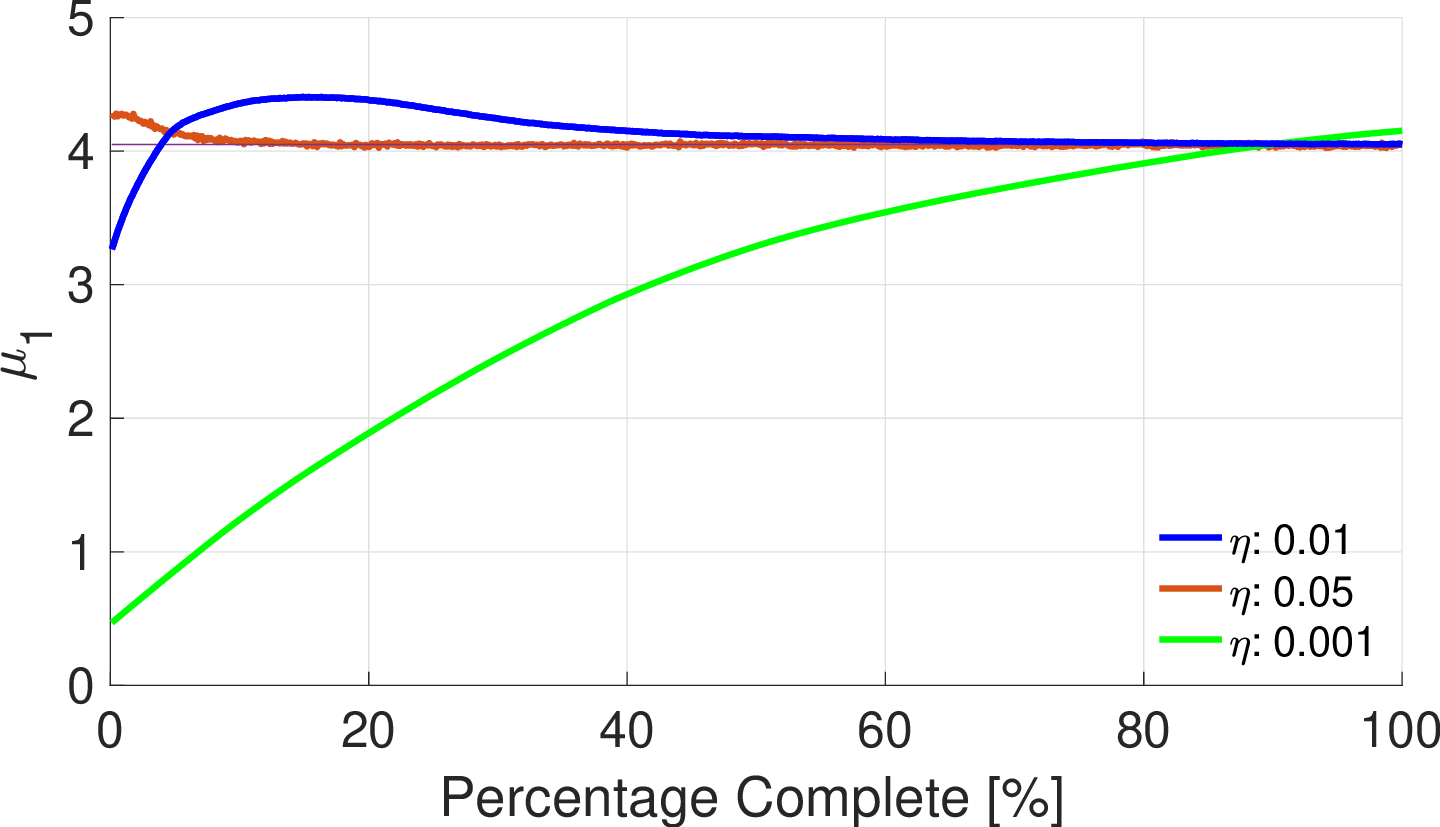}\label{subfig:mu_comparison}}
        \hfil
        \subfloat[]{\includegraphics[width=\if\mycmd1 0.6\else 0.75\fi\linewidth]{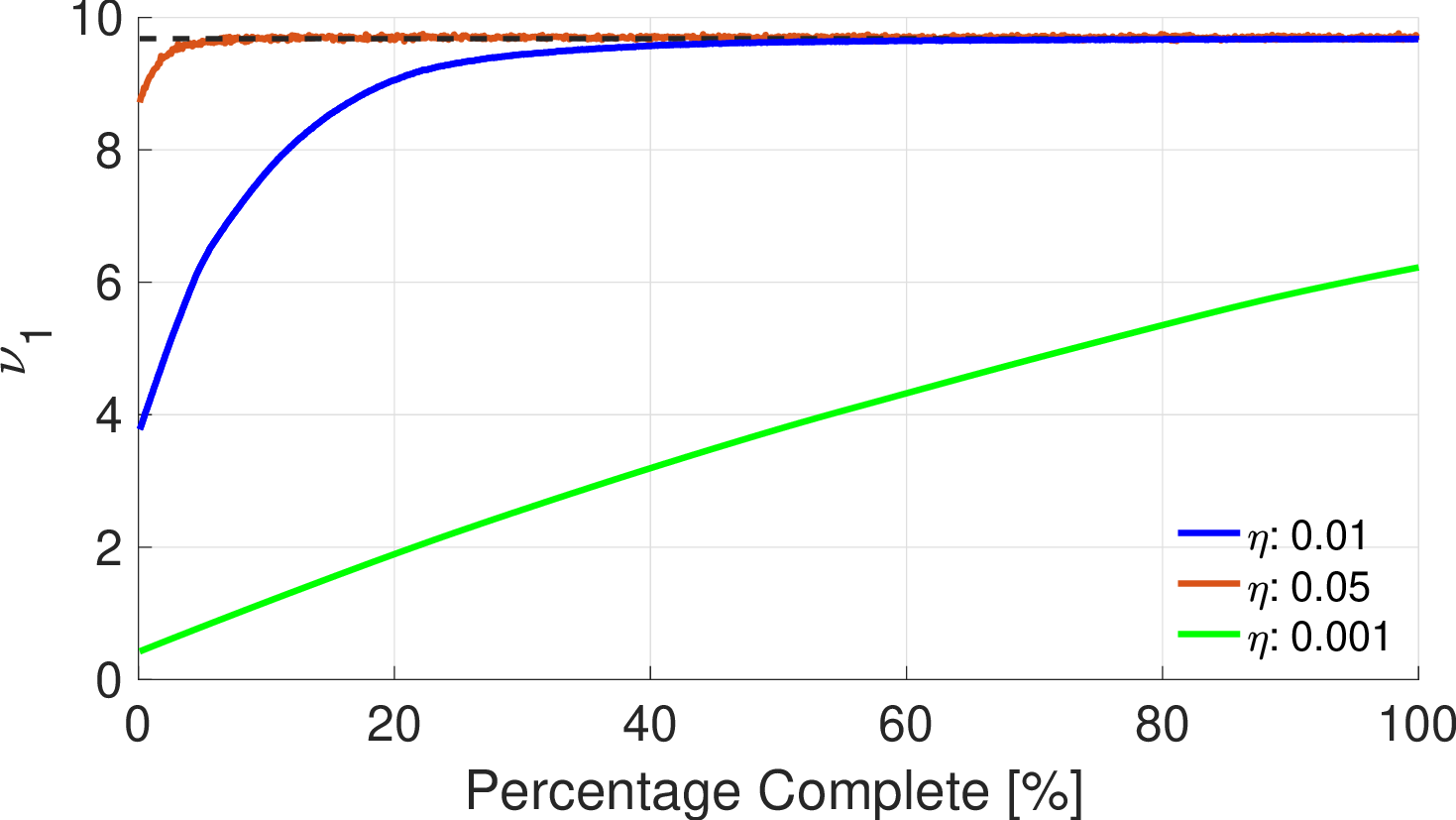}\label{subfig:nu_comparison}}
        \caption{{Convergence graph of the Lagrangian multipliers $\boldsymbol{\mu}$ and $\boldsymbol{\nu}$ at different step sizes, where the network consists of 80 MDs and 4 ESs: (a) Lagrange multiplier $\mu_1$ for ES 1; (b) Lagrange multiplier $\nu_1$ for ES 1.}}
        \label{fig:mu_nu}
    \end{figure}
\fi

\if\mycmd1
    \begin{figure}[!t]
        \centering
        \subfloat[]{\includegraphics[width=0.45\linewidth]{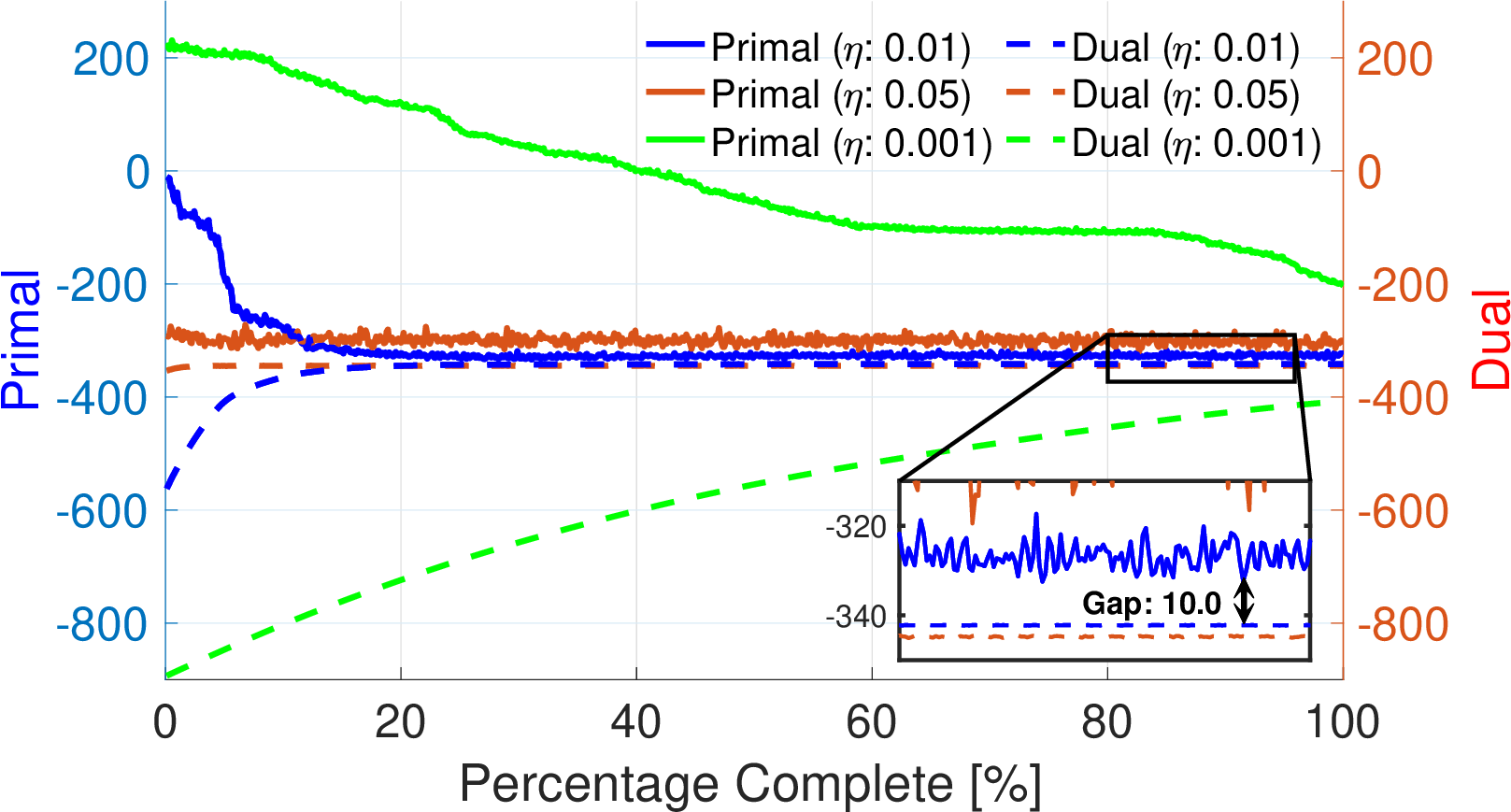}\label{subfig:primal_dual}}
        \subfloat[]{\includegraphics[width=0.45\linewidth]{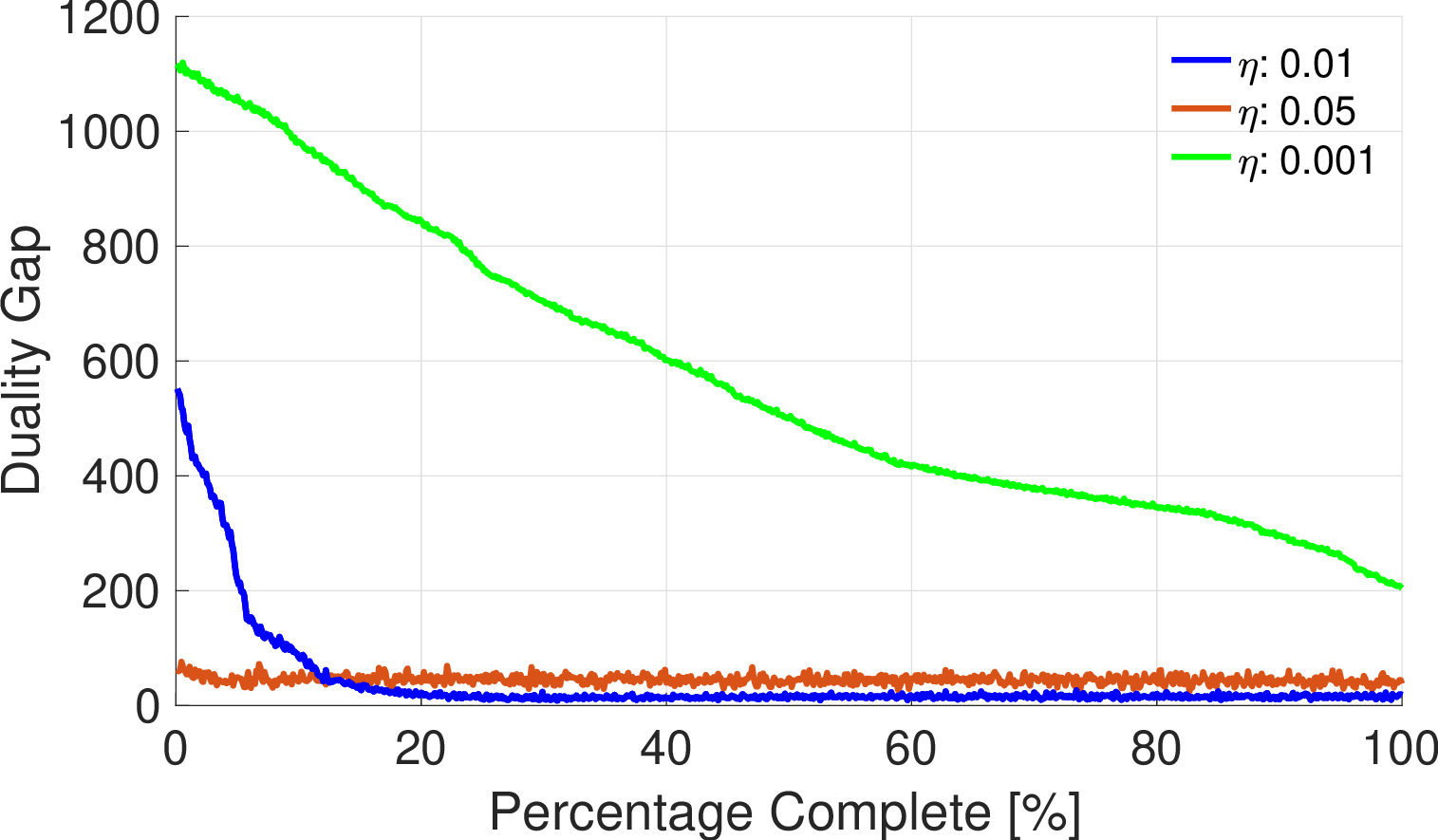}\label{subfig:duality_gap}}
        \caption{\blue{Convergence of the objective functions at different step sizes, where the network consists of 80 MDs and 4 ESs: (a) Convergence graph of primal and dual functions; (b) Convergence graph of the duality gap.}}
        \label{fig:duality}
    \end{figure}
\else
    \begin{figure}[!t]
        \centering
        \subfloat[]{\includegraphics[width=\if\mycmd1 0.6\else 0.75\fi\linewidth]{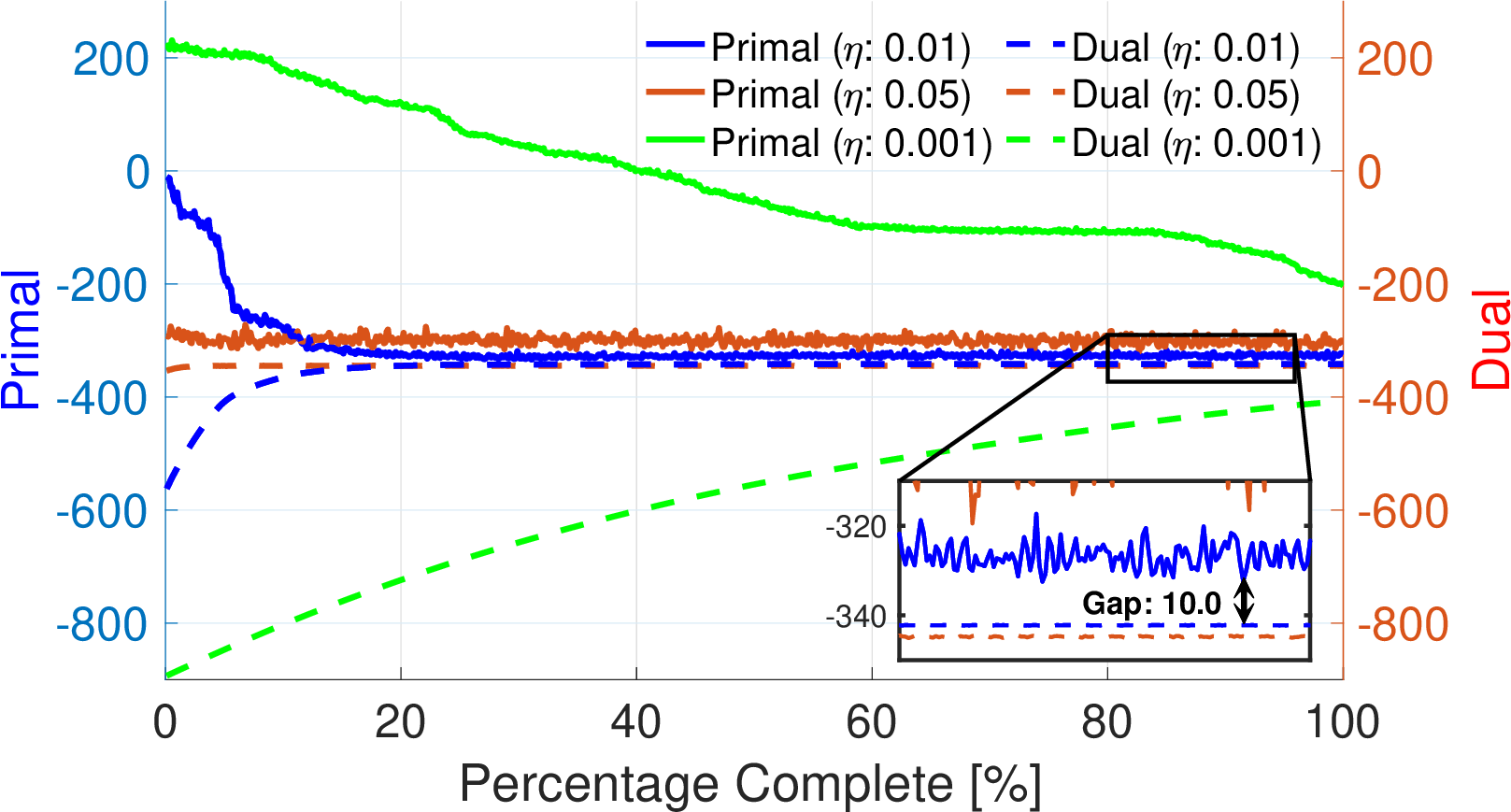}\label{subfig:primal_dual}}
        \hfil
        \subfloat[]{\includegraphics[width=\if\mycmd1 0.6\else 0.75\fi\linewidth]{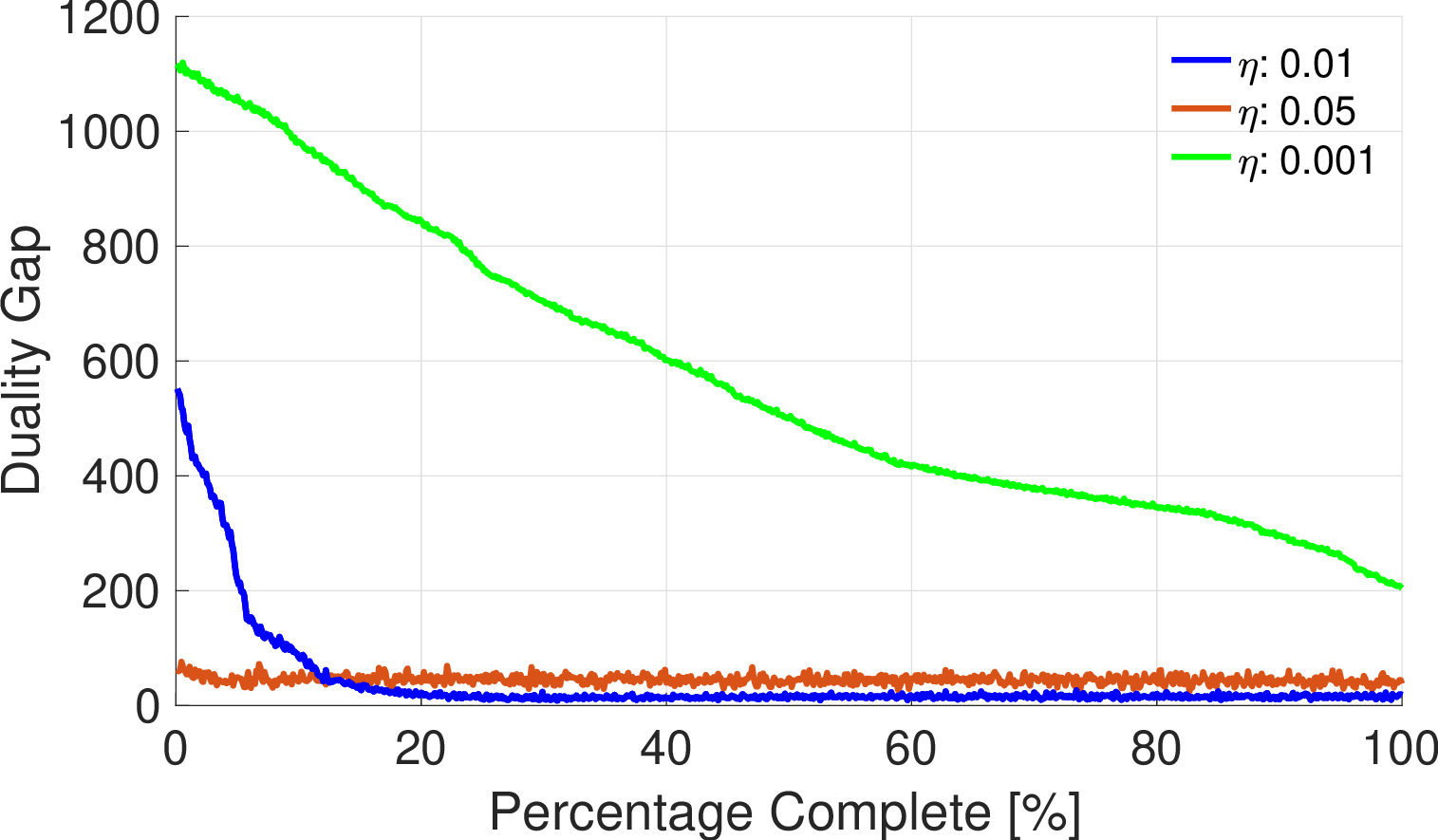}\label{subfig:duality_gap}}
        \caption{{Convergence graph of the objective functions at different step sizes, where the network consists of 80 MDs and 4 ESs: (a) Primal and dual function values of ES 1; (b) Duality gap of ES 1.}}
        \label{fig:duality}
    \end{figure}
\fi

{
Figure \ref{fig:mu_nu} illustrates the effect of different step sizes on the convergence of the Lagrangian multipliers $\boldsymbol{\mu}$ and $\boldsymbol{\nu}$.
Although the experimented step sizes all lead to the gradual convergence of the Lagrangian multipliers, some step sizes showed better convergence performance compared to other values.
As for Fig. \ref{fig:mu_nu}, $\eta=0.01$ and $\eta=0.05$ showed fast convergence compared to $\eta=0.001$.
Figure \ref{fig:duality} exhibits the value of the primal and dual functions of the first ES, as the Lagrangian multipliers $\mu_1$ and $\nu_1$ updates.
Similar to the simulation settings of Fig. \ref{fig:mu_nu}, we compare the results using step sizes of  $0.001$. $0.01$ and $0.05$.
These step sizes are applied to both $\eta_1$ and $\eta_2$.
Consistent with the previous results of Fig. \ref{fig:mu_nu}, $\eta=0.01$ and $\eta=0.05$ demonstrated a good convergence speed.
From Fig. \ref{subfig:primal_dual}, it can be observed that the primal and dual functions corresponding to $\eta=0.01$ has the smallest duality gap.
It can be inferred that a step size of $0.01$ approximates the optimum more closely, whereas a step size of $0.05$ fails to achieve a similar level of precision.
The step size of $0.001$ showed the slowest convergence, due to its small step size.
The duality gap of 10 implies that one MD experienced approximately $10/80=0.125$ seconds of delay difference compared to the ideal latency.
The difference is negligible, given that the average delay of the network was approximately 10 seconds, as can be observed from Fig. \ref{subfig:latency_base}.
Therefore, we conclude that different step sizes affect the rate at which the solution reaches its optimum.
}

\section{Related Works} \label{sec:related_works}

Recent studies related to UARA optimization of MEC networks are categorized into the following subsections, where key characteristics of some of these works are compared with our work in Table \ref{tab:related_works}.

\subsection{Conventional Offloading Approaches: Pricing-Based Approach}  \label{subsec:related_works_conventional_offloading}

Several studies \cite{9738455,8977528,7870686,8796358,9767545,9770179} have been proposed for service/load offloading in cellular networks, characterized by ultra-dense networks, massive MIMO networks, heterogeneous networks (HetNets), and cloud random access networks (C-RAN).
{In particular, in \cite{10243579}, the authors presented an optimization-based approach to minimize delay in multi-cell MEC systems.
Assuming divisible tasks, the solution allocated portions to each ES to minimize delay while considering an energy constraint.}
In \cite{6824825}, the authors proposed a modern approach for network load balancing.
The aim of the approach is distributed optimization of UARA in HetNets.  
In the work, the concept of pricing-based load balancing has been proposed for the first time, which enables distributed network optimization without any additional information exchange between ESs. 
However, such approach was limited to an objective function, specifically proportional fairness, where an artistic formulation led to a pricing framework.
For instance, in \cite{7422839}, a pricing-based UARA optimization is further extended to backhaul-limited cellular networks.

Following \cite{6824825}, recent works have solved general framework for various objective functions such as $\alpha$-fairness \cite{9738455}. 
However, to the best of the authors' knowledge, there has been no previous study using pricing-based optimization for average delay minimization as well as computing/communication offloading in MEC scenarios.

\subsection{Task Offloading in MEC}

Task offloading studies in MEC systems have mainly focused on interpreting the problem as a conventional offloading problem with additional constraints such as user mobility and limited battery.
Typical solution methods include solving optimization problems \cite{8314696, 9525179, 8964328, 8789664, 9050660, 9714257, 9316898, 8745530, 9806318, 8886467, 9745305} to better optimize their objective, mainly task latency and energy consumption.  
Some researchers took a different approach and used Lyapunov optimization method to solve UARA problems in MEC system \cite{9712216, 9087909, 8854900, 10102429}.
The authors of \cite{8962353} used both Karush-Kuhn-Tucker (KKT) methods and Lyapunov optimization method to solve their problem.
Growing number of researchers are implementing deep learning techniques such as deep reinforcement learning (DRL) to optimize task offloading in MEC systems \cite{8771176, 9253665, 9678008, 10024305, 8854270, 9146372, 8941121, 9149124, 9214878}, due to DRL's potential to calculate optimal policies with comparably low computational load compared to calculating heuristic solutions.
Additionally, DRL may provide an optimal solutions in cases where other optimization methods can only yield sub-optimal solutions.
In \cite{10102429}, researchers used DRL after applying conventional solving methods to find the optimal solution.

Properties unique to AI-native MEC systems include energy constraints on MDs and full task offloading.
Many studies have considered energy of MDs as a constraint \cite{9712216, 8314696}.
The authors of \cite{10138567} have shown the fundamental trade-off between the delay and resource used in computing.
Few studies have regarded energy of MDs as their objective \cite{8886467, 9149124, 10102429}.
Full offloading of tasks has also been dealt with in previous research \cite{9712216, 8771176, 9745305}.
The authors of \cite{8886467} proposed a partial offloading design and compared the sum-energy consumption performance with other full offloading schemes.


\subsection{{Age-of-Information (AoI) minimization}}\label{subsec:AoI_related_works}

{
The concept of age-of-information (AoI) is a quantitative metric for the freshness of data, defined as time elapsed since the data was generated. 
In Remark \ref{remark:AoI}, we demonstrate that the objective function in our proposed system model is equivalent to AoI minimization. 
Because AoI is closely related to latency minimization, we here introduce the related studies regarding AoI. 
}

{
In \cite{9681851}, a resource optimization problem for energy harvesting networks is addressed using a difference of convex method to minimize AoI. The study in \cite{9256982} tackles a joint user association and resource allocation (UARA) problem by transforming a discrete user association variable into a stochastic policy, aiming to maximize the sum-rate under an AoI constraint. Similarly, the authors of \cite{10155465} propose an AoI-based offloading method for single ES networks, where a stochastic policy is used to minimize AoI.
Further, the study in \cite{10559839} optimizes single-ES task offloading for opportunistic channel access. Multi-base station (BS) task offloading is explored in \cite{9605668} using a deep Q-network (DQN) approach in a centralized setting. In \cite{9723643}, a multi-BS task offloading method is proposed, where a location-based pricing vector is used to minimize monetary cost under an AoI constraint, employing simulated annealing without resource optimization.
}


Yet, none of the research have focused on jointly optimizing UARA and energy consumption, while considering full offloading scenario.
Considering all of the factors is a complicated problem, but is necessary for achieving a solution that can be implemented in real life.
In this work, we find a new solution to delay-minimizing UARA pricing scheme for multi-cell, multi-user, and multi-resource MEC networks.

\renewcommand{\arraystretch}{1.3}

\begin{table}[t]
    \centering
    \caption{Interpretation of the Notations Used in this Paper}
    \adjustbox{width=\if\mycmd1 0.6\else 1\fi\linewidth}{
    \begin{tabular}{p{1cm}p{6.5cm}}
        \toprule
        \multicolumn{1}{c}{Notation} & \multicolumn{1}{c}{Description} \\
        \midrule
        $\mathcal{M}$ & Set of MDs \\
        $\mathcal{E}$ & Set of ESs \\
        $N_{\text{MD}}$ & Number of MDs \\
        $N_{\text{ES}}$ & Number of ESs \\
        $x_{ij}$ & Association indicator of MD $i$ and ES $j$ \\
        $y_{ij}$ & Frequency resource allocation ratio between MD $i$ and ES $j$\\
        $z_{ij}$ & Computing resource allocation ratio between MD $i$ and ES $j$\\
        $R_{ij}$ & Achievable data rate between MD $i$ and ES $j$\\
        $d_i$ & Number of bits of the task generated by MD $i$\\
        $f_i$ & Number of flops required by MD $i$'s task\\
        $\rho_i$ & Ratio of task by MD $i$ dependent on core count\\
        $B_i$ & Remaining battery of MD $i$\\
        $D_{\text{MD},i} $ & Local computing delay of MD $i$\\
        $D_{\text{ES},i} $ & Edge computing delay of MD $i$ associated with ES $j$\\
        $G_i$ & Local computation penalty term of MD $i$\\
        $\alpha$ & Balance coefficient between MD energy preservation and average latency\\
        \bottomrule
    \end{tabular}
    }
    \label{tab:notations}
\end{table}

\section{Conclusion}

\paragraph{Summary}
This paper investigates AI-native MEC systems consisting of energy constrained users, and design a problem to optimize both the system latency and the user energy usage.
Due to the coupled variables, the problem is decomposed into $N_{\text{ES}}$ sub-problems, which is solved to provide an optimal solution.
Through simulations, we show that the proposed scheme outperforms other baseline methods in terms of latency and energy usage. 
Our simulation results illustrate the relationship between the task loads of the MEC network, offloading policy and the energy usage, giving valuable insights on the optimal offloading and resource allocation methods.
As a result, our method achieved a latency improvement of 1.62 times while reducing the energy consumption by 41.2\%.
Our best experiment in latency reduction exhibited a decrease in latency by 5.29 times.
Also, the best experiment in energy consumption reduced energy consumption by up to 64.0\%. 

\paragraph{Limitations}
Although our method can cope with many types of network scenario in the long term, the optimization of users with high mobility still remains unsolved.
Mobile device users in particular, may have high mobility which could be put into consideration in order to better optimize the offloading policy and tackle the handover problems between the ESs.
In future works, minimizing energy usage, latency and the number of handovers between ESs in such scenarios by jointly optimizing resource allocation and user association could provide promising results.

\bibliographystyle{IEEEtran}
\bibliography{refs}

\begin{thebibliography}{10}
\providecommand{\url}[1]{#1}
\csname url@samestyle\endcsname
\providecommand{\newblock}{\relax}
\providecommand{\bibinfo}[2]{#2}
\providecommand{\BIBentrySTDinterwordspacing}{\spaceskip=0pt\relax}
\providecommand{\BIBentryALTinterwordstretchfactor}{4}
\providecommand{\BIBentryALTinterwordspacing}{\spaceskip=\fontdimen2\font plus
\BIBentryALTinterwordstretchfactor\fontdimen3\font minus
  \fontdimen4\font\relax}
\providecommand{\BIBforeignlanguage}[2]{{%
\expandafter\ifx\csname l@#1\endcsname\relax
\typeout{** WARNING: IEEEtran.bst: No hyphenation pattern has been}%
\typeout{** loaded for the language `#1'. Using the pattern for}%
\typeout{** the default language instead.}%
\else
\language=\csname l@#1\endcsname
\fi
#2}}
\providecommand{\BIBdecl}{\relax}
\BIBdecl

\bibitem{10243579}
M.~Diamanti, C.~Pelekis, E.~E. Tsiropoulou, and S.~Papavassiliou, ``Delay
  minimization for rate-splitting multiple access-based multi-server {MEC}
  offloading,'' \emph{{IEEE/ACM} Trans. Networking}, vol.~32, no.~2, pp.
  1035--1047, 2024.

\bibitem{10024305}
F.~Chai, Q.~Zhang, H.~Yao, X.~Xin, R.~Gao, and M.~Guizani, ``Joint multi-task
  offloading and resource allocation for mobile edge computing systems in
  satellite {IoT},'' \emph{{IEEE} Trans. Veh. Technol.}, vol.~72, no.~6, pp.
  7783--7795, 2023.

\bibitem{9712216}
H.~Jiang, X.~Dai, Z.~Xiao, and A.~Iyengar, ``Joint task offloading and resource
  allocation for energy-constrained mobile edge computing,'' \emph{{IEEE}
  Trans. Mobile Comput.}, vol.~22, no.~7, pp. 4000--4015, 2023.

\bibitem{9806318}
A.~Mohajer, M.~Sam~Daliri, A.~Mirzaei, A.~Ziaeddini, M.~Nabipour, and
  M.~Bavaghar, ``Heterogeneous computational resource allocation for {NOMA}:
  Toward green mobile edge-computing systems,'' \emph{{IEEE} Trans. Serv.
  Comput.}, vol.~16, no.~2, pp. 1225--1238, 2023.

\bibitem{10102429}
L.~T. Hoang, C.~T. Nguyen, and A.~T. Pham, ``Deep reinforcement learning-based
  online resource management for {UAV}-assisted edge computing with dual
  connectivity,'' \emph{{IEEE/ACM} Trans. Networking}, pp. 1--16, 2023.

\bibitem{9714257}
C.~Deng, X.~Fang, and X.~Wang, ``{UAV}-enabled mobile-edge computing for ai
  applications: Joint model decision, resource allocation, and trajectory
  optimization,'' \emph{{IEEE} Internet Things J.}, vol.~10, no.~7, pp.
  5662--5675, 2023.

\bibitem{9745305}
W.~He, Y.~Zhang, Y.~Huang, D.~He, Y.~Xu, Y.~Guan, and W.~Zhang, ``Integrated
  resource allocation and task scheduling for full-duplex mobile edge
  computing,'' \emph{{IEEE} Trans. Veh. Technol.}, vol.~71, no.~6, pp.
  6488--6502, 2022.

\bibitem{9496271}
T.~Bahreini, H.~Badri, and D.~Grosu, ``Mechanisms for resource allocation and
  pricing in mobile edge computing systems,'' \emph{{IEEE} Trans. Parallel
  Distrib. Syst.}, vol.~33, no.~3, pp. 667--682, 2022.

\bibitem{9712282}
X.~Zhong, X.~Wang, T.~Yang, Y.~Yang, Y.~Qin, and X.~Ma, ``Potam: A parallel
  optimal task allocation mechanism for large-scale delay sensitive mobile edge
  computing,'' \emph{{IEEE} Trans. Commun.}, vol.~70, no.~4, pp. 2499--2517,
  2022.

\bibitem{9678008}
Y.~Liu, J.~Yan, and X.~Zhao, ``Deep reinforcement learning based latency
  minimization for mobile edge computing with virtualization in maritime uav
  communication network,'' \emph{{IEEE} Trans. Veh. Technol.}, vol.~71, no.~4,
  pp. 4225--4236, 2022.

\bibitem{9214878}
Y.~Chen, Z.~Liu, Y.~Zhang, Y.~Wu, X.~Chen, and L.~Zhao, ``Deep reinforcement
  learning-based dynamic resource management for mobile edge computing in
  industrial internet of things,'' \emph{{IEEE} Trans. Ind. Informat.},
  vol.~17, no.~7, pp. 4925--4934, 2021.

\bibitem{9525179}
M.~Zhao, J.-J. Yu, W.-T. Li, D.~Liu, S.~Yao, W.~Feng, C.~She, and T.~Q.~S.
  Quek, ``Energy-aware task offloading and resource allocation for
  time-sensitive services in mobile edge computing systems,'' \emph{{IEEE}
  Trans. Veh. Technol.}, vol.~70, no.~10, pp. 10\,925--10\,940, 2021.

\bibitem{9087909}
J.~Feng, Q.~Pei, F.~R. Yu, X.~Chu, J.~Du, and L.~Zhu, ``Dynamic network slicing
  and resource allocation in mobile edge computing systems,'' \emph{{IEEE}
  Trans. Veh. Technol.}, vol.~69, no.~7, pp. 7863--7878, 2020.

\bibitem{6497017}
Q.~Ye, B.~Rong, Y.~Chen, M.~Al-Shalash, C.~Caramanis, and J.~G. Andrews, ``User
  association for load balancing in heterogeneous cellular networks,''
  \emph{{IEEE} Trans. Wireless Commun.}, vol.~12, no.~6, pp. 2706--2716, 2013.

\bibitem{8789664}
P.~Wang, Z.~Zheng, B.~Di, and L.~Song, ``Hetmec: Latency-optimal task
  assignment and resource allocation for heterogeneous multi-layer mobile edge
  computing,'' \emph{{IEEE} Trans. Wireless Commun.}, vol.~18, no.~10, pp.
  4942--4956, 2019.

\bibitem{3gpp.36.932}
3GPP, ``{Scenarios and requirements for small cell enhancements for E-UTRA and
  E-UTRAN},'' {3rd Generation Partnership Project (3GPP)}, Technical Report
  (TR) 36.932, 07 2018, version 15.0.0.

\bibitem{korthikanti2023reducing}
V.~A. Korthikanti, J.~Casper, S.~Lym, L.~McAfee, M.~Andersch, M.~Shoeybi, and
  B.~Catanzaro, ``Reducing activation recomputation in large transformer
  models,'' \emph{Proceedings of Machine Learning and Systems}, vol.~5, 2023.

\bibitem{9380899}
R.~D. Yates, Y.~Sun, D.~R. Brown, S.~K. Kaul, E.~Modiano, and S.~Ulukus, ``Age
  of information: An introduction and survey,'' \emph{{IEEE} J. Select. Areas
  Commun.}, vol.~39, no.~5, pp. 1183--1210, 2021.

\bibitem{9738455}
J.~Jang and H.~J. Yang, ``$\alpha$-fairness-maximizing user association in
  energy-constrained small cell networks,'' \emph{{IEEE} Trans. Wireless
  Commun.}, vol.~21, no.~9, pp. 7443--7459, 2022.

\bibitem{3gpp_38_214}
3GPP, ``{NR; Physical layer procedures for data},'' 3rd Generation Partnership
  Project (3GPP), TS 38.214, Jun. 2021.

\bibitem{3gpp_36_872}
------, ``Small cell enhancements for {E-UTRA} and {E-UTRAN} - physical layer
  aspects,'' 3rd Generation Partnership Project (3GPP), TR 36.872, Dec. 2013.

\bibitem{9705104}
T.~Liu, S.~Ni, X.~Li, Y.~Zhu, L.~Kong, and Y.~Yang, ``Deep reinforcement
  learning based approach for online service placement and computation resource
  allocation in edge computing,'' \emph{{IEEE} Trans. Mobile Comput.}, vol.~22,
  no.~7, pp. 3870--3881, 2023.

\bibitem{9547734}
X.~Ma, A.~Zhou, S.~Zhang, Q.~Li, A.~X. Liu, and S.~Wang, ``Dynamic task
  scheduling in cloud-assisted mobile edge computing,'' \emph{{IEEE} Trans.
  Mobile Comput.}, vol.~22, no.~4, pp. 2116--2130, 2023.

\bibitem{8977528}
A.~Khalili, S.~Akhlaghi, H.~Tabassum, and D.~W.~K. Ng, ``Joint user association
  and resource allocation in the uplink of heterogeneous networks,''
  \emph{{IEEE} Commun. Lett.}, vol.~9, no.~6, pp. 804--808, 2020.

\bibitem{7870686}
N.~Trabelsi, C.~S. Chen, R.~El~Azouzi, L.~Roullet, and E.~Altman, ``User
  association and resource allocation optimization in lte cellular networks,''
  \emph{{IEEE} Trans. Netw. Service Manag.}, vol.~14, no.~2, pp. 429--440,
  2017.

\bibitem{8796358}
N.~Zhao, Y.-C. Liang, D.~Niyato, Y.~Pei, M.~Wu, and Y.~Jiang, ``Deep
  reinforcement learning for user association and resource allocation in
  heterogeneous cellular networks,'' \emph{{IEEE} Trans. Wireless Commun.},
  vol.~18, no.~11, pp. 5141--5152, 2019.

\bibitem{9767545}
S.~K. Singh, V.~S. Borkar, and G.~S. Kasbekar, ``User association in dense
  mmwave networks as restless bandits,'' \emph{{IEEE} Trans. Veh. Technol.},
  vol.~71, no.~7, pp. 7919--7929, 2022.

\bibitem{9770179}
X.~Zhang, Z.~Zhang, and L.~Yang, ``Learning-based resource allocation in
  heterogeneous ultradense network,'' \emph{{IEEE} Internet Things J.}, vol.~9,
  no.~20, pp. 20\,229--20\,242, 2022.

\bibitem{6824825}
K.~Shen and W.~Yu, ``Distributed pricing-based user association for downlink
  heterogeneous cellular networks,'' \emph{{IEEE} J. Select. Areas Commun.},
  vol.~32, no.~6, pp. 1100--1113, 2014.

\bibitem{7422839}
Q.~Han, B.~Yang, G.~Miao, C.~Chen, X.~Wang, and X.~Guan, ``Backhaul-aware user
  association and resource allocation for energy-constrained hetnets,''
  \emph{{IEEE} Trans. Veh. Technol.}, vol.~66, no.~1, pp. 580--593, 2017.

\bibitem{8314696}
M.~Chen and Y.~Hao, ``Task offloading for mobile edge computing in software
  defined ultra-dense network,'' \emph{{IEEE} J. Select. Areas Commun.},
  vol.~36, no.~3, pp. 587--597, 2018.

\bibitem{8964328}
M.~Li, N.~Cheng, J.~Gao, Y.~Wang, L.~Zhao, and X.~Shen, ``Energy-efficient
  {UAV}-assisted mobile edge computing: Resource allocation and trajectory
  optimization,'' \emph{{IEEE} Trans. Veh. Technol.}, vol.~69, no.~3, pp.
  3424--3438, 2020.

\bibitem{9050660}
J.~Feng, F.~R. Yu, Q.~Pei, J.~Du, and L.~Zhu, ``Joint optimization of radio and
  computational resources allocation in blockchain-enabled mobile edge
  computing systems,'' \emph{{IEEE} Trans. Wireless Commun.}, vol.~19, no.~6,
  pp. 4321--4334, 2020.

\bibitem{9316898}
H.~Xu, W.~Huang, Y.~Zhou, D.~Yang, M.~Li, and Z.~Han, ``Edge computing resource
  allocation for unmanned aerial vehicle assisted mobile network with
  blockchain applications,'' \emph{{IEEE} Trans. Wireless Commun.}, vol.~20,
  no.~5, pp. 3107--3121, 2021.

\bibitem{8745530}
J.~Zhao, Q.~Li, Y.~Gong, and K.~Zhang, ``Computation offloading and resource
  allocation for cloud assisted mobile edge computing in vehicular networks,''
  \emph{{IEEE} Trans. Veh. Technol.}, vol.~68, no.~8, pp. 7944--7956, 2019.

\bibitem{8886467}
W.~Wu, F.~Zhou, R.~Q. Hu, and B.~Wang, ``Energy-efficient resource allocation
  for secure {NOMA}-enabled mobile edge computing networks,'' \emph{{IEEE}
  Trans. Commun.}, vol.~68, no.~1, pp. 493--505, 2020.

\bibitem{8854900}
Y.~Deng, Z.~Chen, X.~Yao, S.~Hassan, and A.~M.~A. Ibrahim, ``Parallel
  offloading in green and sustainable mobile edge computing for
  delay-constrained iot system,'' \emph{{IEEE} Trans. Veh. Technol.}, vol.~68,
  no.~12, pp. 12\,202--12\,214, 2019.

\bibitem{8962353}
Q.~Zhang, L.~Gui, F.~Hou, J.~Chen, S.~Zhu, and F.~Tian, ``Dynamic task
  offloading and resource allocation for mobile-edge computing in dense cloud
  {RAN},'' \emph{{IEEE} Internet Things J.}, vol.~7, no.~4, pp. 3282--3299,
  2020.

\bibitem{8771176}
L.~Huang, S.~Bi, and Y.-J.~A. Zhang, ``Deep reinforcement learning for online
  computation offloading in wireless powered mobile-edge computing networks,''
  \emph{{IEEE} Trans. Mobile Comput.}, vol.~19, no.~11, pp. 2581--2593, 2020.

\bibitem{9253665}
M.~Tang and V.~W. Wong, ``Deep reinforcement learning for task offloading in
  mobile edge computing systems,'' \emph{{IEEE} Trans. Mobile Comput.},
  vol.~21, no.~6, pp. 1985--1997, 2022.

\bibitem{8854270}
N.~Kiran, C.~Pan, S.~Wang, and C.~Yin, ``Joint resource allocation and
  computation offloading in mobile edge computing for sdn based wireless
  networks,'' \emph{J. Commun.Netw.}, vol.~22, no.~1, pp. 1--11, 2020.

\bibitem{9146372}
S.~Wang, M.~Chen, X.~Liu, C.~Yin, S.~Cui, and H.~Vincent~Poor, ``A machine
  learning approach for task and resource allocation in mobile-edge
  computing-based networks,'' \emph{{IEEE} Internet Things J.}, vol.~8, no.~3,
  pp. 1358--1372, 2021.

\bibitem{8941121}
J.~Feng, F.~Richard~Yu, Q.~Pei, X.~Chu, J.~Du, and L.~Zhu, ``Cooperative
  computation offloading and resource allocation for blockchain-enabled
  mobile-edge computing: A deep reinforcement learning approach,'' \emph{{IEEE}
  Internet Things J.}, vol.~7, no.~7, pp. 6214--6228, 2020.

\bibitem{9149124}
S.~Nath, Y.~Li, J.~Wu, and P.~Fan, ``Multi-user multi-channel computation
  offloading and resource allocation for mobile edge computing,'' in \emph{ICC
  2020 - 2020 IEEE International Conference on Communications (ICC)}, 2020, pp.
  1--6.

\bibitem{10138567}
K.~Wang, D.~Niyato, W.~Chen, and A.~Nallanathan, ``Task-oriented delay-aware
  multi-tier computing in cell-free massive {MIMO} systems,'' \emph{{IEEE} J.
  Select. Areas Commun.}, vol.~41, no.~7, pp. 2000--2012, 2023.

\bibitem{9681851}
Z.~Fang, J.~Wang, Y.~Ren, Z.~Han, H.~V. Poor, and L.~Hanzo, ``Age of
  information in energy harvesting aided massive multiple access networks,''
  \emph{{IEEE} J. Select. Areas Commun.}, vol.~40, no.~5, pp. 1441--1456, 2022.

\bibitem{9256982}
M.~Li, C.~Chen, H.~Wu, X.~Guan, and X.~Shen, ``Age-of-information aware
  scheduling for edge-assisted industrial wireless networks,'' \emph{{IEEE}
  Trans. Ind. Informat.}, vol.~17, no.~8, pp. 5562--5571, 2021.

\bibitem{10155465}
Y.~Jiang, J.~Liu, I.~Humar, M.~Chen, S.~A. AlQahtani, and M.~S. Hossain,
  ``Age-of-information-based computation offloading and transmission scheduling
  in mobile-edge-computing-enabled iot networks,'' \emph{{IEEE} Internet Things
  J.}, vol.~10, no.~22, pp. 19\,782--19\,794, 2023.

\bibitem{10559839}
L.~Wang, R.~Fan, H.~Hu, G.~Wang, and J.~Cheng, ``Age of information
  minimization for opportunistic channel access,'' \emph{{IEEE} Trans.
  Commun.}, pp. 1--1, 2024.

\bibitem{9605668}
X.~Chen, C.~Wu, T.~Chen, Z.~Liu, H.~Zhang, M.~Bennis, H.~Liu, and Y.~Ji,
  ``Information freshness-aware task offloading in air-ground integrated edge
  computing systems,'' \emph{{IEEE} J. Select. Areas Commun.}, vol.~40, no.~1,
  pp. 243--258, 2022.

\bibitem{9723643}
N.~Modina, R.~El-Azouzi, F.~De~Pellegrini, D.~S. Menasche, and R.~Figueiredo,
  ``Joint traffic offloading and aging control in 5g iot networks,''
  \emph{{IEEE} Trans. Mobile Comput.}, vol.~22, no.~8, pp. 4714--4728, 2023.

\end{thebibliography}

\clearpage

\newpage

\if\mycmd1  

\begin{center}{\textsc{\normalsize Supplemental Material of} \\  \textbf{\huge \mytitle }} \end{center}
\vspace{12pt}
\begin{center}{\myauthor} \end{center}
\vspace{12pt}

\else

\begin{center}{\textsc{\normalsize Supplemental Material of} \\  \textbf{\large \mytitle }} \end{center}

\begin{center}{\myauthor} \end{center}

\fi

\pagenumbering{arabic}
\setcounter{page}{1}

\renewcommand{\theequation}{S\arabic{equation}}
\setcounter{equation}{0}

\appendices

\section{Optimal Solution of Communication / Computation Resource Allocation}\label{app:RA_sol}

\subsection{Optimal Solution of Communication Resource Allocation}\label{app:RA_cm}


We first relax problem \ref{eq:P1} by dividing it into $N_{\text{ES}}$ sub-problems, each problem corresponding to an ES, where
\begin{align}
    & \sum_{i\in \mathcal{M}}\sum_{j\in \mathcal{E}}\left( D_{\text{cm}, ij} + D_{\text{ES}, i} \right)x_{ij} \nonumber \\
    &\hspace{50pt minus 1fil}+ \sum_{i\in \mathcal{M}} \left(1-\sum_{j\in \mathcal{E}} x_{ij}\right) \left( D_{\text{MD},i}  + \alpha \frac{G_i}{B_i} \right) \label{P2:obj_2}\\ 
    & = \sum_{j\in \mathcal{E}}\sum_{i\in \mathcal{M}_j} \left( D_{\text{cm}, ij} + D_{\text{ES}, i} \right) \nonumber \\
    &\hspace{50pt minus 1fil}+ \sum_{i\in \mathcal{M}} \left(1-\sum_{j\in \mathcal{E}} x_{ij}\right) \left( D_{\text{MD},i}  + \alpha \frac{G_i}{B_i} \right) \label{P2:obj_3}
\end{align}
Here, $\mathcal{M}_j$ is a subset of $\mathcal{M}$ such that $\mathcal{M}_j=\{i\in \mathcal{M}|x_{ij}=1\}$.
In equation (\ref{P2:obj_3}), we select $x_{ij}$ that belongs in the set $\mathcal{M}_j$.
Hence, the terms behind, which describes the latency and the cost occurring when we compute locally, can be ignored when considering the users from set $\mathcal{M}_j$.
In other words, we're looking at the cases of $i$ such that $x_{ij}=1$.

As mentioned above, ignoring the terms behind, we can choose one of $N_{\text{ES}}$ sub-problems from equation (\ref{P2:obj_3}).
Using the modified objective function (\ref{P2:obj_3}), we can formulate the Lagrangian as

\begin{gather}
    \mathcal{L}_1=\sum_{i\in\mathcal{M}_j}\left( D_{\text{cm}, i} + D_{\text{ES}, i} \right) + \lambda^{(1)} (\sum_{i\in\mathcal{M}_j} y_{ij} - 1) \nonumber \\
    + \lambda^{(2)}\sum_{i\in\mathcal{M}_j}(z_{ij}-1) - \sum_{i\in\mathcal{M}_j} ( \lambda^{(3)}_{i}y_{ij} + \lambda^{(4)}_{i}z_{ij}).  \label{p2:lagrangian}
\end{gather}

From the Lagrangian (\ref{p2:lagrangian}), we derive the KKT conditions of the problem as follows:
\begin{subequations}
    \begin{empheq}[left=\empheqlbrace]{align}
        & \frac{\partial \mathcal{L}_1}{\partial y_{ij}} = -\frac{d_{i}R_{ij}}{(R_{ij} y_{ij})^2} + \lambda^{(1)} - \lambda^{(3)}_{i} = 0 \label{eq:kkt1} \\
        & \frac{\partial \mathcal{L}_1}{\partial z_{ij}} = -\frac{f_i F_{\text{ES},j} Z_{\text{ES},j}}{(F_{\text{ES},j} Z_{\text{ES},j} ze_a)^2}\rho + \lambda^{(1)} - \lambda^{(3)}_{i} = 0 \label{eq:kkt3} \\
        & \lambda^{(1)} (\sum_{i\in\mathcal{M}_j} y_{ij} - 1) = 0 \label{eq:kkt2} \\
        & \lambda^{(2)} (\sum_{i\in\mathcal{M}_j} z_{ij} - 1) = 0 \label{eq:kkt8} \\
        & \lambda^{(3)}_{i}y_{ij} = 0 \label{eq:kkt4} \\
        & \lambda^{(4)}_{i}z_{ij} = 0 \label{eq:kkt5} \\
        & \lambda^{(1)} \geq 0 \label{eq:kkt6} \\
        & \lambda^{(3)}_{i} \geq 0, \; \lambda^{(4)}_{i} \geq 0, \; \lambda^{(2)}_{i} \geq 0 \; \forall i \in \mathcal{M}_j.\label{eq:kkt7}
    \end{empheq}
\end{subequations}

From condition (\ref{eq:kkt1}), we obtain a solution of $y_{ij}$, consisting of the Lagrangian multipliers $\lambda^{(1)}$ and $\lambda^{(3)}_i$.
We divide the cases of possible $y_{ij}$ according to the values of $\lambda^{(1)}$, and examine the cases to solve for the optimal $y_{ij}$.

\textit{1) Case 1 ($\lambda^{(1)}=0$; Invalid Case):} In this case,
\begin{equation}
    y_{ij}=\left( \sqrt{\frac{d_{i} R_{ij}}{-\lambda^{(3)}_{i}}} \right) / R_{ij}. \label{eq:invalid}
\end{equation}
From equation (\ref{eq:invalid}) and condition (\ref{eq:kkt7}), there exists no feasible solution $y_{ij}$.


\textit{2) Case 2 ($\lambda^{(1)}>0$; Valid Case):} In this case, we have
\begin{equation}
    y_{ij}=\left( \sqrt{\frac{d_{i} R_{ij}}{\lambda^{(1)} - \lambda^{(3)}_{i}}} \right) / R_{ij}. \label{eq:valid}
\end{equation}
Because $y_{ij}>0$, $\lambda^{(3)}_{i}=0$ by condition (\ref{eq:kkt4}).
$y_{ij}$ has a feasible solution in this case, and $y_{ij}=\left( \sqrt{\frac{d_{i} R_{ij}}{\lambda^{(1)}}} \right) / R_{ij}$.

Next, to solve for $\lambda^{(1)}$, we use the condition of $y_{ij}$, i.e., $\sum_{i\in\mathcal{M}_j}y_{ij}-1=0$.
Then, we have
\begin{align}
    \sum_{i\in\mathcal{M}_j}y_{ij} ={}& \sum_{i\in\mathcal{M}_j}\left( \sqrt{\frac{1}{\lambda^{(1)}}} \sqrt{\frac{d_{i}}{R_{ij}}} \right) \\
    ={}&\sqrt{ \frac{1}{\lambda^{(1)}} } \sum_{i\in\mathcal{M}_j} \sqrt{ \frac{d_i}{R_{ij}} } =1. \label{eq:sum_y}
\end{align}
Therefore, 
\begin{equation}
    \lambda^{(1)}=\left( \sum_{i\in\mathcal{M}_j}\sqrt{ \frac{d_i}{R_{ij}} } \right)^2. \label{eq:lambda1_y}
\end{equation}
Inserting the value above to equation (\ref{eq:valid}), we obtain the optimal $y_{ij}$ as follows:

\begin{align}
    y_{ij}={}&\left(\sqrt{d_i/R_{ij}} \right) \Big/ \left(\sum_{k\in\mathcal{M}_j}\sqrt{d_k/R_{kj}}\right) \\
    ={}&\left(\sqrt{d_i/R_{ij}} \right) \Big/ \left(\sum_{k\in\mathcal{M}}\sqrt{d_k/R_{kj} } x_{ij}\right).
\end{align}

\subsection{Optimal Solution of Computation}\label{app:RA_cp}

We use the Lagrangian (\ref{p2:lagrangian}) derived from Appendix \ref{app:RA_cm}.
After solving for $ze_a$, we divide the cases of $\lambda^{(1)}$.

\textit{1) Case 1 ($\lambda^{(1)}=0$; Invalid Case):} In this case,
\begin{equation}
    z_{ij}=\left( \sqrt{\frac{f_i F_{\text{ES},j} Z_{\text{ES},j} \rho}{-\lambda^{(3)}_{i}}} \right) \Big/ F_{\text{ES},j} Z_{\text{ES},j}.\label{eq:invalid2}
\end{equation}
From equation (\ref{eq:invalid2}) and condition (\ref{eq:kkt7}), there exists no feasible solution $z_{ij}$, as $z_{ij}$ either becomes imaginary or diverges.

\textit{2) Case 2 ($\lambda^{(1)}>0$; Valid Case):}
As for this case, $z_{ij}$ is expressed as follows:
\begin{equation}
    z_{ij}=\left( \sqrt{\frac{f_i F_{\text{ES},j} Z_{\text{ES},j} \rho}{\lambda^{(1)}}} \right) \Big/ F_{\text{ES},j} Z_{\text{ES},j}. \label{eq:valid2}
\end{equation}
Solving equation (\ref{eq:valid2}) gives $z_{ij}>0$, which leads to $\lambda^{(3)}_{i}=0$ by condition (\ref{eq:kkt5}).
No contradiction occurs in this case, and $z_{ij}=\left( \sqrt{\frac{f_iF_{\text{ES},j} Z_{\text{ES},j} \rho}{\lambda^{(1)}}} \right) \Big/ F_{\text{ES},j} Z_{\text{ES},j}$.

Next, we use the resource allocation condition of $z_{ij}$, i.e., $\sum_{i\in\mathcal{M}_j}z_{ij}-1=0$.
Then, we have the following condition:
\begin{align}
    \sum_{i\in\mathcal{M}_j}z_{ij} ={}& \sum_{i\in\mathcal{M}_j}\left( \sqrt{\frac{1}{\lambda^{(1)}}} \sqrt{\frac{f_{i}\rho}{F_{\text{ES},j} Z_{\text{ES},j}}} \right) \\
    ={}&\sqrt{ \frac{1}{\lambda^{(1)}} } \sum_{i\in\mathcal{M}_j} \sqrt{ \frac{f_i \rho}{F_{\text{ES},j} Z_{\text{ES},j}} } =1. \label{eq:sum_z}
\end{align}
Therefore,
\begin{equation}
    \lambda^{(1)}=\left( \sum_{i\in\mathcal{M}_j}\sqrt{ \frac{f_i \rho}{F_{\text{ES},j} Z_{\text{ES},j}} } \right)^2. \label{eq:lambda1_z}
\end{equation}
Inserting the value above to equation (\ref{eq:valid}), we can obtain the optimal $y_{ij}$ as follows:

\begin{align}
    y_{ij}={}&\left(\sqrt{f_i \rho/F_{\text{ES},j} Z_{\text{ES},j}} \right) \Big/ \left(\sum_{k\in\mathcal{M}_j}\sqrt{f_k/F_{\text{ES},j} Z_{\text{ES},j}}\right)\nonumber \\
    ={}&\left(\sqrt{f_i \rho/F_{\text{ES},j} Z_{\text{ES},j}} \right) \Big/ \left(\sum_{k\in\mathcal{M}}\sqrt{f_k/F_{\text{ES},j} Z_{\text{ES},j} } x_{ij}\right).
\end{align}

\section{Proof of Theorem \ref{thm:duality} \label{sec:appendix_A} }

In the Lagrangian duality, the weak duality holds for all feasible solutions, i.e., $d^*\le f^*$ (minimization problem), where $d^*$ and $f^*$ denote the optimal dual function and optimal primal function, respectively. 
However, in the proposed method, the variable $\mathbf{X}$ recovered by the dual approach cannot meet the auxiliary constraints \eqref{eq:P6_const_d} and \eqref{eq:P6_const_e}. 
Thus, there is a gap between the objective function of the primal problem \ref{eq:P6} and the actual delay corresponding to $\mathbf{X}$, which are denoted as:
\begin{itemize}
    \item The objective function of the primal problem \ref{eq:P6}: $\sum_{j\in\mathcal{E}} (a_j + b_j)+ \sum_{i\in\mathcal{M}}\sum_{j\in\mathcal{E}}c_{ij}x_{ij}$.
    \item Actual delay with penalized energy consumption corresponding to $\mathbf{X}$: 
    $$
    \begin{aligned}
        &\sum_{j\in\mathcal{E}} \left(\sum_{i\in\mathcal{M}}\sqrt{\hat{D}_{\text{ES}(s),ij}}x_{ij}\right)^2  + \sum_{j\in\mathcal{E}}\left(\sum_{i\in\mathcal{M}}\sqrt{\hat{D}_{\text{cm},ij}}x_{ij}\right)^2\\
        &+ \sum_{i\in\mathcal{M}}\sum_{j\in\mathcal{E}}c_{ij}x_{ij}.
    \end{aligned}
    $$
\end{itemize}

Here, we first show that the weak duality still holds for the actual delay if we use this instead of the objective function of the problem \ref{eq:P6}. 

\subsection{Conservation of the Weak Duality Condition}

Denoting the solution we obtained from Lagrangian dual problem \ref{eq:P7} as $\boldsymbol{\mu}^*$ and $\boldsymbol{\nu}^*$, we can represent the maximum value of the Lagrangian dual function as $d^* = g(\boldsymbol{\mu}^*,\boldsymbol{\nu}^*)$.
Here, we denote the global optimal solution minimizing the objective function in \ref{eq:P5} as  $\hat{f}^*$.
Denoting the solution we obtained as $\hat{f}$, the following inequality holds: 
\begin{equation}
    \begin{aligned}\label{eq:weak_dual_appendix}
        d^* &= \max_{\boldsymbol{\mu},\boldsymbol{\nu}}\min_{\mathbf{a}, \mathbf{b}, \mathbf{X}} \mathcal{L}(\mathbf{a}, \mathbf{b}, \mathbf{X}, \boldsymbol{\mu},\boldsymbol{\nu})\\
        & \le \max_{\boldsymbol{\mu},\boldsymbol{\nu}}\min_{\mathbf{a},\mathbf{b},\mathbf{X}\in\mathcal{F}}\mathcal{L}(\mathbf{a}, \mathbf{b}, \mathbf{X}, \boldsymbol{\mu},\boldsymbol{\nu}) \\
        & = \min_{\mathbf{a},\mathbf{b}, \mathbf{X}\in\mathcal{F}} 
        \sum_{j\in\mathcal{E}} \left(\sum_{i\in\mathcal{M}}\sqrt{\hat{D}_{\text{ES}(s),ij}}x_{ij}\right)^2 \\
        & ~~~~ + \sum_{j\in\mathcal{E}}\left(\sum_{i\in\mathcal{M}}\sqrt{\hat{D}_{\text{cm},ij}}x_{ij}\right)^2\\
        & ~~~~ + \sum_{i\in\mathcal{M}}\sum_{j\in\mathcal{E}}c_{ij}x_{ij}\\
        & = \hat{f}^*  \le \hat{f}.
    \end{aligned}
\end{equation}
where $\mathcal{F}$ denotes a space where the constraints \eqref{eq:P6_const_d} and \eqref{eq:P6_const_e} hold.

\subsection{Duality Gap\label{subsec:duality_gap}}

In the previous section, we show that the weak duality condition still holds for the solution with recovered $\mathbf{X}$. 
Then, inspired by the inequality in \eqref{eq:weak_dual_appendix}, we show the duality gap of the solution for recovered $\mathbf{X}$.
Let us represent the solution $\hat{f}$ as 
\begin{align}
    \hat{f} = &\sum_{j\in\mathcal{E}} \left(\sum_{i\in\mathcal{M}}\sqrt{\hat{D}_{\text{ES}(s),ij}}x_{ij}\right)^2  \nonumber\\
     & + \sum_{j\in\mathcal{E}}\left(\sum_{i\in\mathcal{M}}\sqrt{\hat{D}_{\text{cm},ij}}x_{ij}\right)^2 + \sum_{i\in\mathcal{M}}\sum_{j\in\mathcal{E}}c_{ij}x_{ij} \\ 
     =&\sum_{j\in\mathcal{E}}(\sqrt{a_j} - \Delta_j^{(1)})^2+\sum_{j\in\mathcal{E}}(\sqrt{b_j} - \Delta_j^{(2)})^2 + \sum_{i\in\mathcal{M}}\sum_{j\in\mathcal{E}}c_{ij}x_{ij}\nonumber,
\end{align}
where $\Delta_j^{(1)} = \sqrt{a_j}-\left(\sum_{i\in\mathcal{M}}\sqrt{\hat{D}_{\text{cm},ij}}x_{kj}\right)$ and $\Delta_j^{(2)} = \sqrt{b_j}-\left(\sum_{i\in\mathcal{M}}\sqrt{\hat{D}_{\text{ES}(s),ij}}x_{ij}\right)$.
Then, we have the optimality gap as below:
\begin{align}\label{eq:duality_gap}
    & \hat{f} - \hat{f}^* \\
    \le & \hat{f} - d^* \\
    = & \sum_{j\in\mathcal{E}}\left(\sqrt{a_j} - \Delta_j^{(1)}\right)^2 + \sum_{j\in\mathcal{E}}\left(\sqrt{b_j} - \Delta_j^{(2)}\right)^2 \\ 
    & ~~~ -\sum_{j\in\mathcal{E}}a_j -\sum_{j\in\mathcal{E}}b_j \\ 
    \le & \sum_{j\in\mathcal{E}}\left(\left(\Delta_j^{(1)}\right)^2 + \left(\Delta_j^{(2)}\right)^2 - 2\sqrt{a_j}\Delta_j^{(1)} - 2\sqrt{b_j}\Delta_j^{(2)}\right).\nonumber
\end{align}
Finally, we confirm that the optimality gap of the proposed scheme is bounded by $\mathcal{O}\left(\left(\Delta_j^{(1)}\right)^2 +\left(\Delta_j^{(2)}\right)^2\right)$.

\section{Proof of Proposition \ref{prop:A} \label{sec:appendix_B} }

Here, we prove the convergence of the proposed scheme. 
Let us define the optimal solution to Problem \ref{eq:P7} as $\boldsymbol{\mu}^*$ and $\boldsymbol{\nu}^*$.
To solve the problem \ref{eq:P7}, we apply the sub-gradient ascent in \eqref{eq:sub-grad-descent}. which is represented by 
\begin{equation}
    \begin{aligned}
        \boldsymbol{\mu}^{(t+1)} = \boldsymbol{\mu}^{(t)} + \eta_1\nabla_{\boldsymbol{\mu}} g(\boldsymbol{\mu},\boldsymbol{\nu}), \forall t\in\{0, 1, ... \}\\
        \boldsymbol{\nu}^{(t+1)} = \boldsymbol{\nu}^{(t)} + \eta_2\nabla_{\boldsymbol{\nu}} g(\boldsymbol{\mu},\boldsymbol{\nu}), \forall t\in\{0, 1, ... \}.
    \end{aligned}
\end{equation}

To show the convergence, we assume $\Vert \nabla_{\boldsymbol{\mu}}g(\boldsymbol{\mu},\boldsymbol{\nu})\Vert_2\le \delta_1$, $\Vert \nabla_{\boldsymbol{\nu}}g(\boldsymbol{\mu},\boldsymbol{\nu})\Vert_2\le \delta_1$, $\Vert \boldsymbol{\mu}^{(0)} - \boldsymbol{\mu}^*\Vert_2 \le \delta_2$, and $\Vert \boldsymbol{\nu}^{(0)} - \boldsymbol{\nu}^*\Vert_2 \le \delta_2$. For brevity of the notation, we denote $\nabla_{\boldsymbol{\mu}}g(\boldsymbol{\mu},\boldsymbol{\nu})$ and $\nabla_{\boldsymbol{\nu}}g(\boldsymbol{\mu},\boldsymbol{\nu})$ as $\Delta_{\boldsymbol{\mu}}$ and $\Delta_{\boldsymbol{\nu}}$, respectively. Then, we have
\begin{align}
    &\Vert \boldsymbol{\mu}^{(t+1)}-\boldsymbol{\mu}^*\Vert_2 - \Vert \boldsymbol{\mu}^{(t)}-\boldsymbol{\mu}^*\Vert_2 \\
    ={}& \eta_1^2 \Vert \Delta_{\boldsymbol{\mu}} \Vert_2  - 2\eta_1(\Delta_{\boldsymbol{\mu}})^\mathrm{T} (\boldsymbol{\mu}^*-\boldsymbol{\mu}^{(t)}) \nonumber
\end{align}
and
\begin{align}
    &\Vert \boldsymbol{\nu}^{(t+1)}-\boldsymbol{\nu}^*\Vert_2 - \Vert \boldsymbol{\nu}^{(t)}-\boldsymbol{\nu}^*\Vert_2 \\
    ={}& \eta_2^2 \Vert \Delta_{\boldsymbol{\nu}} \Vert_2  - 2\eta_2(\Delta_{\boldsymbol{\nu}})^\mathrm{T} (\boldsymbol{\nu}^*-\boldsymbol{\nu}^{(t)} ), \nonumber
\end{align}
where the inequality holds due to the convexity.  
Let us define $g_{\text{max},T}=\max_{t=1,...,T}g(\boldsymbol{\mu},\boldsymbol{\nu})$. 
By combining the inequalities from $t=0$ to $T$, we have 
\begin{align} \label{eq:ineq_prop1}
    &2(T+1) \left(g^* - g_{\text{max},T} \right)\nonumber \\
    &\le 2\sum_{t=0}^{T} \left(g^*-g(\boldsymbol{\mu},\boldsymbol{\nu}) \right)\nonumber\\
    &\le 2\sum_{t=0}^{T} \left( (\nabla_{\boldsymbol{\mu}})^\mathrm{T}(\boldsymbol{\mu}^* - \boldsymbol{\mu}^{(t)}) + (\nabla_{\boldsymbol{\nu}})^\mathrm{T}(\boldsymbol{\nu}^* - \boldsymbol{\nu}^{(t)})  \right)\nonumber\\
    &= \eta_1 \sum_{t=0}^{T}\Vert \Delta_{\boldsymbol{\mu}}\Vert_2 + \eta_2 \sum_{t=0}^{T}\Vert \Delta_{\boldsymbol{\nu}}\Vert_2 \\ 
    & ~~ + \frac{1}{\eta_1}\Vert \boldsymbol{\mu}^{(0)} - \boldsymbol{\mu}^*\Vert_2 - \frac{1}{\eta_1}\Vert \boldsymbol{\mu}^{(t+1)} - \boldsymbol{\mu}^*\Vert_2\nonumber\\
    & ~~ + \frac{1}{\eta_2}\Vert \boldsymbol{\nu}^{(0)} - \boldsymbol{\nu}^*\Vert_2 - \frac{1}{\eta_2}\Vert \boldsymbol{\nu}^{(t+1)} - \boldsymbol{\nu}^*\Vert_2\nonumber\\
    &\le (\eta_1+\eta_2)(T+1)\delta_1 + \left(\frac{1}{\eta_1}+\frac{1}{\eta_2}\right)\delta_2. \nonumber
\end{align}
Finally, the inequality in \eqref{eq:ineq_prop1} can be rewritten as
\begin{equation}
    g^*  - g_{\text{max},T} \le \frac{(\eta_1+\eta_2)\delta_1}{2} + \frac{\delta_2}{2(T+1)}\left(\frac{1}{\eta_1}+\frac{1}{\eta_2}\right).
\end{equation}
By choosing $\eta_1 = \eta_2 = \sqrt{\frac{\delta_2}{(T+1)\delta_1}}$, the following condition holds: $g^* - g_{\text{max},T}  \le \epsilon$, where $\epsilon =\sqrt{\frac{\delta_1\delta_2}{(T+1)}}$. Thus, for any $\epsilon>0$, there exist $\eta_1$, $\eta_2$, and $T\in\mathcal{O}(\frac{1}{\epsilon^2})$ satisfying $g^* - g_{\text{max},T}   \le \epsilon$.

\end{document}